\documentclass[12pt, aps, superscriptaddress, nofootinbib, onecolumn, tightenlines]{revtex4-2}
\setcitestyle{super}
\usepackage[utf8]{inputenc}

\usepackage{graphicx}
\usepackage{bm}
\graphicspath{ {./figures/} }
\usepackage{textcomp,gensymb}
\usepackage{lineno}

\usepackage{prettyref}
\newrefformat{fig}{Fig.~\ref{#1}}
\newrefformat{exfig}{Extended Data Fig.~\ref{#1}}

\usepackage{ltablex}  %
\usepackage[normalem]{ulem}
\usepackage{color}
\usepackage{units}
\usepackage{times}
\usepackage{amsmath,amssymb}
\usepackage{comment}
\usepackage{hyperref}
\hypersetup{
    colorlinks=true,
    linkcolor=blue,
    filecolor=magenta,      
    urlcolor=cyan,
    citecolor=red,
    pdftitle={Observation of Cooper-pair density modulation state},
}
\usepackage{wasysym}
\usepackage{dsfont}
\usepackage{xcolor}

\usepackage{amsmath,amssymb}
\usepackage{mathtools}
\usepackage{booktabs} 
\usepackage{longtable}

\renewcommand{\figurename}{\textbf{Fig.}}
\newcommand{\beginsupplement}{%
        \setcounter{table}{0}
        \renewcommand{\table}{\arabic{table}|}%
        \renewcommand{\thesection}{\arabic{section}}
        \setcounter{figure}{0}
        \renewcommand{\figurename}{\textbf{Supplementary Fig.}}
     }

\newcommand{\caltechPH}{Department of Physics, California Institute of Technology, Pasadena, California 91125, USA}
\newcommand{\caltechAPH}{T. J. Watson Laboratory of Applied Physics, California Institute of Technology,
  1200 East California Boulevard, Pasadena, California 91125, USA}

\newcommand{\caltechIQIM}{Institute for Quantum Information and Matter, California Institute of Technology, Pasadena, California 91125, USA}

\newcommand{\nims}{National Institute for Materials Science, Namiki 1-1, Tsukuba, Ibaraki 305 0044, Japan}

\begin{document}

\title{Observation of Cooper-pair density modulation state}

\author{Lingyuan Kong}
\email{Correspondence: lykong@caltech.edu}
\affiliation{\caltechAPH}
\affiliation{\caltechIQIM}

\author{Michał Papaj}
\affiliation{Department of Physics, University of California, Berkeley, CA 94720, USA}

\author{Hyunjin Kim}
\affiliation{\caltechAPH}
\affiliation{\caltechIQIM}
\affiliation{\caltechPH}

\author{Yiran Zhang}
\affiliation{\caltechAPH}
\affiliation{\caltechIQIM}
\affiliation{\caltechPH}

\author{Eli Baum}
\affiliation{\caltechAPH}
\affiliation{\caltechIQIM}

\author{Hui Li}
\affiliation{Department of Materials Science and Engineering, Northwestern University, 2220 Campus Drive, Evanston, Illinois 60208, USA.}

\author{Kenji Watanabe}
\affiliation{\nims}

\author{Takashi Taniguchi}
\affiliation{\nims}

\author{Genda Gu}
\affiliation{Condensed Matter Physics and Materials Science Department, Brookhaven National Laboratory, Upton, NY 11973, USA.}

\author{Patrick A. Lee}
\affiliation{Department of Physics, Massachusetts Institute of Technology, Cambridge, MA 02139, USA}

\author{Stevan Nadj-Perge}
\email{Correspondence: s.nadj-perge@caltech.edu}
\affiliation{\caltechAPH}
\affiliation{\caltechIQIM}

\begin{abstract}
\end{abstract}
\maketitle

\textbf{Superconducting states that break space-group symmetries of the underlying crystal can exhibit nontrivial 
spatial modulation of the order parameter. Previously, such remarkable states were intimately associated with the 
breaking of translational symmetry\cite{fulde1964superconductivity,larkin1964nonuniform}, giving rise to the density-wave orders\cite{himeda2002stripe,berg2007dynamical,lee2014amperean,fradkin2015colloquium,fernandes2019intertwined,agterberg2020physics}, with wavelengths spanning several unit cells\cite{hamidian2016detection,ruan2018visualization,du2020imaging,liu2021discovery,chen2021roton,wang2021scattering, 
chen2022identification,gu2023detection,zhao2023smectic,liu2023pair,wei2023discovery}. 
However, a related basic concept has been long overlooked: when only intra-unit-cell symmetries of 
the space group are broken, the superconducting states can display a distinct type of nontrivial modulation 
preserving long-range lattice translation. Here, we refer to this new concept as the pair density modulation (PDM), and report the first observation of a PDM state in exfoliated thin flakes of iron-based superconductor FeTe$_{\text{0.55}}$Se$_{\text{0.45}}$. 
Using scanning tunneling microscopy, we discover robust superconducting gap modulation with the wavelength corresponding 
to the lattice periodicity and the amplitude exceeding 30$\bf \%$ of the gap average. 
Importantly, we find that the observed modulation originates 
from the large difference in superconducting gaps on the two nominally equivalent iron sublattices. 
The experimental findings, backed up by model calculations, suggest that in contrast to the density-wave orders, 
the PDM state is driven by the interplay of sublattice symmetry breaking and a peculiar nematic distortion specific to the thin flakes.
Our results establish new frontiers 
for exploring the intertwined orders in strong-correlated electronic systems and open a new chapter for iron-based superconductors.}

\begin{figure}[p]
\begin{center}
   \includegraphics[width=15cm]{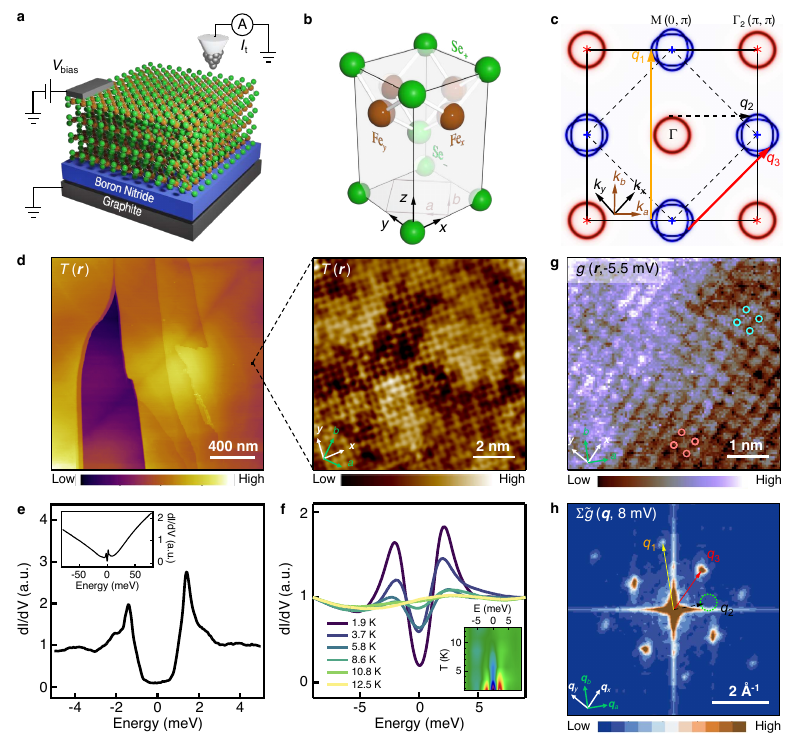}
\end{center}
\caption{{\bf Electronic structure and superconductivity of Fe(Te,Se) thin flakes.} {\bf a}, Schematic of the experiment. 
{\bf b}, A unit cell of Fe(Te,Se) crystal structure. It contains two selenium atoms,  Se$_{+}$ and Se$_{-}$, placed within the top and bottom layers of the ``sandwich" block, and two middle-layer iron atoms, Fe$_{x}$ and Fe$_{y}$, placed along the $\textit{x}$- and $\textit{y}$- bonds of the top selenium atoms. Tellurium atoms substitute for selenium sites (not shown). Axis $\textit{x}/\textit{y}$ (or $\textit{a}/\textit{b}$) are along the nearest Se$_{+}$-Se$_{+}$ (or  Fe$_{x}$- Fe$_{y}$) directions. {\bf c}, Typical Fermi surface of bulk FeTe$_{\text{0.55}}$Se$_{\text{0.45}}$ 
single crystals. The solid and dashed squares indicate 1-Fe and 2-Fe Brillouin zone (BZ), respectively; the BZ orientation 
corresponds to the lattice orientation in ({\bf b}). The electron and hole Fermi pockets are depicted in blue and red 
colors. Three dominant scattering vectors are marked as $\boldsymbol{q}_{1,2,3}$. 
Note that $\boldsymbol{q}_{3}$ is a reciprocal lattice vector, with the Bragg peak marked as ($\pi$,$\pi$). {\bf d}, Left 
panel: large STM topography (1.9$\times$1.9 $\mu$m$^{\text{2}}$) measured on Fe(Te,Se) thin flakes showing several flat and clean surfaces
as well as multiple atomically-sharp monolayer steps. Right panel: atom-resolved STM topography (30$\times$30 
nm$^{\text{2}}$), showing a checkerboard pattern formed by Se$_{+}$ or Te$_{+}$ atoms. {\bf e}, A dI/dV spectrum shows U-shaped hard 
SC gap. Inset: wide-bias range dI/dV spectrum. {\bf f}, Temperature dependence of spatially-averaged dI/dV spectra. Inset: an 
interpolated false-color plot of ({\bf f}). {\bf g}, A typical conductance map, $g(\boldsymbol{r},E)$, for energy above SC gap. 
Solid circles mark the positions of Se$_{+}$ atoms.  Conductance maxima appear at Se$_{-}$ sites. 
{\bf h}, Magnitude of energy-integrated Fourier transform (FT) of conductance map within $\pm$8 mV. 
The absence of prominent features around $(\pi,0)$ indicates that there is no scattering corresponding 
to $\boldsymbol{q}_{2}$. See definitions of all the symbols in Supplementary Table~\ref{table: tabs1}.
}

\label{fig: fig1}
\end{figure}

Iron-based superconductors (FeSC) exhibit multiple electronic orders involving spin, orbital, and charge degrees of 
freedom that are, in many instances, intertwined with superconductivity\cite{fernandes2022iron,fernandes2014drives}. 
Moreover, the electrons in FeSCs are strongly correlated\cite{yin2011kinetic,checkelsky2024flat}, leading to shallow electronic 
bands, especially in an archetypal compound FeTe$_{\text{0.55}}$Se$_{\text{0.45}}$, where the Fermi energy was observed
within only several millielectronvolt\cite{lubashevsky2012shallow}. These salient features make this family of 
compounds a good candidate for exploring new superconducting phenomena\cite{shibauchi2020exotic,kong2019half}. 
While previous scanning tunneling microscopy (STM) measurements focused on the bulk iron-based superconductors and films grown by molecular beam epitaxy, 
here, for the first time, we study this material system in the form of exfoliated thin flakes. 
The schematic of the experiment 
and unit cell of Fe(Te,Se) are shown in \prettyref{fig: fig1}a,b. A large flake of Fe(Te,Se) is placed on a hexagonal 
boron nitride, contacted by a graphite electrode, and its surface is probed using the STM tip. 
The main technical challenge in studying flakes of these materials arises due to their air sensitivity 
and the great care that has to be taken during the sample preparation (see Methods for details, see also Extended Data \prettyref{fig: figdeviceconfig}, Supplementary \prettyref{fig: figfab} and Supplementary \prettyref{fig: fighistory}). 
Large area topographic scans reveal clean micron size surfaces accompanied by multiple step edges 
(\prettyref{fig: fig1}d). Remarkably, the measured step edge heights are
about 18$\%$ larger compared to the bulk measurements 
(Supplementary \prettyref{fig: figcheight}d), implying the nontrivial role of the 
exfoliation process, inducing effective negative pressure (Supplementary \prettyref{fig: figcheight}f) 
and driving the system into an unexplored phase space of FeSC compounds. Here, we report measurements 
primarily focused on the 50-nm flake; however, the properties of flakes in the 25-210 nm range were also examined, the details are shown in Methods and Supplementary Information (SI).

\noindent {\bf Modified microscopic properties}

First, we focus on basic spectroscopic characterization. The wide-bias-range differential conductance (dI/dV) spectrum 
(inset of \prettyref{fig: fig1}e) shows a characteristic V-shaped local density of states (LDOS) within ±80 meV, consistent with the 
semimetallic band structure. A high-resolution dI/dV spectrum focusing on the small energies around the Fermi level 
reveals a U-shaped hard superconducting (SC) gap, indicating a nodeless SC order parameter (\prettyref{fig: fig1}e, 
Extended Data \prettyref{fig: figHRSCgap}, see also SI Section \ref{SI:multipeak}) and temperature-dependent measurements 
show the SC gap collapsing around 11 K (\prettyref{fig: fig1}f). Moreover, the SC coherence length ($\xi_{\Delta_{0}}$) 
of thin flakes was determined to be around 4.4 nm (\prettyref{fig: fig3}k). These observations are quantitatively in line with 
previous measurements on bulk FeTe$_{\text{0.55}}$Se$_{\text{0.45}}$\cite{hanaguri2010unconventional,wang2020evidence}, 
as well as recent transport measurements of thin flakes that show no dramatic differences from bulk 
properties\cite{tang2019quasi}. 

However, quasiparticle interference (QPI) reveals the striking difference in electronic structure 
between the flakes measured here and the bulk material. These measurements are carried out by collecting a dI/dV 
spatial map alongside the topography (\prettyref{fig: fig1}g) and performing a two-dimensional 
Fourier transform (FT) of the map (\prettyref{fig: fig1}h). The QPI measurements typically reveal the details 
of the electronic structure, such as dominant scattering vectors in the Brillouin zone, which are not accessible 
by transport measurements.
Note that in bulk FeTe$_{\text{0.55}}$Se$_{\text{0.45}}$, Fermi surface 
consists of one hole pocket and two electron pockets\cite{morfoot2023resurgence} and is characterized by three dominant scattering vectors 
(\prettyref{fig: fig1}c), as established in previous STM experiments\cite{hanaguri2010unconventional,chen2019direct}.

Surprisingly, in contrast to bulk, our measurements show that the $\boldsymbol{q}_{2}$ vector, connecting $\Gamma$ and M pockets, 
disappears within ±8 meV bias voltage range (\prettyref{fig: fig1}h), indicating no coexisting electron and hole pockets 
within this energy window. The Fermi level only crosses either hole-like $\Gamma$ pocket or electron-like M pockets, 
and as a consequence, the shortest scattering vector is $\boldsymbol{q}_{3}$, instead of $\boldsymbol{q}_{2}$.
Despite the modified Fermi surface, the  SC  order parameter appears still to be sign-changing, as suggested by measurements 
on the monolayer step edge (Supplementary \prettyref{fig: figedgekillsc}). In addition, phase-referenced QPI measurement (see SI Section \ref{SI:PRQPI} for further details) reveals a negative signal around (0,0), implying that sign changing appears for a small 
$\boldsymbol{q}$ (Supplementary \prettyref{fig: figPRQPI}a-c). These observations place constraints on the SC order parameter in thin flakes
especially as some of the thermodynamic quantities remain comparable to bulk crystals while several microscopic properties are 
strongly modified (see Supplementary \prettyref{fig: figPRQPI}d-i for further discussions).

\noindent {\bf Robust gap modulation} 

We now focus on the spatial characterization of the SC gap. In bulk FeTe$_{\text{0.55}}$Se$_{\text{0.45}}$, 
previous experiments observed that the gap is strongly inhomogeneous at a length scale of $\xi_{\Delta_{0}}$ 
(about ten times lattice constant)\cite{cho2019strongly}, but roughly uniform below that scale (Supplementary 
\prettyref{fig: figbulk}). In contrast, here we find the dI/dV spectrum of the thin flakes exhibits 
substantial modulation of the superconducting gap. The modulation is observed almost everywhere in the thin flake regions 
(Supplementary \prettyref{fig: fignevig}), including the areas with high impurity concentrations (see Methods and Extended Data \prettyref{fig: figlinecutdirty}), 
but disappears at the sufficiently thick flakes (Supplementary \prettyref{fig: figthickflake}). Also, the 
periodicity is present even at high magnetic fields close to pinned vortices (see Methods and Extended Data \prettyref{fig: figlinecutfield}).  

\begin{figure}[p]
\begin{center}
    \includegraphics[width=15cm]{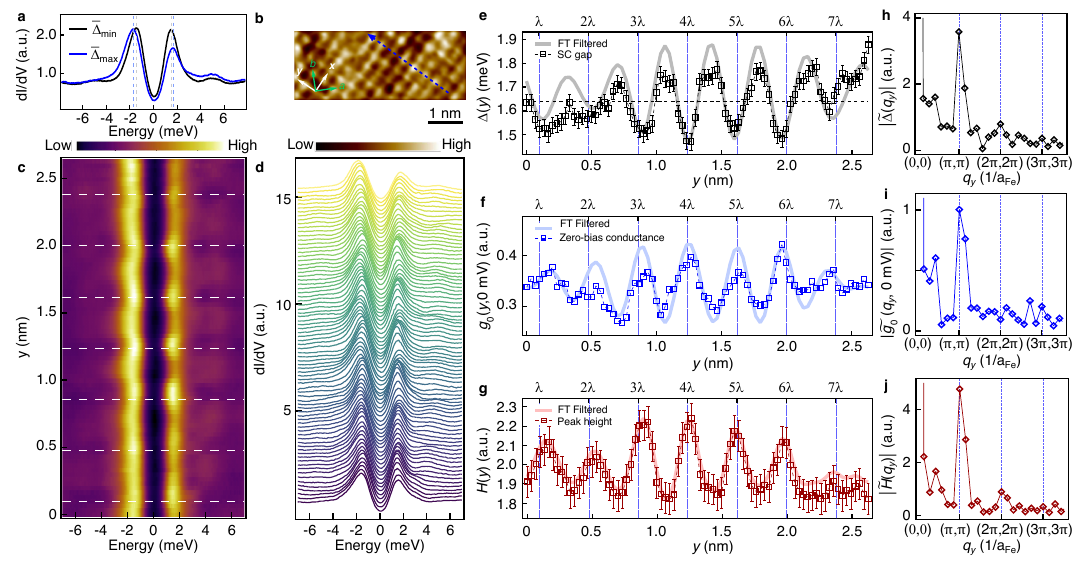}
\end{center}
\caption{
{\bf  Periodic modulation of superconducting gap.} {\bf a}, Spatially-averaged dI/dV spectra at gap maximum (blue) and minimum 
(black) sites of the linecut. {\bf b}, Atom-resolved topography (5 $\times$ 2 nm$^{\text{2}}$) at the same area dI/dV 
spectra were measured. {\bf c}, False-color plot of a dI/dV linecut measured along $\textit{y}$-axis [dashed blue line in 
({\bf b})]. The SC gap is minimized at Se$_{+}$ atom sites (dashed white lines). The linecut was measured on a Y-Domain (see 
\prettyref{fig: fig3} for the definition). {\bf d}, The waterfall plot of ({\bf c}). The curves are offset for clarity. 
{\bf e-g}, The spatial variation of SC gap, $\Delta(\textit{y})$; the zero-bias conductance, $g_{0}(\textit{y})$; and the coherence 
peak height, $H(\textit{y})$, respectively. The semi-transparent solid lines are obtained by FT filtering of the raw data. 
The horizontal dashed line in ({\bf e}) indicates the average value of SC gap in the linecut ($\bar{\Delta}$). {\bf h-j}, 
Magnitude of FT of raw data in ({\bf e-g}). The modulation ratio $|\Delta_{\boldsymbol{\rm P}_{\rm Y}}|/\bar{\Delta}$ is 7$\%$ in this linecut.}  
\label{fig: fig2}
\end{figure}

\begin{figure}[p]
\begin{center}
    \includegraphics[width=15cm]{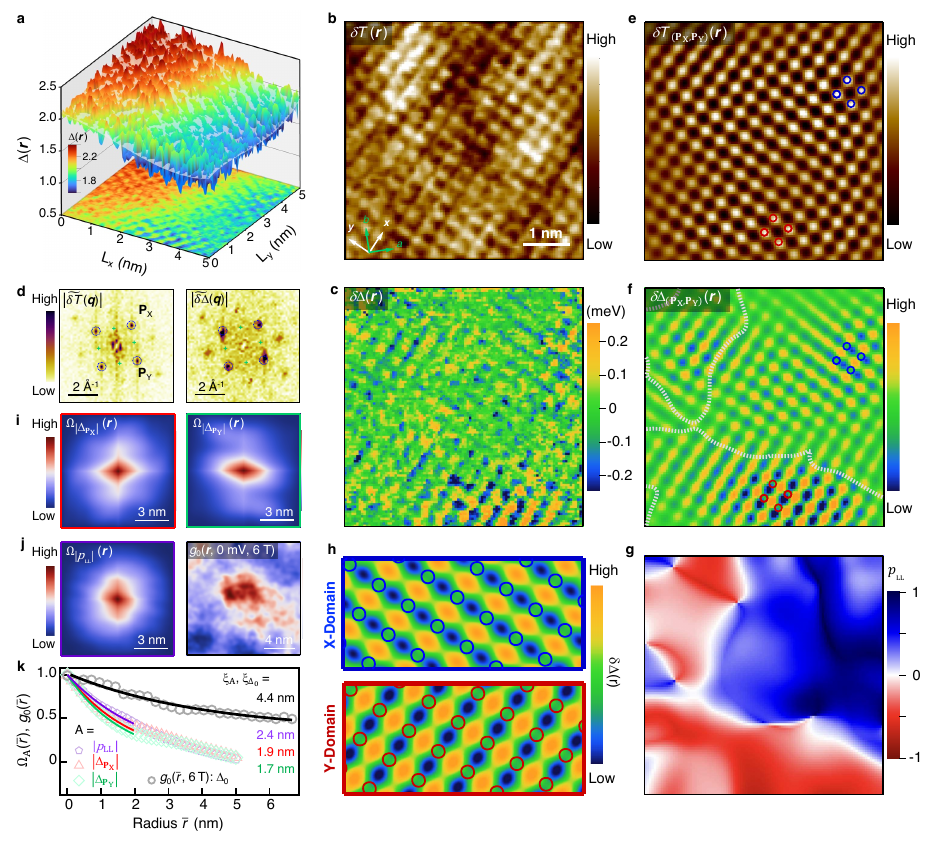}
\end{center}
\caption{
{\bf  Imaging the pair density modulation state.} SI-STM data on a 5$\times$5 nm$^{\text{2}}$ area with a 100$\times$100 grid. At each point, a dI/dV spectrum is measured within $\pm$8 mV. {\bf a}, The SC gap map, $\Delta(\boldsymbol{r})$. 
The modulating PDM component, $\delta\Delta(\boldsymbol{r})$, coexists with non-modulating component, $\Delta_{0}(\boldsymbol{r})$. {\bf b}, Topographic image, $\delta T(\boldsymbol{r})$, with atomic resolution acquired simultaneously with data in ({\bf a}). {\bf c}, The PDM modulation component, $\delta\Delta(\boldsymbol{r})$. {\bf d}, The FT magnitude of ({\bf b}) and ({\bf c}). The green crosses indicate $(\pi,0)$ and equivalent locations. The two PDM vectors are marked as $\pm$$\boldsymbol{\rm P}_{\rm X}$ and  $\pm$$\boldsymbol{\rm P}_{\rm Y}$.  {\bf e, f}, FT filtered images at  $\pm$$\boldsymbol{\rm P}_{\rm X}$ and  $\pm$$\boldsymbol{\rm P}_{\rm Y}$, of ({\bf b}) and ({\bf c}) respectively. 
The red and blue circles mark the positions of Se$_{+}$ atoms (the same circles are plotted in \prettyref{fig: fig1}{g} and \prettyref{fig: fig4}{d,e}). Gray lines in ({\bf f}) show domain boundaries. Within each domain, the gap maxima are pinned to 
one iron sublattice (Fe$_x$, or Fe$_y$), while the gap minima appear at the other (Fe$_y$, or Fe$_x$). We define these as X-Domain (or Y-Domain) (see also panel {\bf h}). {\bf g}, Lattice-lock-in polarization of the PDM state, $p_{_{\text{LL}}}(\boldsymbol{r})$ (see the definition in Method, and the calculation details in Supplementary \prettyref{fig: figpll}). $p_{_{\text{LL}}}(\boldsymbol{r})$ is +1 (or -1), if gap maxima are perfectly locked to Fe$_{x}$ (or Fe$_{y}$) (see also Supplementary \prettyref{fig: figpll_demo}). The domain boundaries can be traced from the zero contour, shown as the gray dashed lines in ({\bf f}). The Gaussian cut-off length, $\sigma$, was set to 1.3 nm for ({\bf e-g}). {\bf h}, Schematic of PDM state in X- and Y-domain. The circles mark the positions of Se$_{+}$ atoms. The PDM state also shows the oval shape that points along \textit{x}- or \textit{y}- directions (see details in Supplementary \prettyref{fig: figHist_pnem}). {\bf i}, Autocorrelation of the PDM modulation amplitude (Supplementary \prettyref{fig: figHist_pnem}a,b). {\bf j}, Left Panel: Autocorrelation of $|p_{_{\text{LL}}}(\boldsymbol{r})|$. Right panel: zero-bias conductance map of a vortex. {\bf k}, Angle-averaged radial distribution of ({\bf i}) and ({\bf j}) showing the corresponding coherence lengths.}
\label{fig: fig3}
\end{figure}

\prettyref{fig: fig2} shows an example of high-spatial-resolution dI/dV linecut measured 
along the crystallographic \textit{y}- axis. Importantly, the observed modulation of the coherence peak positions 
matches with the wavelength of the crystal lattice. The locations of the top selenium atoms (Se$_{+}$) are indicated by the white 
dashed line in \prettyref{fig: fig2}c corresponding to local minima in the linecut, while local maxima are reached at the 
positions of iron atoms (Fe$_{y}$), located at the middle of the \textit{y}- bonds between two adjacent Se$_{+}$ atoms
(see the crystal structure in \prettyref{fig: fig1}b). The spatially averaged dI/dV curves of the positions of local maxima and minima sites are shown 
in \prettyref{fig: fig2}a. Three different quantities related to the superconductivity, the spatial variation of the 
gap $\Delta$(\textit{y}), zero-bias conductance $g_{0}$(\textit{y}) and height of the coherence peak 
$H$(\textit{y}) (\prettyref{fig: fig2}e-g), all show the same periodicity. Moreover, the observation of the sharp 
Bragg peak in the Fourier transform further verifies well-ordered modulation (\prettyref{fig: fig2}h-j, see also 
Extended Data \prettyref{fig: figlinecut2}, Extended Data \prettyref{fig: figlinecutHR} and 
Supplementary \prettyref{fig: figlinecut3} for more examples).  

The observation of the strongly modulated gap is highly unexpected, as it was not reported before either for
the bulk crystal, whose surfaces are also created by the exfoliation-like cleaving process, or in the previously studied 
films synthesized by molecular beam epitaxy. Naively, the observed modulation resembles the well-studied density-wave orders\cite{
du2020imaging,liu2021discovery,chen2021roton,wang2021scattering, 
gu2023detection,zhao2023smectic,liu2023pair,wei2023discovery}; however, with an important distinction: the wavelength matches the crystal lattice, implying that the long-range 
lattice translational symmetry is preserved. In the following, we investigate the properties of this PDM state characterized by the intra-unit-cell gap modulation.  

\noindent {\bf Imaging the PDM state}

To get further insights, we performed spectroscopic imaging (SI) STM measurements on a 5$\times$5 nm$^{\text{2}}$ area 
(\prettyref{fig: fig3}). A map of local superconducting gap value $\Delta(\boldsymbol{r})$ (\prettyref{fig: fig3}a) can 
be extracted from LDOS data collected alongside topography (\prettyref{fig: fig3}b, see Methods for 
details about SC gap extraction). The observed $\Delta(\boldsymbol{r})$ modulates along two lattice directions (\prettyref{fig: fig3}d) on top of a non-modulating component $\Delta_{0}(\boldsymbol{r})$ (Extended Data \prettyref{fig: figcompete}e). We can express the total gap as,
$\Delta(\boldsymbol{r}) = \Delta_{0}(\boldsymbol{r})+\delta \Delta(\boldsymbol{r})$,
with the PDM modulating component
$\delta \Delta(\boldsymbol{r})=\sum_{\boldsymbol{\rm{Q}} =  \boldsymbol{\rm P}_{\rm X},\boldsymbol{\rm P}_{\rm Y}}\left|\Delta_{\boldsymbol{{\rm Q}}}(\boldsymbol{r})\right| \text{cos} \left[\boldsymbol{{\rm Q}} \cdot \boldsymbol{r}+\phi_{\boldsymbol{{\rm Q}}}^{\Delta}(\boldsymbol{r})\right]$, where $\left|\Delta_{\boldsymbol{{\rm Q}}}(\boldsymbol{r})\right|$ and $\phi_{\boldsymbol{{\rm Q}}}^{\Delta}(\boldsymbol{r})$ are PDM modulation amplitude and modulation phase at the two Bragg vectors: $\boldsymbol{\rm Q} = \pm\boldsymbol{\rm P}_{\rm X}$ or $\pm\boldsymbol{\rm P}_{\rm Y}$. 
These quantities can be extracted individually by a standard 2D lock-in method\cite{lawler2010intra,fujita2014direct} (see description in SI Section \ref{SI:2dlockin}, phase in Extended Data \prettyref{fig: topodefect}a,b, and amplitude in Supplementary \prettyref{fig: figHist_pnem}a,b). In general, we find that the gap modulation in the thin flakes is very strong, with $\left|\Delta_{\boldsymbol{{\rm Q}}}\right|/\Delta_{0}$ reaching 20$\%$ in the current device (Extended Data \prettyref{fig: figcompete}b,c), and an even stronger modulation of 40$\%$ in another device with thinner flakes (Extended Data \prettyref{fig: figlinecutdevice2}). In order to compare with the previously reported density-wave modulations, we plot the $\left|\Delta_{\boldsymbol{{\rm Q}}}\right|/\Delta_{0}$ versus superconducting critical temperature (\textit{T}$_{\text{c}}$) for several materials in Extended Data \prettyref{fig: figcompete}a.

Now we focus on the observed $\delta\Delta(\boldsymbol{r})$ (\prettyref{fig: fig3}c). By comparing it to the atomic topography 
$\delta T(\boldsymbol{r})$ (\prettyref{fig: fig3}b), it is clear that the gap modulation does not trivially follow the 
surface corrugation, i.e., heights of the Se$_{+}$ atoms. FT filtered images (see Methods) of raw data (\prettyref{fig: 
fig3}b,c), surprisingly, show that while the atomic lattice is well-ordered (\prettyref{fig: fig3}e), the gap map shows 
strong distortion, breaking the area into small domains (\prettyref{fig: fig3}f). An auto-correlation analysis of the 
extracted modulation amplitude $\left|\Delta_{\boldsymbol{{\rm Q}}}(\boldsymbol{r})\right|$ shows that the corresponding coherence length of the PDM state (\prettyref{fig: fig3}i,k), is very short  
(about half of the SC coherence length, measured from a zero-bias conductance map of an SC vortex core, \prettyref{fig: fig3}j,k). 
Further analysis of the $\phi_{\boldsymbol{{\rm Q}}}^{\Delta}(\boldsymbol{r})$ maps shows the existence of topological defects around the domain boundaries, where both half and integer dislocations of the PDM state were observed (see Extended Data \prettyref{fig: topodefect}, and details in Methods).

\begin{figure}[p]
\begin{center}
    \includegraphics[width=15cm]{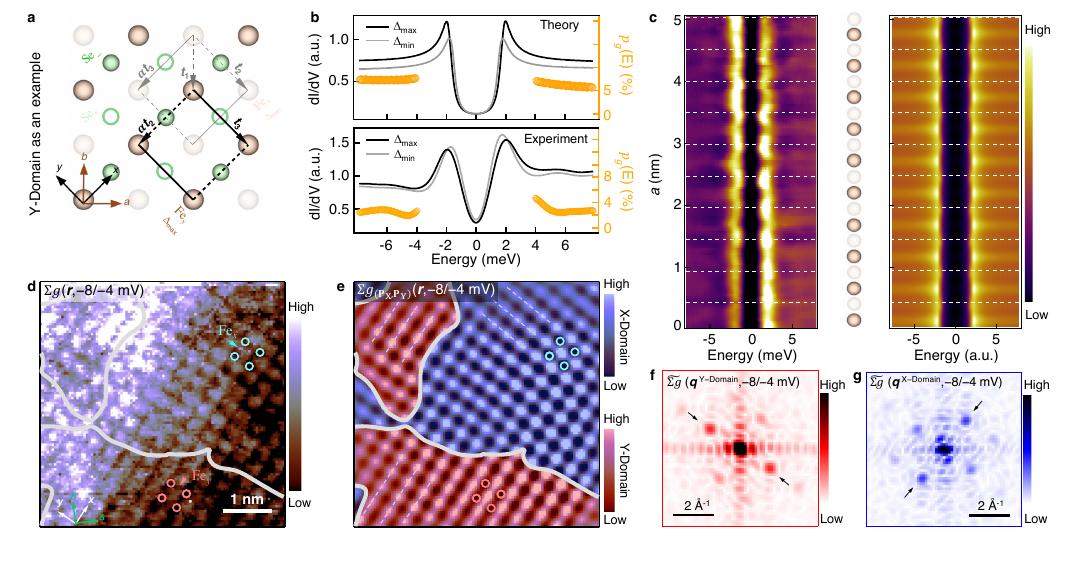}
\end{center}
\caption{
    {\bf  Origin of the pair density modulation state.} {\bf a}, Our model proposes nonequivalent hopping amplitude on the two iron sublattices Fe$_{x}$ and Fe$_{y}$. This breaks sublattice symmetry, leading to higher LDOS and a larger SC gap simultaneously at one of the sublattices ($\alpha$ is the nematic distortion parameter, $t_{1}$ is the nearest neighbor hopping integral and $t_{2,3}$ are the next-nearest neighbor hopping integrals, see further details in SI Section~\ref{SI:theory}). The schematic takes a Y-Domain as an example (gap maxima appear at Fe$_{y}$). Here, the faded and open-circle atoms show the positions with lower LDOS, leading to the formation of a stripe-like pattern.  {\bf b}, Experimental (lower panel) and theoretical (upper panel) comparison between iron sites with gap maxima and minima. The experimental data are averaged over all the spatial locations of the two kinds of iron atoms respectively (Supplementary \prettyref{fig: figLseg2}a,b). The LDOS imbalance ratio ($p_{g}$) was calculated at $|E|>$~4 meV, in order to avoid the influence from the modulation of coherence peaks (see SI Section \ref{SI:induce} and Supplementary \prettyref{fig: figPDMinduceCDW}). The theoretical results qualitatively match the experimental observations. {\bf c}, Left panel: a dI/dV linecut extracted from the SI-STM data along \textit{a}-axis (linecut position is shown in Supplementary \prettyref{fig: figLseg2}a). This linecut is along Fe$_{x}$-Fe$_{y}$ direction, avoiding the LDOS from Se$_{+}$/Se$_{-}$ sites. Right panel: the simulated gap modulation along the same Fe$_{x}$-Fe$_{y}$ direction.  {\bf d}, Energy-integrated  $g(\boldsymbol{r},E)$ between -8 and -4 meV (see also the results at positive bias voltage in Supplementary \prettyref{fig: figLseg2}{c}). The arrows indicate the Fe sites with higher LDOS. {\bf e}, FT filtered image at  $\pm$$\boldsymbol{\rm P}_{\rm X}$ and  $\pm$$\boldsymbol{\rm P}_{\rm Y}$ of ({\bf d}). The dashed lines indicate the directions of the stripes that change orientation across the domain wall (marked by the gray lines tracing $p_{_{\text{LL}}}=0$). The pink and cyan circles indicate Se$_{+}$ atoms on two different domains. {\bf f, g}, Selected-area FT magnitude of ({\bf d}) on Y-Domain and X-Domain, respectively.}
\label{fig: fig4}
\end{figure}

The presence of domains provides further insights into the origin of the observed modulation. We first note that the 
crystal structure of FeSC is characterized by a nonsymmorphic space group. The up-down staggered distribution 
of selenium atoms enlarges the crystallographic unit cell (\prettyref{fig: fig1}{b}), and as a result, it includes two iron atoms, which sit below 
top selenium atoms Se$_{+}$ along the \textit{x-} and \textit{y-} direction (Fe$_{x}$ and Fe$_{y}$). Structure-wise, the equivalence 
of the two iron sublattices is provided by a glide-mirror symmetry of the space group $P4/nmm$. However, remarkably, this symmetry is broken in our PDM state. 
Focusing on the observed 
behavior within the domain, we find that they correspond to regions where the gap maxima are observed on the Fe$_{x}$ 
(or Fe$_{y}$) sublattice. This is clearly resolved by comparing \prettyref{fig: fig3}e and \prettyref{fig: fig3}f where we marked four Se$_{+}$ atoms on each of the two adjacent domains. We find that within a domain, the gap maxima are locked to one of the iron sublattices, while the gap minima to the other. The gap difference between these neighboring extrema on iron sites, i.e. $|\Delta_{\text{Fe}_{x}}-\Delta_{\text{Fe}_{y}}|/(\Delta_{\text{Fe}_{x}}+\Delta_{\text{Fe}_{y}})$, reaches 32$\%$ (Extended Data \prettyref{fig: figcompete}d, see also SI Section~\ref{SI:extraction} for details).
The gap maxima/minima positions swap from Fe$_{x}$ to Fe$_{y}$ when the domain wall is crossed. A schematic for the two kinds of domain was shown in \prettyref{fig: fig3}h. 
To quantitatively determine the domain boundaries, we define PDM polarization $p_{_{\text{LL}}}$ that captures whether the maxima positions are on Fe$_x$ or Fe$_y$ sites corresponding to values of $p_{_{\text{LL}}}=\pm$1 (\prettyref{fig: fig3}g, see Methods for definition, Supplementary \prettyref{fig: figpll} for calculation details, and Supplementary \prettyref{fig: figpll_demo} for a simulation). The domain boundaries can be extracted by tracing the value of $p_{_{\text{LL}}}=$~0 (shown as gray lines in \prettyref{fig: fig3}f and \prettyref{fig: fig4}d,e).

\noindent {\bf Origin of the PDM state} 

Having identified that the gap modulation is linked to the iron sublattices, the key question arises: why the PDM state has not 
been seen in the bulk FeSC or molecular beam epitaxy synthesized films. 
In FeSCs, when the glide-mirror symmetry, composed of $z \rightarrow -z$ reflection and a translation by the nearest neighbor Fe-Fe lattice distance, is preserved, 
all iron atoms are equivalent. However, in the vicinity of the surface, the chalcogenide atoms above and below iron plane can reside at a different distances from the plane (\prettyref{fig: fig1}b), breaking the glide-mirror symmetry. Since 
the hoppings between the next-nearest neighbors in iron lattice (i.e. the $\boldsymbol{q}_3$ direction) are facilitated by the chalcogenide atoms, this results in two 
different next-nearest hopping integrals ($t_2$ and $t_3$) as indicated in \prettyref{fig: fig4}a. However, though the breaking of glide-mirror symmetry necessitates using a two-iron unit cell, the two iron sublattices can still map to each other by 90$^{\circ}$ rotation, another ingredient is needed to explain our observations.

Taking a careful look into the real-space resolved dI/dV, we now focus on the normal state properties above the 
SC gap (\prettyref{fig: fig4}d-g). Here the previously identified domains appear as stripes along \textit{x-} or \textit{y-}
directions (\prettyref{fig: fig4}d,e). The measured LDOS reflects contributions from both Se$_{+}$/Se$_{-}$ and Fe$_{x}$/Fe$_{y}$ sublattices.
While the LDOS imbalance from Se$_{+}$/Se$_{-}$ is expected owing to the two atoms being in different planes (see \prettyref{fig: fig1}b), it alone can not reproduce the observed stripe behavior. We then concluded that a small LDOS imbalance between Fe$_{x}$/Fe$_{y}$ sublattices in the normal state has to exist, implying normal state nematicity with the director pointing alongside $\boldsymbol{q}_3$ directions (\prettyref{fig: fig4}f,g), which is 45$\degree$ rotated from the nematic director of the bulk (Supplementary \prettyref{fig: fignematicFeSC}c,d). The origin of this nematic distortion is likely driven by electronic instability (see Methods for further discussion). Based on these observations, we built a simple model with glide-mirror symmetry breaking and a nematic 
director being along the next-nearest neighbor of iron atoms ($\boldsymbol{q}_3$ direction). This model results in the LDOS imbalance between the iron sublattices and a different superconducting gap on Fe$_x$ and Fe$_y$ atoms, thus capturing the main features of our experiments (see also SI Section \ref{SI:theory} for more realistic modeling and further details).

To compare directly the findings of the model and our experiments, we eliminated the influence from Se$_{+}$/Se$_{-}$ sublattices, and directly visualized the LDOS imbalance on Fe$_{x}$/Fe$_{y}$. This was done in three different ways. First, we traced the dI/dV linecut along \textit{a}-axis from the SI-STM data (left panel in \prettyref{fig: fig4}c, see also linecut locations in Supplementary \prettyref{fig: figLseg2}a). This linecut follows the Fe$_{x}$-Fe$_{y}$ direction and avoids the positions of selenium atoms. We simulated the SC gap along the same direction (right panel in \prettyref{fig: fig4}c), and both the gap size and LDOS above SC gap match well with the experimental measurements.
Second, we averaged the dI/dV curves on iron sites with gap maxima or gap minima among the SI-STM data (lower panel in \prettyref{fig: fig4}b). The normal-state LDOS was always larger at the iron sites with gap maxima, and vice versa, matching 
the ratio from calculations (see Methods and Supplementary \prettyref{fig: figLseg2}a,b). Third, we performed lattice segregation analysis, and the extracted LDOS maps of Fe$_{x}$/Fe$_{y}$ sublattice further supporting the existence of sublattice LDOS imbalance in the PDM state (see Methods and Extended Data \prettyref{fig: figLseg}).

\noindent {\bf Discussion}  

We emphasize that the direction of the nematic distortion is crucial for the observation of the PDM state. 
In most bulk FeSCs, both $\Gamma$ and M Fermi pockets are present\cite{fernandes2022iron}. The 
dominating nesting vector of the Fermi surface is the $\boldsymbol{q}_2$ vector [($\pi$,0)] connecting the two 
pockets (see Supplementary \prettyref{fig: fignematicFeSC}a,c). When a nematic phase appears, the nematic director always follows the nesting vector pointing along 
\textit{a}- or \textit{b}- direction so that the 1-Fe unit cell is stretched, but the Fe$_{x}$ and Fe$_{y}$ sublattice 
remain equivalent\cite{fernandes2014drives,bohmer2022nematicity}. As a result, these compounds cannot exhibit the PDM state described here. In FeTe$_{\text{0.55}}$Se$_{\text{0.45}}$ thin flakes, one of the 
pockets is missing (\prettyref{fig: fig1}h) and the nematic director is observed along \textit{x}- or \textit{y}- direction 
($\boldsymbol{q}_3$ direction, see \prettyref{fig: fig4}f,g), compatible with the 45$\degree$-rotation of the nesting vector (Supplementary \prettyref{fig: fignematicFeSC}c,d). Combined with the 
breaking of glide-mirror symmetry, this peculiar nematic distortion distinguishes the two iron sublattices, and leads to 
the formation of the PDM state (see \prettyref{fig: fig4}a). Note that a similar arrangement of Fermi surfaces appears in monolayer Fe(Te,Se)/SrTiO$_{\text{3}}$ films where only the M pockets are present. However, it was well-characterized that 
the nematic instability is absent in these previously measured monolayer films\cite{wang2012interface,lee2018routes,yuan2021incommensurate}, therefore not satisfying conditions for the PDM state. 

In summary, we observed the PDM state in thin flakes of Fe(Te,Se), highlighting the nontrivial role of the exfoliation process. In this context, one interesting avenue of future explorations would be to study other correlated materials in this form, where the interplay between strong electronic interaction and quantum confinement could potentially modify the underlying symmetries and electronic structure, thereby giving rise to novel phenomena. Finally, we note that 
the observed PDM state is likely not limited to the thin flakes studied here; instead, it may occur more generally in 
other unconventional superconductors with sublattice degeneracy and intertwined orders. The PDM state 
might be resolved in some compounds of cuprates\cite{keimer2015quantum}, heavy fermion\cite{ronning2017electronic} and kagome\cite{jiang2023kagome} superconductors by revisiting STM measurement with super-high spatial 
resolution. Some rarely studied FeSC compounds are also promising candidates, especially the heavily hole-doped RbFe$_{\text{2}}$As$_{\text{2}}$ and 
CsFe$_{\text{2}}$As$_{\text{2}}$\cite{li2016reemergeing,liu2019evidence,ishida2020novel}, where the nematic directors aligns along the $\boldsymbol{q}_3$ vector, favors the formation of the PDM state (Supplementary \prettyref{fig: fignematicFeSC}b). 
Other candidates, such as the superconductors based on van der Waals materials and heterostructures exhibiting a plethora of symmetry-breaking states, call for a careful investigation across the phase diagram.

\let\oldaddcontentsline\addcontentsline%
\renewcommand{\addcontentsline}[3]{}%

\let\addcontentsline\oldaddcontentsline%

\clearpage

\clearpage

\clearpage

\clearpage

\clearpage

\begin{linenumbers}

\vspace{5pt}
\noindent {\bf \large Methods:}

\vspace{5pt}
\noindent {\bf Device fabrication:} High-quality single crystals of FeTe$_{\text{0.55}}$Se$_{\text{0.45}}$ were synthesized 
using the self-flux method, and their values of \textit{T}$_{\text{c}}$ were determined to be 14.5 K from magnetization 
measurements\cite{wen2009short}. Although FeTe$_{\text{0.55}}$Se$_{\text{0.45}}$ single crystal is stable in air, exfoliated 
thin flakes are very air-sensitive. STM measurements require open surface geometry that excludes the full encapsulation as a 
method for sample protection\cite{wang2013one}. To overcome this challenge, we combined ultra-pure argon environment 
fabrication, suitcase transfer, and optimized fabrication process to minimize the pre-vacuum contamination of the surface. 
In Supplementary \prettyref{fig: fighistory}, we show a history of device quality optimization. Even though all the five 
devices are fabricated under the same ultra-pure argon environments, minimizing the time of Fe(Te,Se) thin flake being exposed in the argon environment ($t_{\text{Ar}}$) was crucial in reaching sufficiently high surface quality. 

The optimized fabrication procedure is demonstrated in Supplementary \prettyref{fig: figfab}. To minimize $t_{\text{Ar}}$, 
thin flakes of bottom graphite gate, boron nitride dielectric (h-BN) layer, and top graphite contact were prepared in 
advance on polydimethylsiloxane (PDMS) films. After the bottom graphite and h-BN are transferred onto a pre-patterned gold 
electrodes, the FeTe$_{\text{0.55}}$Se$_{\text{0.45}}$ single crystal was exfoliated using Scotch tape directly onto the PDMS film. Thin flakes ($d_{\text{t}}$ = 6 - 50 nm) were obtained using this method\cite{tang2019quasi}. Afterward, the Fe(Te,Se) flake was transferred onto h-BN under 60-90 $^{\circ}$C\cite{castellanos2014deterministic}. The electric contact 
between the sample and the bias electrode was established by utilizing top graphite, and the complete device was transferred into STM ultra-high vacuum (UHV) chamber immediately using home-made air-tight suitcase. The optical images of each step of device fabrication can be found in the lower panels of Supplementary \prettyref{fig: figfab}. With this particular fabrication scheme, we achieved clean surface roughly comparable to in-situ cleaved single crystals in a field of view extending to 30$\times$30 nm$^{\text{2}}$ (\prettyref{fig: fig1}d, Extended Data \prettyref{fig: figlinecutdirty}{a}, Supplementary \prettyref{fig: figorthtopo}{a}). The remaining contamination comprised of small, 1-2 Å high impurities that are mobile and can be removed by constant-current-mode scanning (Supplementary \prettyref{fig: figremovedirt}). 
The PDM state was confirmed to exist, even in the areas with high impurity concentrations, see Extended Data \prettyref{fig: figlinecutdirty}.

The locations of interest were identified using a combination of optical imaging, atomic force microscopy (AFM) and STM measurements (Extended Data \prettyref{fig: figdeviceconfig}). The thickness of each flake was determined by AFM 
after the STM experiments were completed. In device $\text{\#}$1, the thickness of h-BN dielectric layer is 50 nm, and the Fe(Te,Se) thin flake is 25 nm - 50 nm from left to right. There is 
also a 210 nm-thick flake attached to the left-bottom corner of the sample. Using precise tip navigation (Supplementary \prettyref{fig: fignevig}), we can investigate flake regions with specific thicknesses. The averaged single step height was found to be 7.1 Å (Supplementary \prettyref{fig: figcheight}). This value is about 18$\%$ larger to its bulk counterpart (c = 6.0 Å), effectively acting as negative chemical pressure, and has not been realized in bulk materials\cite{ishida2022pure,hou2023review}.

\noindent {\bf STM measurement:} STM measurements were performed in a commercial Unisoku USM 1300J STM/AFM setup using a 
Pt/Ir tip in an ultra-high vacuum environment. All features are measured by using several different micro-tips. 
The differential conductance (dI/dV) spectra were acquired by a standard lock-in amplifier at a frequency of 973 Hz, 
under lock-in modulation voltage $V_{\text{mod}}$ = 0.1 - 1 mV, setpoint voltage $V_{\text{bias}}$ = -5 mV and tunneling 
current $I_{\text{t}}$= 100 pA. STM images were acquired in the constant-current mode under the same setpoint parameters 
of dI/dV spectra, except the micrometer-sized topographies (left panel of \prettyref{fig: fig1}{d}, Extended Data \prettyref{fig: 
figdeviceconfig}{e}, Supplementary \prettyref{fig: fighistory} and Supplementary \prettyref{fig: fignevig}) which were 
applied a larger tunneling barrier ($V_{\text{bias}}$ = -100 mV, $I_{\text{t}}$= 20 pA).  
Unless otherwise specified, the data was measured in a zero magnetic field. The precise tip 
navigation (Supplementary \prettyref{fig: fignevig}) is facilitated by the fine calibration of walking piezos on pre-patterned reference markers. Unless specified otherwise, data was acquired under a system temperature of 300 mK and electron temperature ranging from 1-4 K depending on the lock-in amplitude and circuit configuration as determined by the 
sharpness of the gap edge. At lower electron temperatures, multiple features near the coherence peak are identified (SI Section \ref{SI:multipeak}). All these features exhibit similar spatial behavior (Extended Data \prettyref{fig: figlinecutHR}). In order to simplify the data analysis, we focused on the high electron temperature limit where the features are merged into a single peak.

\noindent {\bf Summary of the STM data:} In this work, we measured five Fe(Te,Se) thin-flake devices (Supplementary \prettyref{fig: fighistory}), and 
a FeTe$_{\text{0.55}}$Se$_{\text{0.45}}$ bulk single crystal cleaved in-situ for comparison. The results of single crystal are 
shown in Supplementary \prettyref{fig: figcheight}{a,c} and  Supplementary \prettyref{fig: figbulk}. The 
devices $\text{\#}$3- $\text{\#}$5 are test devices (the spectroscopic data not shown). The results of 
device $\text{\#}$2 are shown in Extended Data \prettyref{fig: figlinecutdevice2}. All the other figures were 
measured on device $\text{\#}$1. The PDM state was observed on the thin part of devices $\text{\#}$1 and $\text{\#}$2, 
but disappeared at 210-nm-thick part of the flake (Supplementary \prettyref{fig: figthickflake}). Unless specified otherwise, the data from device $\text{\#}$1 was measured at 50 nm-thick flakes. An example of 35-nm thick flake is shown in Extended Data \prettyref{fig: figlinecut2}{b}, and data of 30-nm-thick flake from device $\text{\#}$2 is shown in Extended Data \prettyref{fig: figlinecutdevice2}.

\noindent {\bf Robustness of the PDM state:} The PDM state is robust against influence of magnetic field and nearby impurities. By applying a 6 T magnetic field perpendicularly to the thin flakes, a single SC vortex core was identified in Extended Data \prettyref{fig: figlinecutfield}a. With dI/dV linecuts measured along crystallographic directions, we observed well-ordered gap modulation at the area far away from the core (Extended Data \prettyref{fig: figlinecutfield}b-f). At the core center, we observed bound states oscillated within ±1 meV with a same periodicity of lattice (Extended Data \prettyref{fig: figlinecutfield}g-l). In the areas with high impurity concentration, due to pair breaking scattering, higher zero-bias conductance and lower SC coherence peak height were observed (Extended Data \prettyref{fig: figlinecutdirty}b,c). Even in these areas, periodic gap modulation is well resolved at the edges (Extended Data \prettyref{fig: figlinecutdirty}d,e). Inside these areas, new sub-gap states, induced by disorder, appear 
and modulate with the same spatial wavelength (Extended Data \prettyref{fig: figlinecutdirty}f-k). 

\noindent {\bf Gap map extraction:} The SI-STM data shown in the main text was measured under 300 mK, with an effective electron temperature around 4 K (spectral broadening: 1.2 meV). For this electron temperature, only a single peak is observed around SC gap. The local spectral gap is defined as ($\Delta_{\text{+}}$-$\Delta_{\text{-}}$)/2, where $\Delta_{\text{+,-}}$ are defined as the energies at the coherence peak maxima determined by Gaussian fitting. In rare cases (less than 0.1$\%$ of data) when the fitting failed, the left and right peak energy were determined by the positions of local maximum after a slope function subtracted from the raw data (see also SI Section~\ref{SI:multipeak} for the discussion of low electron temperature data).

\noindent {\bf FT filtering:} A reciprocal-vector-locked FT filtering enhances the visualization of modulation by filtering out irrelevant signals (\prettyref{fig: fig3}e,f, \prettyref{fig: fig4}e and Extended Data \prettyref{fig: topodefect}c,d). First, the vector-selecting Gaussian windows was applied to the Fourier transform of the image. Then, an inverse Fourier transform construct the filtered image. The FT filtering function is expressed as,
\begin{equation}
    \mathcal{F}lt_{_{(\boldsymbol{{\rm Q}}_{\text{1}}... \boldsymbol{{\rm Q}}_{\text{n}})}}[A(\boldsymbol{r})]\equiv \mathcal{F}^{-1}\left\{\frac{\mathcal{F}[A(\boldsymbol{r})]}{\sqrt{2 \pi}\sigma_q} \sum_{\boldsymbol{Q}= \pm\boldsymbol{{\rm Q}_{1}}...\pm\boldsymbol{{\rm Q}_{n}}} \mathrm{e}^{-\frac{(\boldsymbol{q}-\boldsymbol{Q})^2}{2 \sigma_q^2}}\right\}
\end{equation}
where $\mathcal{F}$ and $\mathcal{F}^{-1}$ are Fourier transform and inverse Fourier transform operation respectively. $\sigma_{q}=1/\sigma$ is the Gaussian cutoff in the reciprocal space, it was set to be 1/1.3 nm$^{-1}$ throughout this work (see also a related method in SI Section \ref{SI:2dlockin}).

\noindent {\bf Origin of the observed PDM domains:} As discussed in the main text, the PDM state is intimately related to the nematic distortion in thin flakes. 
In bulk, the composition studied here (FeTe$_{\text{0.55}}$Se$_{\text{0.45}}$) is at the boundary of the nematic phase\cite{fernandes2014drives,ishida2022pure,hou2023review}. While the long-range order is absent, the elastoresistance experiments observed diverging nematic fluctuations\cite{kuo2016ubiquitous,ishida2022pure}, and a recent STM measurement showed puddles of short-range order pinned by inhomogeneity\cite{zhao2021nematic}. We expect that this behavior is also present in thin flakes but with a different nematic director, giving rise to the PDM state, as explained in the main text. The appearance of small domains of the PDM state, as well as various topological defects around the boundaries, are direct consequence of these fluctuations.

\noindent {\bf Topological defects around the domain walls:} We found that the distortions around the PDM domain walls (\prettyref{fig: fig3}f) originate from topological defects. In Extended Data \prettyref{fig: topodefect}c,d, we disentangle the gap modulation along the two Bragg directions ($\pm$$\boldsymbol{\rm P}_{\rm X}$ and $\pm$$\boldsymbol{\rm P}_{\rm Y}$) by using a vector selective FT filtering, that allows us to analyze the spatial dependence of the two unidirectional modulations individually. Several PDM dislocations were observed for both unidirectional modulations. Since an integer (or half) dislocation displays as a single (or half) vortex at the spatial map of the modulation phase $\phi_{\boldsymbol{{\rm Q}}}^{\Delta}(\boldsymbol{r})$, the phase analysis can quantitatively identify the properties of these defects (Extended Data \prettyref{fig: topodefect}a,b). By tracing the phase around a vortex center, we identified a half dislocation (Extended Data \prettyref{fig: topodefect}e) and three integer dislocations (Extended Data \prettyref{fig: topodefect}f), where two of them formed a dislocation-antidislocation pair (Supplementary \prettyref{fig: figvortexpair}). A zoom-in into two of these defects (marked by dashed boxes in Extended Data \prettyref{fig: topodefect}c) is shown in Extended Data \prettyref{fig: topodefect}g and Extended Data \prettyref{fig: topodefect}j. These defects, corresponding to half-integer and integer dislocations, can be fully reproduced from a simple simulation (see Extended Data \prettyref{fig: topodefect}h,i,k,l and SI Section \ref{SI:naivesimu} for details). 
The related modulation amplitude of the PDM state is suppressed in the centers of these defects (Supplementry \prettyref{fig: figHist_pnem}a,b), while the non-modulating component is not affected (Extended Data \prettyref{fig: figcompete}e). The presence of these unusual topological defects\cite{agterberg2008dislocations,berg2009charge}, implies 
a nontrivial origin of the small domains of the PDM 
state\cite{radzihovsky2009quantum,mross2015spin,aishwarya2024melting}.

\noindent {\bf Lattice-lock-in polarization of the PDM state:} In order to quantitatively describe the orthogonal domains of the PDM state, we introduce the lattice-lock-in polarization ($p_{_{\text{LL}}}$). The status of lattice lock-in can be resolved by comparing the difference of modulation phase between topography and gap map along the directions of the two modulation vectors,  $\pm$$\boldsymbol{\rm P}_{\rm X}$ and $\pm$$\boldsymbol{\rm P}_{\rm Y}$. In Supplementary \prettyref{fig: figpll}g,h, we show the phase difference, i.e. $\delta\phi_{\boldsymbol{{\rm P}}_{\rm X,Y}}^{\Delta/T}(\boldsymbol{r})\equiv \phi_{\boldsymbol{{\rm P}}_{\rm X,Y}}^{\Delta}(\boldsymbol{r})-\phi_{\boldsymbol{{\rm P}}_{\rm X,Y}}^{T}(\boldsymbol{r})$, extracted and converted by the 2D lock-in method (see SI Section \ref{SI:2dlockin}, and definitions in Supplementary Table~\ref{table: tabs1} and Supplementary Eq.~\ref{eq:15}). Then the lattice-lock-in polarization can be expressed as,
\begin{equation}
    p_{_{\text{LL}}}(\boldsymbol{r})\equiv \left[|\delta \phi_{\boldsymbol{{\rm P}}_{\rm X}}^{\Delta/T}(\boldsymbol{r})|-|\delta \phi_{\boldsymbol{{\rm P}}_{\rm Y}}^{\Delta/T}(\boldsymbol{r})|\right]/\pi
\end{equation}
It is +1 or -1 if gap maxima fully polarize at Fe$_{x}$ or Fe$_{y}$ positions (Supplementary \prettyref{fig: figpll_demo}).

\noindent {\bf Estimation of the lattice distortions:} The lattice distortion in our thin flakes was 
tested using a large topography image with atomic resolution, measured on a 100$\times$100 nm$^{\text{2}}$ area (Supplementary \prettyref{fig: figorthtopo}a-d). The orthorhombic distortion is defined as, $\delta \equiv (x_{\rm o}-y_{\rm o})/(x_{\rm o}+y_{\rm o})$ where $x_{\rm o}$ and $y_{\rm o}$ are lattice constants of orthorhombic unit cell, which is the 2-Fe unit cell on thin flake. We obtained the lattice constants by converting the reciprocal vector from FT magnitude image. This leads to $\pm\text{0.1}\%$ resolution on $\delta$ for a 100$\times$100 nm$^{\text{2}}$ topography, when considering the uncertainty as the size of reciprocal pixel. Following this method, we resolved $\delta = \text{(0.34}\pm\text{0.1)}\%$. This vanishing distortion is comparable to the values of nematoelastic coupling induced distortion in bulk crystals, which can be up to 0.6$\%$ in literature\cite{prozorov2009intrinsic,rossler2022nematic}, indicating an electronic-driven nematic distortion in the thin flakes. Additionally, we applied the same method to the topography of SI-STM data (\prettyref{fig: fig3}b, Supplementary \prettyref{fig: figorthtopo}e-h). Even though the small image size leads to a larger uncertainty, we obtained zero average distortion $\delta = \text{(0}\pm\text{1.8)}\%$ for this particular scan, consistent with the analysis on large topography. 
Moreover, the electronic deformations on the LDOS and gap map were found to be much larger than the lattice distortion, which further supports an electronic origin of the nematic distortion in our thin flakes (see SI Section \ref{SI:distortion} and Supplementary \prettyref{fig: figcorrect}{d}). 
Note that the sample fabrication process inevitably introduces strain in some areas. We show an example of a strained area in Supplementary \prettyref{fig: figstraintopo}, where dense stripe-like feature was observed on the topography. In this work, we only focused on areas that do not show strain.

\noindent {\bf Iron sites with gap maxima or minima:} 
The lower panel of \prettyref{fig: fig4}b shows the extracted dI/dV spectra from the iron sites with gap maxima or minima separately. In order to obtain this plot, we first use the normalized gap map  $\Delta_{\text{temp}}\equiv\text{2}\delta\Delta(\boldsymbol{r})/[(\delta\Delta)_{\text{max}}-(\delta\Delta)_{\text{min}}]$ as a template, to locate the positions of iron sites with gap maxima and minima. The positions for gap maxima and minima were determined by $\Delta_{\text{temp}} >$ 0.2 and $\Delta_{\text{temp}} <$ -0.2 respectively, and small variation of this threshold do not have considerable effects on the extraction results. The segregation results are shown in Supplementary \prettyref{fig: figLseg2}a,b and the averaged spectrum is shown in the main figure. 

\noindent {\bf Lattice segregation:} 
Here we introduce the method for isolating the signal of Fe$_{x}$/Fe$_{y}$ from the measured LDOS map. 
The positions of the Se$_{+}$ sites are determined by first fitting Supplementary Eq.~\ref{eq:14} to $\delta T_{_{(\boldsymbol{{\rm P}}_{\text{1}}, \boldsymbol{{\rm P}}_{\text{2}})}}(\boldsymbol{r})$ (\prettyref{fig: fig3}{e}), while keeping a unitary amplitude. Then, the Se$_{+}$ positions are determined as areas where the value of the fitted function exceeds 0.9. 
In the second step, the iron sites (Fe$_x$ and Fe$_y$) are identified by shifting the Se$_{+}$ positions 
by half a wavelength in \textit{x}- or \textit{y}- direction. 
Then the LDOS of each sublattice can be segregated by picking up the dI/dV signals from the corresponding positions determined above. Finally, we rebuild a continuous image for each sublattice by a standard \textit{Voronoi} image interpolation method.
The segregated Fe$_{x}$/Fe$_{y}$ images show lattice modulation due to sublattice symmetry breaking, while segregated Se$_{+}$ images show no modulation as expected (Extended Data \prettyref{fig: figLseg}, see also an example in Supplementary \prettyref{fig: segsimu}).

\vspace{5pt}

{%
\noindent {\bf Acknowledgments:} We thank Sankar Das Sarma, Ruixing Zhang, Yahui Zhang, Constantin Schrade, Fangdong Tang, He Zhao, Shangfei Wu, Xiong Huang, Hu Miao, Jiaqi Cai, Jianfeng Ge, Jia-Xin Yin and Zhengran Wu for helpful discussions. We thank Haoxin Zhou for technical support on sample fabrication. This work has been primarily
supported by the Institute for Quantum Information and Matter (IQIM) an NSF Physics Frontiers Center, with support of
the Gordon and Betty Moore Foundation through Grant GBMF1250. Part of the work was supported by National Science Foundation (grant no. DMR-2005129); P.A.L acknowledges support by DOE (USA) office of Basic Sciences Grant No. DE-FG02-03ER46076.; L.K. acknowledges support from an IQIM-AWS Quantum postdoctoral fellowship. Portions of this work were supported by the U.S. Department of Energy, Office of Science, National Quantum Information Science Research Centers, Quantum Science Center (M.P.). M.P. received additional fellowship support from the Emergent Phenomena in Quantum Systems program of the Gordon and Betty Moore Foundation. H.K. acknowledges support from the Kwanjeong fellowship. The work at BNL was supported by the US Department of Energy, oﬃce of Basic Energy Sciences, contract no. DOE-sc0012704.

\noindent {\bf Author Contribution:} L.K. designed the experiments; L.K. fabricated samples with assistance of H.K., Y.Z. and S.N.-P.; L.K. performed STM measurements with the assistance of H.K., E.B. and S.N.-P.; L.K. analysed data with the assistance of M.P., H.L. and S.N.-P.; L.K. developed data analysis codes on Igor Pro; M.P. and P.A.L. provided theoretical explanations and model simulations; G.G. provided FeTe$_{\text{0.55}}$Se$_{\text{0.45}}$ single crystals; K.W. and T.T. provided the h-BN crystals; Y.Z performed AFM measurements; L.K., M.P., P.A.L. and S.N.-P. wrote the manuscripts with input from other authors.
S.N.-P. supervised the project.

\noindent {\bf Data availability:} The data that support the findings of this
study are available from the corresponding authors on reasonable request.

\noindent {\bf Code availability:} The code that supports the findings of this study is available from the corresponding authors upon reasonable request.

\noindent {\bf Competing interests:} The authors declare no competing interests.

}
\end{linenumbers}

\clearpage

\renewcommand{\figurename}{\textbf{Extended Data Fig.}}
\renewcommand{\theHfigure}{Extended.\thefigure}
\setcounter{figure}{0}

\begin{figure}[p]
\begin{center}
    \includegraphics[width=15cm]{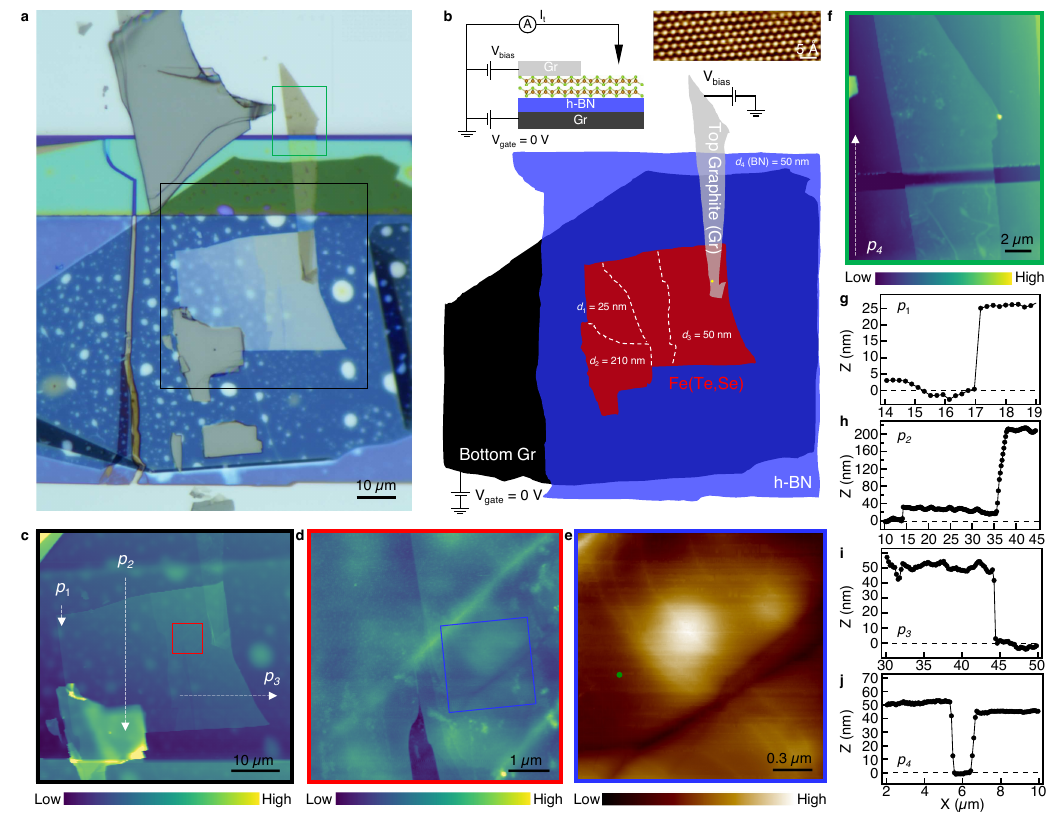}
\end{center}
\caption{{\bf Sample details.}
{\bf a}, Optical image of device $\text{\#}$1. {\bf b}, Schematic of the van der Waals stacks and the circuits. 
The FeTe$_{\text{0.55}}$Se$_{\text{0.45}}$ thin flakes were stacked on top of hexagonal boron nitride dielectric 
layer, with a graphite gate at the bottom. Bias voltage ($V_{\text{bias}}$) was applied to the Fe(Te,Se) flakes, and the 
graphite gate was grounded. Top-right inset: Atom-resolved STM topography of the graphite top contact (setpoints: 
$V_{\text{bias}}$ = 100 mV, $I_{\text{t}}$= 20 pA). The yellow spot on the bottom panel indicates the position where the 
graphite topography was measured. Top-left inset: side view of the STM device circuit. {\bf c, d, f}, AFM images of the 
regions of interest. Panels ({\bf c}) and ({\bf d}) show the zoomed-out view of the region where STM data has been taken (shown in {\bf e}). Panel ({\bf f}) shows the region of top graphite contact. The green dot in ({\bf e}) indicates the position where
the SI-STM data for the main figures was measured. {\bf g-i}, AFM line-profiles along the traces in ({\bf c}), showing the thicknesses of the different regions of the Fe(Te,Se) flake. From left to right, the thickness increases from 25 nm to 50 nm (see also Supplementary \prettyref{fig: fignevig}, the STM topographies mapping through the Fe(Te,Se) flakes). A 210-nm-thick flake is attached in the left-bottom corner. {\bf j}, AFM line profile along the trace in ({\bf f}), used to determine the thickness of h-BN dielectric layer (50 nm). Note that the AFM measurements were performed after the STM measurements were completed.}
\label{fig: figdeviceconfig}
\end{figure}
\clearpage

\begin{figure}[p]
\begin{center}
    \includegraphics[width=15cm]{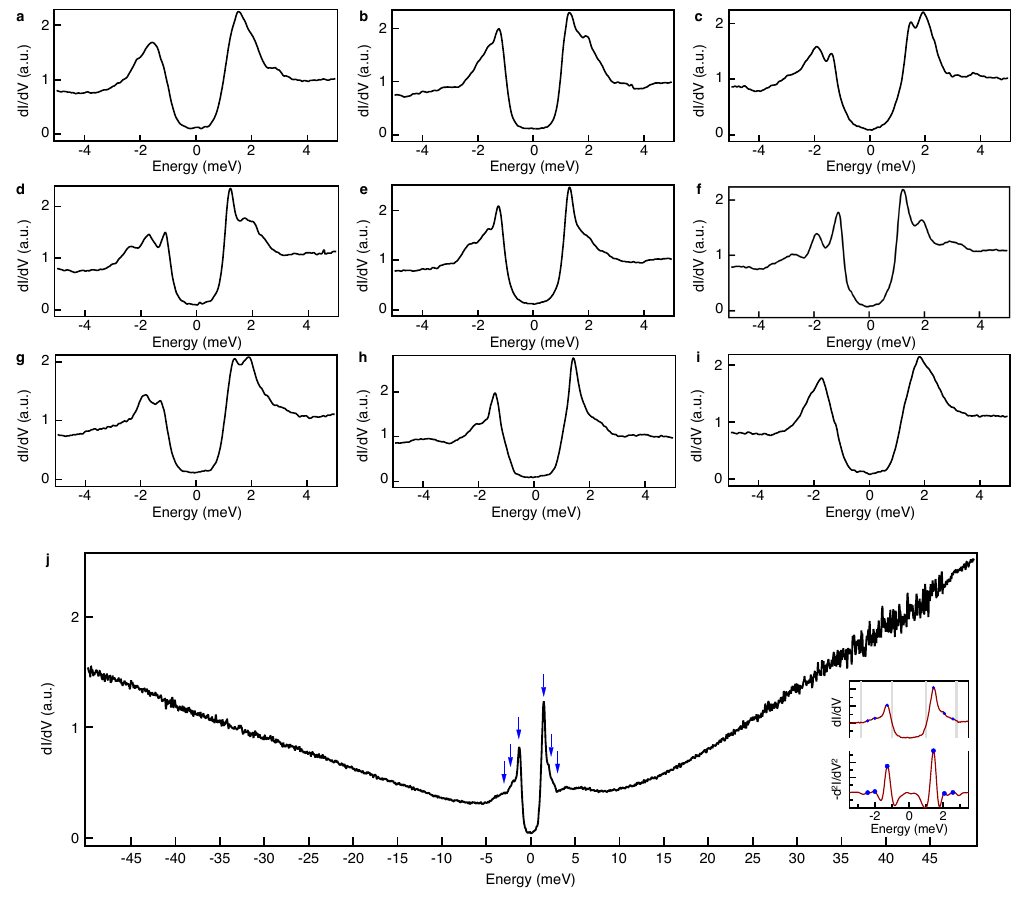}
\end{center}
\caption{{\bf Examples of dI/dV spectra on Fe(Te,Se) thin flakes.}
dI/dV spectra taken at different positions showing one- ({\bf a, b, i}), two- ({\bf c, g, h}) or 
three- ({\bf d, e, f}) peak-like spectral features. Panel ({\bf h}) is also shown in \prettyref{fig: fig1}{e}. 
{\bf j}, A high-resolution dI/dV spectrum measured with a wide bias range. Here we can resolve three spectral 
features around the superconducting gap by the second derivative analysis (lower inset, see also SI Section \ref{SI:multipeak}), although the second and 
the third peaks are hardly distinguishable in the raw data (higher inset).}
\label{fig: figHRSCgap}
\end{figure}
\clearpage

\begin{figure}[p]
\begin{center}
    \includegraphics[width=15cm]{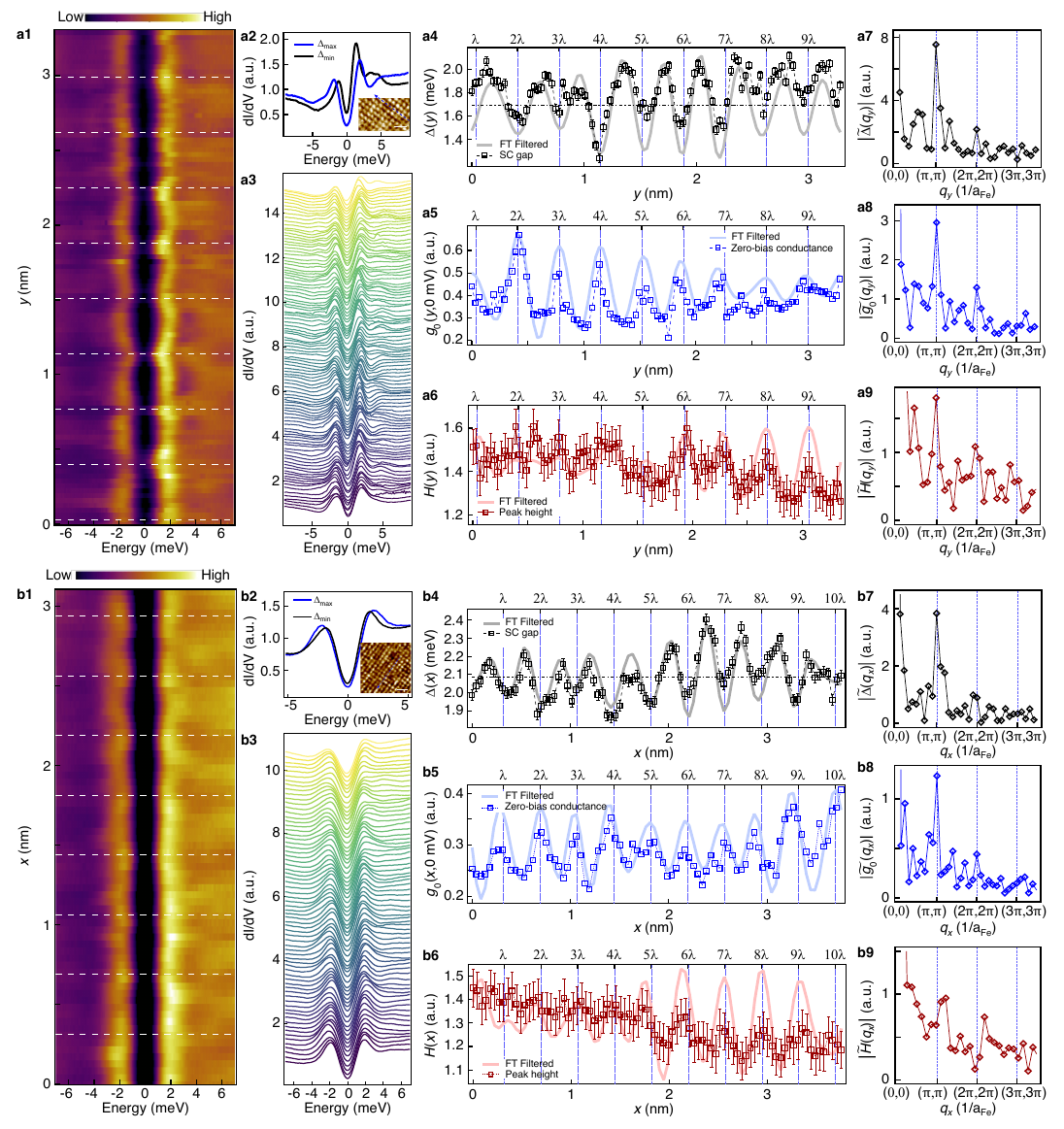}
\end{center}
\caption{{\bf Two more examples of periodic gap modulation.} The panel arrangements are the same as \prettyref{fig: fig2}. {\bf a}, Flake thickness $d_{\text{t}}$ is 50 nm. The modulation ratio $|\Delta_{\boldsymbol{\rm P}_{\rm Y}}|/\bar{\Delta}$ is 13.4$\%$. {\bf b}, Flake thickness $d_{\text{t}}$ is 35 nm. The modulation ratio $|\Delta_{\boldsymbol{\rm P}_{\rm X}}|/\bar{\Delta}$ is 6.7$\%$. The scale bars in the inset of ({\bf a2}) and ({\bf b2}) are 1 nm.}
\label{fig: figlinecut2}
\end{figure}
\clearpage

\begin{figure}[p]
\begin{center}
    \includegraphics[width=15cm]{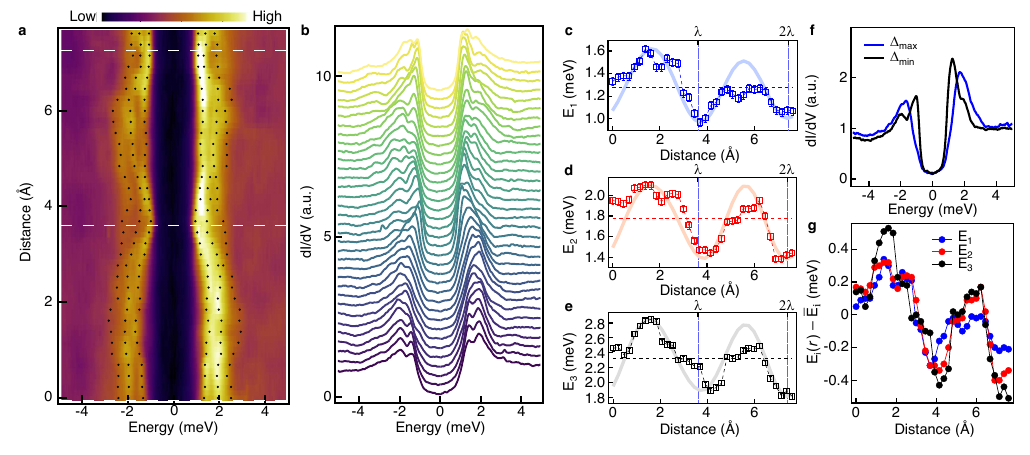}
\end{center}
\caption{{\bf Gap modulation at lower electron temperature.} {\bf a}, False-color plot of a dI/dV linecut measured along $\textit{x}$-axis, across three nearest Se$_{+}$ atoms. The SC gaps are minimized at Se$_{+}$ atom sites (dashed white lines). {\bf b}, The waterfall plot of ({\bf a}). The curves are offset for clarity. On each dI/dV spectrum, three features can be identified by fitting the second derivative of the dI/dV curves (see SI Section \ref{SI:multipeak} and Supplementary \prettyref{fig: figMPfit}). The fitting results are shown in ({\bf a}) as black dots. All the three features (located at energies $E_{1}$, $E_{2}$ and $E_{3}$) modulate in phase and are individually plotted in {\bf c-e}. The semi-transparent solid lines are extracted by FT filtering of the raw data. The horizontal dashed lines indicate the average energy of the spectral features ($\Bar{E}_{\rm i}$ = 1.28 meV, 1.78 meV, 2.32 meV). {\bf f}, Spatially-averaged dI/dV spectra at gap maximum (blue) and minimum (black) sites. {\bf g}, Comparison of the modulation amplitude corresponding to the three features, where the curves from ({\bf c-e}) have their average value subtracted. The feature at higher energy has a larger modulation amplitude ($|\delta E_{\rm i}|$ = 0.22 meV, 0.31 meV, 0.41 meV), but the modulation ratio ($|\delta E_{\rm i}|/\Bar{E}_{\rm i}$ with  i = 1,2,3) is roughly constant among these features, i.e. 17.2$\%$, 17.4$\%$ and 17.7$\%$, respectively.}
\label{fig: figlinecutHR}
\end{figure}
\clearpage

\begin{figure}[p]
\begin{center}
    \includegraphics[width=15cm]{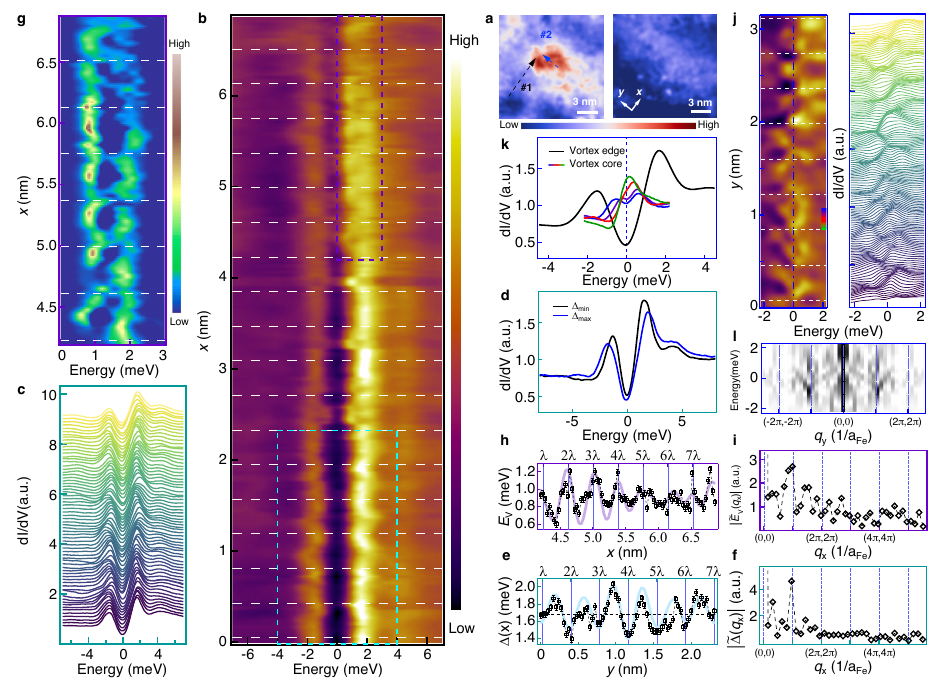}
\end{center}
\caption{{\bf Presence of the PDM state in high magnetic fields. }
{\bf a}, Zero-bias conductance map of a single vortex (15$\times$15 nm$^{\text{2}}$, see also a 
vortex lattice in Supplementary \prettyref{fig: figvortexlattice}). The left and right panels are
measured at the same field of view but under different magnetic fields: 6 T (out of plane) and 0 T, respectively. 
The left panel is also shown in \prettyref{fig: fig3}{j}.
We measured two dI/dV linecuts in this field of view, and both of them follow the Se$_{+}$ lattice. The linecut $\text{\#}$1 (along $\textit{x}$-axis) starts far away from the vortex and goes across the edge of the vortex core, with the corresponding dI/dV data shown in {\bf b}. The periodic modulation of SC gap was observed outside the vortex core (cyan dashed box). The waterfall plot of the cyan region is shown in {\bf c}. Spatially-averaged dI/dV spectra at the gap maximum (blue) and minimum (black) sites are shown in {\bf d}. The extraction of the gap modulation is shown in {\bf e}, and its FT magnitude is shown in {\bf f}. At the edge of the vortex core, in-gap bound states were observed around 1 meV and they exhibit modulation with the same periodicity [purple dashed box in ({\bf b})]. {\bf g}, Negative curvature plot (see SI Section \ref{SI:curvature}) of the purple dashed box in ({\bf b}). The extraction of spatial modulation of the in-gap bound state is shown in {\bf h}, and its FT magnitude is shown in {\bf i}. The linecut $\text{\#}$2 (along $\textit{y}$-axis) was measured at the vortex center [blue arrow in ({\bf a})]. {\bf j}, False-color plot of dI/dV linecut (left) and waterfall spectrum plot (right) of linecut $\text{\#}$2. {\bf k}, Typical spectra inside the vortex core. The positions of each curve are indicated by the color bar in ({\bf j}). {\bf l}, Energy-dependent magnitude of one-dimensional FT of ({\bf j}). Color scale: black for high and white for low intensity. Inside the vortex core, the bound states oscillate from 0 meV to around $\pm$1 meV with a similar wavelength as the PDM state. This observation is consistent with the signal that appears around ($\pi$,$\pi$) in ({\bf l}).
}
\label{fig: figlinecutfield}
\end{figure}
\clearpage

\begin{figure}[p]
\begin{center}
    \includegraphics[width=15cm]{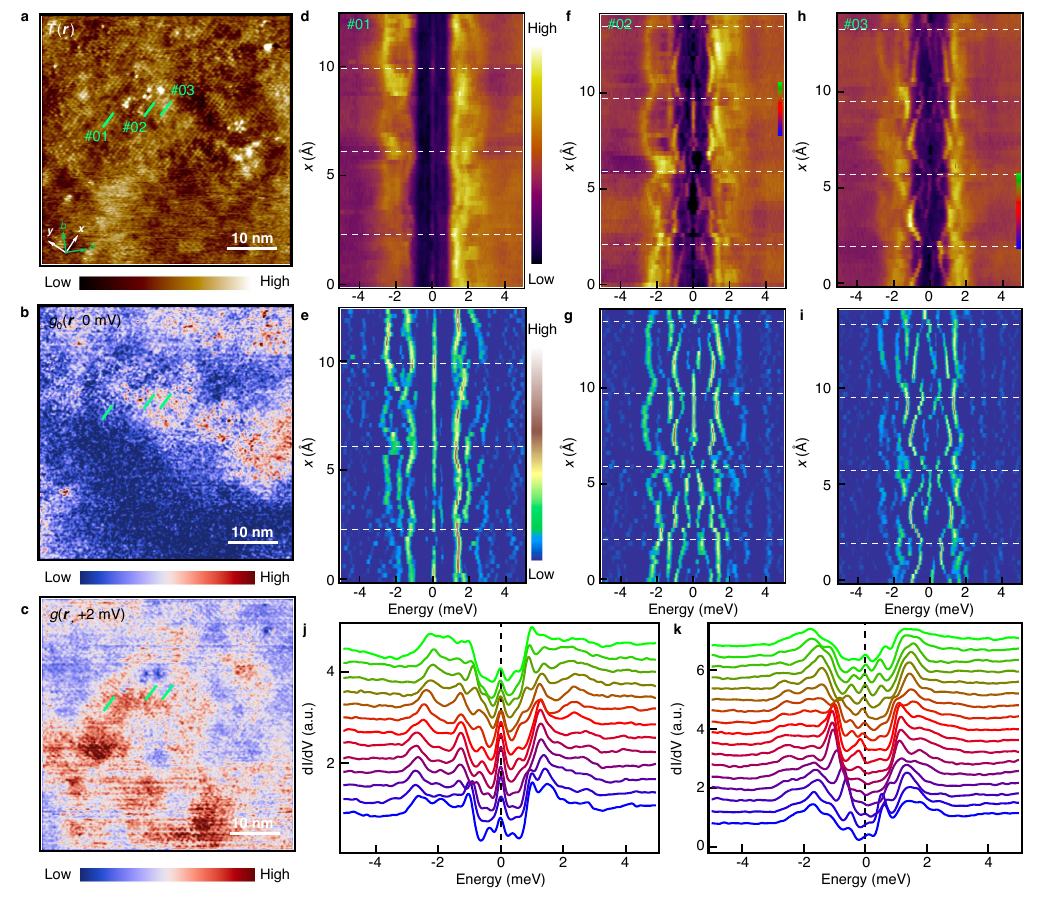}
\end{center}
\caption{{\bf Periodic modulation in the areas with high impurity concentration.} A large area (50$\times$50 nm$^{\text{2}}$) contains regions with low and high impurity concentrations. {\bf a}, Atom-resolved STM topography. Some impurities can be identified in the upper region, resulting in a high zero-bias conductance ({\bf b}) and lower coherence peaks ({\bf c}). In contrast to the interstitial iron impurities on the bulk single crystals without annealing, the impurities observed here are mobile and can be removed by constant-current-mode scanning (Supplementary \prettyref{fig: figremovedirt}).  {\bf b, c}, Differential conductance map at zero bias (0 mV) and the energy of SC gap coherence peak (+2 mV). {\bf d-i}, Three dI/dV linecuts measured in this area. Their locations are marked in ({\bf a-c}). They go gradually deeper inside the region with higher impurity concentration from $\text{\#}$1 to $\text{\#}$3. ({\bf d, f, h}) are false-color plots of dI/dV linecut. ({\bf e, g, i}) are the corresponding negative curvature plots (see SI Section \ref{SI:curvature}). The periodic gap modulation exists throughout. Note that sharp bound states appear in the dI/dV spectrum when measured near the impurities. These states exhibit the same modulation periodicity as the PDM state. {\bf j}, Selected spectra from ({\bf f}). The positions of the curves are indicated in ({\bf f}) by the color bar. {\bf k}, Selected spectra from ({\bf h}) feature the spatial modulation of the bound states. The positions of these curves are indicated in ({\bf h}) by the color bar.}
\label{fig: figlinecutdirty}
\end{figure}
\clearpage

\begin{figure}[p]
\begin{center}
    \includegraphics[width=15cm]{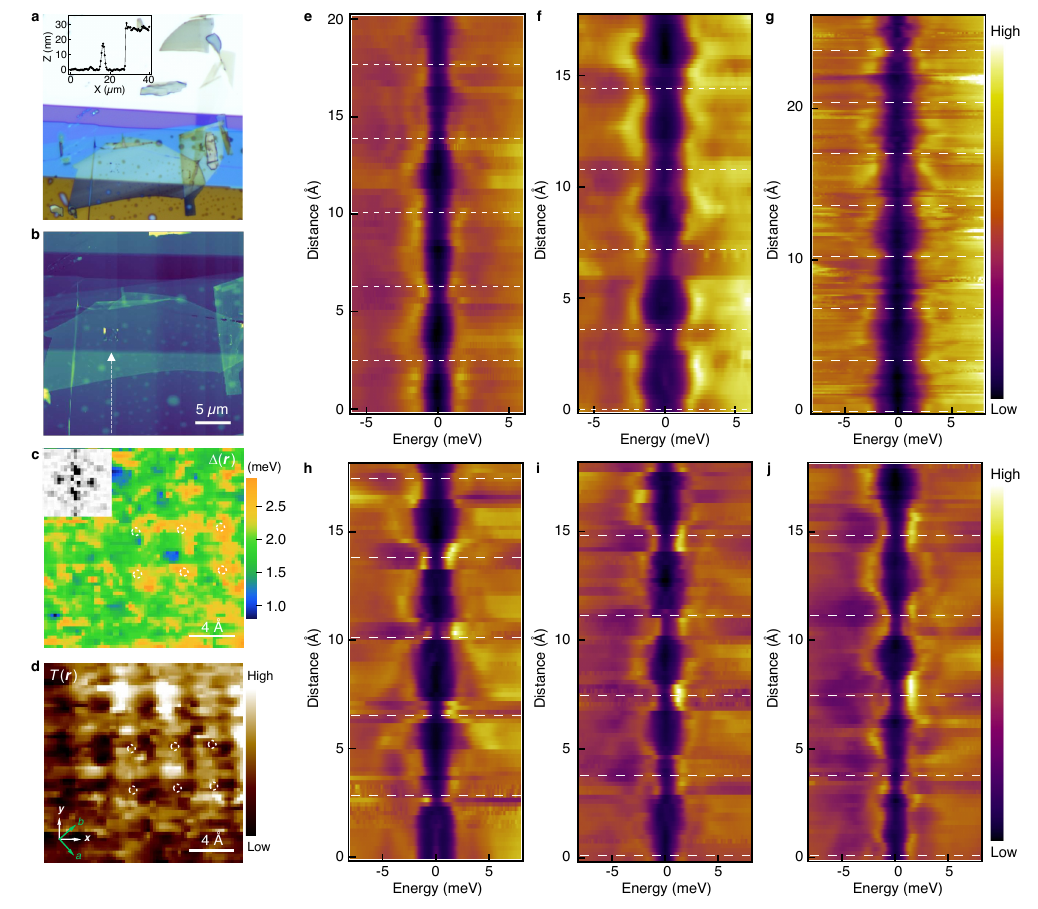}
\end{center}
\caption{{\bf Gap modulation on Device \(\text{\#}\)2.}
{\bf a}, Optical image of Device \(\text{\#}\)2. Inset: AFM line profile measured along the white dashed 
line in ({\bf b}), showing the thickness of the flake (30 nm). {\bf b}, AFM image of Device \(\text{\#}\)2. {\bf c}, 
SC gap map on a 1.7$\times$1.7 nm$^{\text{2}}$ area. Inset: the FT magnitude of the SC gap map. {\bf d}, 
The STM topography acquired simultaneously with the gap map in ({\bf c}). The white circles mark the positions of 
gap maxima appearing at the Fe$_{y}$ sites. {\bf e-j}, dI/dV linecuts measured on device \(\text{\#}\)2 along 
either $\textit{x}$-axis or $\textit{y}$-axis. The average gaps are $\bar{\Delta}$ = 1.30, 2.00, 2.09, 2.03, 1.60, 1.76 meV 
and PDM modulation amplitudes are $|\Delta_{\text{P}}|$ = 0.40, 0.50, 0.61, 0.69, 0.56, 0.72 meV for each corresponding 
linecut. The average modulation ratio is 35 $\%$ in this device, with individual ratios 
being: 31$\%$, 40$\%$, 29$\%$, 34$\%$, 35$\%$, 41$\%$, respectively. 
Setpoints: $V_{\text{bias}}$ = -5 mV; $I_{\text{t}}$: 700 pA for ({\bf e}), 1.3 nA for ({\bf c, d, f}), 1.5 nA 
for ({\bf g-i}).}
\label{fig: figlinecutdevice2}
\end{figure}
\clearpage

\begin{figure}[p]
\begin{center}
    \includegraphics[width=15cm]{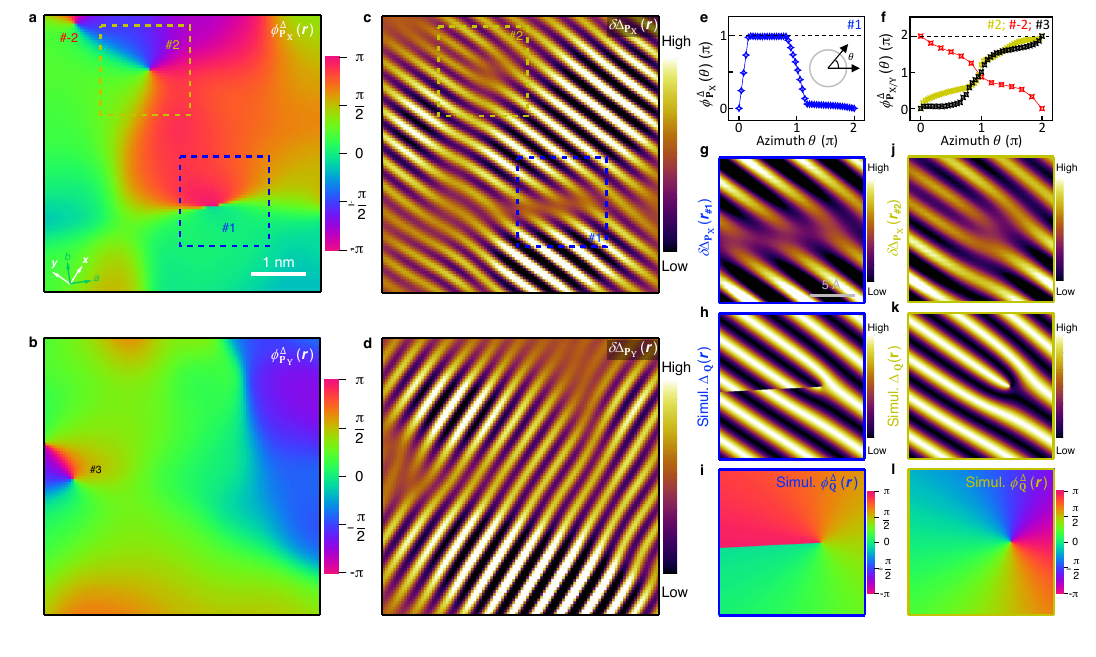}
\end{center}
\caption{
{\bf  Fractional and integer topological defects on the PDM state.} {\bf a, b}, PDM modulation phase at the $\pm$$\boldsymbol{\rm P}_{\rm X}$ and $\pm$$\boldsymbol{\rm P}_{\rm Y}$, calculated by 2D lock-in method (SI Section \ref{SI:2dlockin}). {\bf c, d}, Separate visualization of the modulation at $\pm$$\boldsymbol{\rm P}_{\rm X}$ and $\pm$$\boldsymbol{\rm P}_{\rm Y}$, calculated by FT filtering with only unidirectional vectors selected (see Methods). Multiple topological defects are identified ($\text{\#}$1 - $\text{\#}$3). {\bf e, f}, Phase winding around fractional and integer topological defects. For clarity, the curves are shifted to match their phase minima at zero. In ({\bf f}), a dislocation-antidislocation pair is demonstrated by their opposite phase winding (see also a simulation in Supplementary \prettyref{fig: figvortexpair}).  {\bf g-i}, Zoom-in of an half-integer quantized topological defect. {\bf g}, A zoom-in image of blue dashed square area in ({\bf c}), which contains the topological defect $\text{\#}$1. {\bf h, i}, Simulation of the fractional topological defect $\text{\#}$1 (see SI Section \ref{SI:naivesimu}). A misaligned half dislocation on the $\pm$$\boldsymbol{\rm P}_{\rm X}$ modulation of the PDM state ({\bf h}) appears as an equivalent half vortex in its phase field ({\bf i}). {\bf j-l}, Same as ({\bf g-i}), showing an integer topological defect $\text{\#}$2. A single dislocation on the $\pm$$\boldsymbol{\rm P}_{\rm X}$ modulation ({\bf k}) appears as an equivalent single vortex in its phase field ({\bf l}). The Gaussian cut-off length ($\sigma$) of FT was set to 1.3 nm in ({\bf a-d}).}
\label{fig: topodefect}
\end{figure}

\clearpage

\begin{figure}[p]
\begin{center}
    \includegraphics[width=15cm]{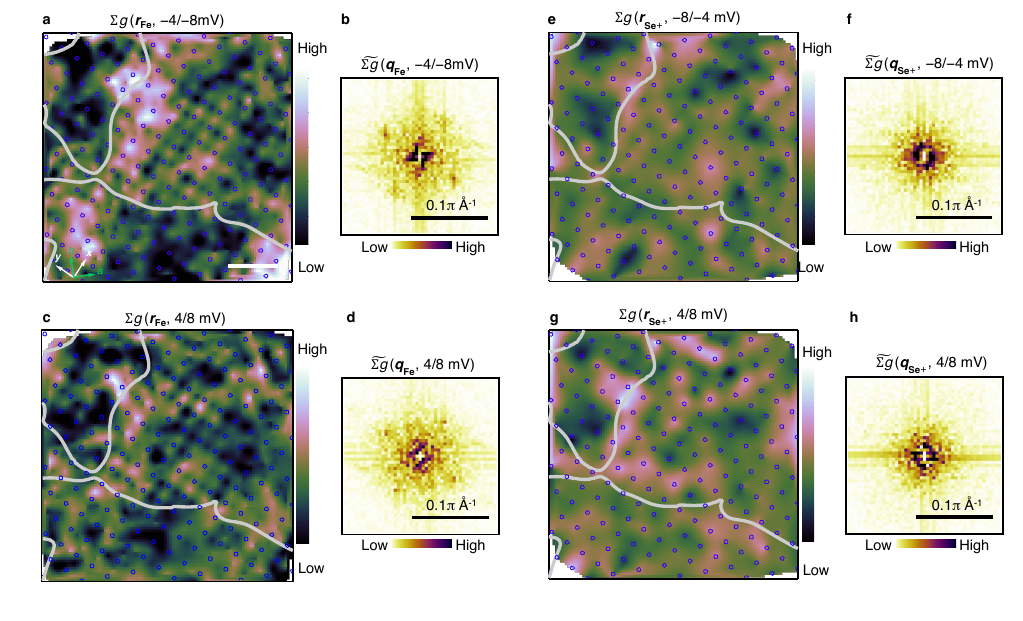}
\end{center}
\caption{{\bf Lattice segregation and LDOS imbalance.}
{\bf a-d}, Energy-integrated differential conductance map of segregated iron sublattice, both Fe$_{x}$ and Fe$_{y}$ sites are selected (see Methods). The integration energy window was selected in order to avoid the influence of SC coherence peak (see SI Section \ref{SI:induce} and Supplementary \prettyref{fig: figPDMinduceCDW}). The gray lines show the PDM domain wall, which traces $p_{_{\text{LL}}}=0$. The segregated LDOS at iron sites, $\Sigma g(\boldsymbol{r}_{\text{Fe}})$, shows the periodic modulation with a wavelength equal to the lattice constant, regardless if positive ({\bf a}) or negative ({\bf b}) energy window is selected. The same modulation can be seen from the FT magnitude [({\bf b}) and ({\bf d})], where the Bragg peaks can be resolved. {\bf e-h}, Same as ({\bf a-d}) but on segregated Se$_{+}$ sublattice. No modulation is seen as expected. The blue circles indicate the positions of Se$_{+}$ sites. See also Supplementary \prettyref{fig: segsimu} for an example of lattice segregation procedure performed on simulated images with and without LDOS imbalance between Fe$_{x}$/Fe$_{y}$ sites.
}
\label{fig: figLseg}
\end{figure}
\clearpage

\begin{figure}[p]
\begin{center}
    \includegraphics[width=15cm]{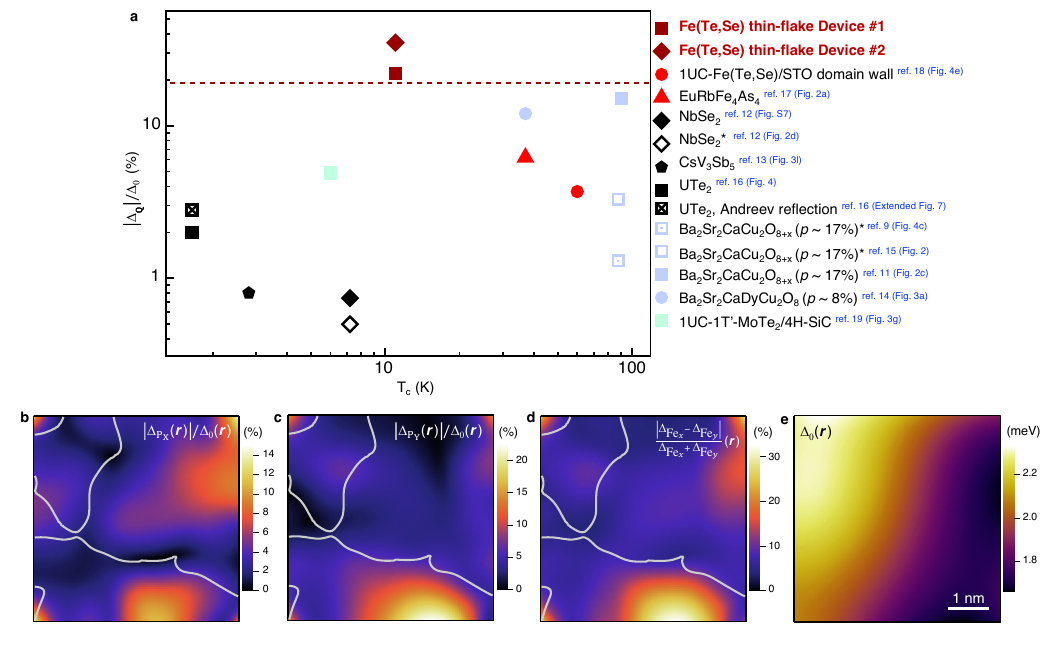}
\end{center}
\caption{{\bf Gap modulation ratio.}
{\bf a}, Literature survey of the materials showing gap modulation\cite{hamidian2016detection,du2020imaging,liu2021discovery,chen2021roton,wang2021scattering, 
chen2022identification,gu2023detection,zhao2023smectic,liu2023pair,wei2023discovery}. Unidirectional gap modulation ratio, $|\Delta_{{\boldsymbol{\rm Q}}}|/\Delta_{0}$, versus SC critical temperature, $T_{c}$, is plotted for multiple materials. The filled symbols are extracted from SC gap measurements, $\Delta (\boldsymbol{r})$; while the open symbols are converted from Cooper pair density map, $n_{\rm s}(\boldsymbol{r})$, which was measured by Josephson SI-STM experiments in the literature (see details in SI Section \ref{SI:extraction}). The current work (dark-red filled symbols) exhibits an unprecedentedly large gap modulation ratio (see also Extended Data \prettyref{fig: figlinecutdevice2}). 
{\bf b, c}, Spatial distribution of unidirectional gap modulation ratio of the SI-STM data in device $\text{\#}$1 (\prettyref{fig: fig3} and Supplementary \prettyref{fig: figHist_pnem}a,b). The ratio is calculated individually along $\pm$$\boldsymbol{\rm P}_{\rm X}$ and $\pm$$\boldsymbol{\rm P}_{\rm Y}$ directions. The gray lines are the lattice-lock-in domain walls, which trace $p_{_{\text{LL}}}=0$. The ratio is above 20$\%$ inside the domain. 
{\bf d}, The gap difference between the neighboring Fe$_{x}$ and Fe$_{y}$ atoms [$p_{_{\Delta}}(\boldsymbol{r})$] which is derived by the sum of ({\bf b}) and ({\bf c}), see details in SI Section~\ref{SI:extraction} and Supplementary Table~\ref{table: tabs1}.  
{\bf e}, Map of the inhomogeneous non-modulating component, $\Delta_{0}(\boldsymbol{r})$. In practice, it was defined as the polynomial background of the total gap, $\Delta(\boldsymbol{r})$ (\prettyref{fig: fig3}{a}), and extracted by function fitting. The $\Delta_{0}(\boldsymbol{r})$ was also depicted as the semi-transparent plane in the three-dimensional plot of \prettyref{fig: fig3}{a}.}
\label{fig: figcompete}
\end{figure}
\clearpage

\clearpage

\beginsupplement
\renewcommand{\theHfigure}{Supplement.\thefigure}
\begin{center}
\noindent \textbf{\large Supplementary Information: Observation of Cooper-pair density modulation state} 

\end{center}

\setcounter{footnote}{0}
\renewcommand{\thefootnote}{\fnsymbol{footnote}}

{\raggedleft Lingyuan Kong$^{\text{1,2}}$ \footnote{Correspondence: lykong@caltech.edu}, Michał Papaj$^{\text{3}}$, Hyunjin Kim$^{\text{1,2,4}}$, Yiran Zhang$^{\text{1,2,4}}$, Eli Baum$^{\text{1,2}}$, Hui Li$^{\text{5}}$, Kenji Watanabe$^{\text{6}}$, Takashi Taniguchi$^{\text{6}}$, Genda Gu$^{\text{7}}$, Patrick A. Lee$^{\text{8}}$, and Stevan Nadj-Perge$^{\text{1,2}}$ \footnote{Correspondence: s.nadj-perge@caltech.edu}}

\tableofcontents

\setcounter{equation}{0}

\section{Models for the pair density modulation state}\label{SI:theory}

We build the model for the pair density modulation (PDM) state based on several experimental observations. The quasiparticle interference patterns observed in our thin flake devices suggest the absence of either $\Gamma$ or $M$ pockets based on the lack of scattering processes with momentum transfer $\boldsymbol{q}_2$ (\prettyref{fig: fig1}h). In the following we will assume that only $M$ pockets are present, similarly to the monolayer FeSe on SrTiO$_{3}$ substrate. We then note that the  SC  order modulation is commensurate with two-iron unit cell and the  SC  gap extrema coincide with the two iron sublattices, with maximum on Fe$_{x}$ site and minimum on Fe$_{y}$ site. Therefore, the model has to make a distinction between the two sublattices. Such a distinction can be achieved by breaking the glide-mirror symmetry present in the bulk crystals, which can happen at the crystal surface due to the different environment of the Se/Te atoms that are below and above the Fe plane (\prettyref{fig: fig1}b). Moreover, our measurements also indicate differences in normal state density of states (for bias voltage far outside of the SC gap, see \prettyref{fig: fig4}), which requires additional symmetry breaking beyond that of the glide-mirror plane. We propose the nematic distortion along one of the sides of the two-iron unit cell to be responsible for differentiation between the two iron sublattices. Short-range nematic order (see Methods for details) also explains the presence of two types of domains, in which the role of Fe$_{x}$ and Fe$_{y}$ switches.

While in general iron-based superconductors are described by multiple orbital models constructed out of the $3d$ shell of iron atoms, here we will focus on a simple model that nevertheless can capture the essence of the PDM state. We base our approach on phenomenological models previously suggested for monolayer FeSe on SrTiO$_{3}$\cite{gao2016hidden,kitamura2022quantum}, with two iron sublattices, and nearest and next-nearest hoppings between them:
\begin{equation}
    H_0(\mathbf{k}) = \begin{pmatrix} \epsilon_{\mathrm{Fe}_{x}}(\mathbf{k}) & \epsilon_T(\mathbf{k}) \\
                                      \epsilon_T(\mathbf{k}) & \epsilon_{\mathrm{Fe}_{y}}(\mathbf{k})
    \end{pmatrix}
\end{equation}
\begin{equation}
    \epsilon_{\mathrm{Fe}_{x}}(\mathbf{k}) = -2\left(\alpha\, t_2 \cos k_x + t_3 \cos k_y \right) - \mu
\end{equation}
\begin{equation}
    \epsilon_{\mathrm{Fe}_{y}}(\mathbf{k}) = -2\left(\alpha\, t_3 \cos k_x  + t_2 \cos k_y \right) - \mu
\end{equation}
\begin{equation}
    \epsilon_T(\mathbf{k}) = -4t_1 \cos \frac{k_x}{2} \cos\frac{k_y}{2}
\end{equation}

As shown in \prettyref{fig: fig4}a, 
the glide-mirror symmetry is broken when the next-nearest neighbor hoppings $t_2$ and $t_3$ have different values as they are facilitated by the chalcogenide atoms that are below and above the iron plane, respectively. However, when only the glide mirror symmetry is broken, the normal state density of states on both iron sublattices remains the same. Therefore, the second crucial ingredient of the model is the appearance of nematic order in a direction of one of the next-nearest neighbor Fe-Fe bonds. We stress this is different from the nematic order commonly observed in other FeSC compounds, which is along the nearest neighbor Fe-Fe bond. In other words, in our case the nematic direction is rotated by 45$\degree$ from the direction usually encountered in other iron-based superconductors (see details in Supplementary \prettyref{fig: fignematicFeSC}). The effect of this nematic order is described 
by the parameter $\alpha$, which changes the values of hopping along $x$ direction, in accordance with experimentally determined nematicity direction. By slightly changing $\alpha$ value from 1, we can obtain normal state density of states ratio between the two sublattices as observed in the experiment, which is on the order of few percent.

With such a normal state Hamiltonian, we can now discuss the  SC  order parameter. In the simplest case, we consider on-site interaction that gives rise to two momentum-independent order parameters:
\begin{equation}
    \Delta_X = V_0 / N_k \sum_\mathbf{k} \langle c_{\mathbf{k}X\uparrow} c_{-\mathbf{k}X\downarrow} \rangle
\end{equation}
\begin{equation}
    \Delta_Y = V_0 / N_k \sum_\mathbf{k} \langle c_{\mathbf{k}Y\uparrow} c_{-\mathbf{k}Y\downarrow} \rangle
\end{equation}
where $V_0$ is the onsite interaction strength and $c_{\mathbf{k}X/Y\uparrow\downarrow}$ are the annihilation operators for electrons on sublattice Fe$_{x}$/Fe$_{y}$ with spin up/down, respectively. Due to the difference in normal state density of states, the self-consistent solutions of the gap equation will lead to different values of $\Delta_X$ and $\Delta_Y$. We then obtain the density of states in the superconducting state by first constructing the Bogoliubov-de Gennes Hamiltonian:

\begin{equation}
    H_\mathrm{BdG}(\mathbf{k}) = \begin{pmatrix}
        H_0(\mathbf{k}) & \Delta \\
        \Delta^\dagger & -H^T_0(-\mathbf{k})
    \end{pmatrix}
    , \quad \quad \Delta = \begin{pmatrix}
        \Delta_X & 0 \\
        0 & \Delta_Y
    \end{pmatrix}
\end{equation}
With such a Hamiltonian, we calculate the spectral function and, from it, the density of states on each sublattice:
\begin{equation}
    G^R(\omega, \mathbf{k}) = (\omega + i \eta - H_\mathrm{BdG}(\mathbf{k}))^{-1}
\end{equation}
\begin{equation}
    \rho_{X/Y}(\omega) = -\frac{1}{\pi} \sum_{\mathbf{k}} \mathrm{Im}\, G^R(\omega, \mathbf{k})_{11/22}
\end{equation}

The difference in on-site order parameters results thus in different positions of the coherence peaks on both sublattices as presented in Fig.~\ref{fig: fig4} of the main text. The parameter values used in calculations were $t_1 =1.6$, $t_2=0.4$, $t_3=-2.0$, $\mu=-2.4$, $\alpha=0.975$, $V_0=12.5$ with cut-off $\pm0.25$. Nevertheless, the obtained results are not strongly dependent on the exact values used for calculations. Mainly, a larger deviation of $\alpha$ from 1 increases the difference in the normal density of states, and in consequence, the difference in order parameters. In general, we can also consider pairing between the nearest and next-nearest sites, which will change the gap structure quantitatively (including some impact on the coherence peak separation), but the main feature of shifted coherence peaks between the two sublattices will remain unchanged.

\begin{figure}
    \includegraphics[width=\linewidth]{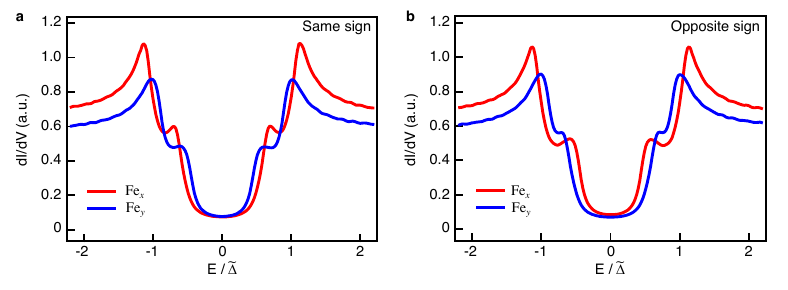}
\caption{
{\bf  Calculated density of states in PDM state model with two Fermi surface pockets at the two iron sublattices.} The two gaps oscillate either in-phase when the SC order parameters have the same sign ({\bf a}), or out-of-phase when they have opposite sign ({\bf b}). The unit $\tilde{\Delta} = \Delta_0 + \Delta_H $ = 0.045, $\Delta_0$ and $\Delta_H$ are defined in the supplementary Information. }
\label{fig: figPDMsign}
\end{figure}

To make our model more realistic, we make an extension by considering two pockets at the $M$ point. These pockets correspond to two orbital degrees of freedom and their presence is found by the first principles calculations and angle-resolved photoemission spectroscopy. The simplest way to incorporate the presence of this second pocket is to include two copies of the single orbital model and introduce hybridization between them to lift the degeneracy between the two bands. In such a case, the normal state Hamiltonian is:
\begin{equation}
    H_0(\mathbf{k}) = \begin{pmatrix} \epsilon_{\mathrm{Fe}_{x}}(\mathbf{k}) & \epsilon_H(\mathbf{k}) & \epsilon_T(\mathbf{k}) & 0 \\
    \epsilon_H(\mathbf{k}) & \epsilon_{\mathrm{Fe}_{x}}(\mathbf{k}) & 0 & \epsilon_T(\mathbf{k}) \\
    \epsilon_T(\mathbf{k}) & 0 & \epsilon_{\mathrm{Fe}_{y}}(\mathbf{k}) & \epsilon_H(\mathbf{k}) \\
    0 & \epsilon_T(\mathbf{k}) & \epsilon_H(\mathbf{k}) & \epsilon_{\mathrm{Fe}_{y}}(\mathbf{k})
    \end{pmatrix}
\end{equation}
where the hybridization term is given by $\epsilon_H(\mathbf{k})=-2t_4(\cos k_x + \cos k_y)$. The inclusion of the second pocket allows us to consider different relationships between the order parameters on each pocket. Theoretical proposals include $s_{++}$ and $s_{+-}$ configurations with the same and opposite phases of the order parameter on the two pockets, and quasi-nodeless $d$-wave order parameter, which also has opposite phase on both pockets, but with additional momentum space structure as shown in Supplementary \prettyref{fig: figPRQPI}. We can now investigate on a phenomenological level various consequences of different possible order parameter types. First of all, the presence of two Fermi surface pockets can lead to the appearance of an additional coherence peak related to different  SC  gaps on both pockets. While exact experimental regime can considerably 
complicate this simple picture, in principle, the behavior of such peaks can serve as an indicator for the nature of the SC state thanks to the emergence of PDM state. To model these differences between the gap sizes and signs of order parameters between the pockets further, we can use the following $\Delta$ matrix in the sublattice and orbital space:
\begin{equation}
    \Delta = \begin{pmatrix}
        \Delta_0 + \Delta_z & \Delta_H & \Delta_T & 0 \\
        \Delta_H & \Delta_0 + \Delta_z & 0 & \Delta_T \\
        \Delta_T & 0 & \Delta_0 - \Delta_z & \Delta_H \\
        0 & \Delta_T & \Delta_H & \Delta_0 - \Delta_z
    \end{pmatrix}
\end{equation}
Here we decomposed $\Delta_X$ and $\Delta_Y$ into $\Delta_0$ and $\Delta_z$. Additionally, we include the possibility of intersublattice intraorbital pairing $\Delta_T$ and intrasublattice interorbital pairing $\Delta_H$. When rotated to the band basis of the normal state Hamiltonian $H_0(\mathbf{k})$, the intrapocket pairing on the two pockets will have the form:

\begin{equation}
    \Delta_{P1/2} = \Delta_0 \pm \Delta_H + \Delta_T \frac{\epsilon_T}{\sqrt{\epsilon_T^2+(\epsilon_{\mathrm{Fe}_{x}}-\epsilon_{\mathrm{Fe}_{y}})^2/4}} + \Delta_z  \frac{(\epsilon_{\mathrm{Fe}_{x}}-\epsilon_{\mathrm{Fe}_{y}})/2}{\sqrt{\epsilon_T^2+(\epsilon_{\mathrm{Fe}_{x}}-\epsilon_{\mathrm{Fe}_{y}})^2/4}}
\end{equation}
where we have suppressed the momentum dependence of the normal state Hamiltonian and order parameter components. In this basis, we can see that the order parameter in the two pockets, and in particular their relative sign can be determined by the relative sign of $\Delta_0 \pm \Delta_H$, since the $\Delta_T$ and $\Delta_z$ terms have the same sign for both pockets. We can qualitatively test the consequence of PDM state for $s_{++}$ (in our model, $\Delta_H < \Delta_0$), $s_{+-}$ ($\Delta_H > \Delta_0$), and nodeless $d$-wave ($\Delta_H = \tilde{\Delta}_H \sin k_x/2 \sin k_y/2$, $\Delta_H > \Delta_0$ on the Fermi surface) pairings.  When the density of states on each sublattice is calculated with such $\Delta$ parametrizations, the relative sign of the order parameters has a qualitative consequence on the spectrum. If the sign is the same on both pockets, the maxima and minima of the coherence peak positions for both pockets coincide with the same iron atoms (Supplementary \prettyref{fig: figPDMsign}a). However, when the order parameter signs are opposite, maximum of one gap will coincide with minimum of the second gap on a given sublattice, with the roles reversed on the other sublattice (Supplementary \prettyref{fig: figPDMsign}b). This will result in either in-phase oscillation of coherence peaks for the same order parameter sign, or out-of-phase oscillation of coherence peaks for the opposite signs of order parameters. As such, the presence of PDM state can thus enable distinguishing between the relative sign of superconductivity in the two pockets.

\section{Phase-referenced quasiparticle interference} \label{SI:PRQPI}

Phase-referenced (PR) QPI can resolve the sign changing between two SC order parameters\cite{hirschfeld2015robust,sprau2017discovery}. Here, we employed a defect-bound-state-based PR-QPI method\cite{chi2017extracting}, which is more applicable in practice. In this method, the PR-QPI signal is expressed as, 
\begin{equation}
g_{_{\text{PR}}}\left(\boldsymbol{q}, E\right) \equiv\left|\widetilde{g}\left(\boldsymbol{q}, E\right)\right| \cos \left(\theta_{\boldsymbol{q}, E} -\theta_{\boldsymbol{q}, -E} \right)    
\end{equation}
where $\left|\widetilde{g}\left(\boldsymbol{q}, E\right)\right|$ and $\theta_{\boldsymbol{q}, E}$ are the FT magnitude and phase of $g\left(\boldsymbol{r}, E\right)$, respectively. The $g_{_{\text{PR}}}\left(\boldsymbol{q}, E\right)$ is negative, if the $\boldsymbol{q}$ connects states with the opposite sign of SC order parameters, $\text{sign}[\Delta_{i}(\boldsymbol{k})]=-\text{sign}[\Delta_{j}(\boldsymbol{k}-\boldsymbol{q})]$, but is positive if the order parameters for the initial and final states have the same sign. This method was originally proposed for a sharp spectral peak of impurity-induced bound states, but it was later extended to the case without a clear subgap peak\cite{gu2019directly,chen2019direct}. Previous experiments showed that the $g_{_{\text{PR}}}$ signal captured the gap structure of FeTe$_{\text{0.55}}$Se$_{\text{0.45}}$\cite{chen2019direct} and Bi$_2$Sr$_2$CaCu$_2$O$_{8+\delta}$\cite{gu2019directly} bulk superconductors correctly even when the subgap peaks are absent. We employed this strategy in our SI-STM data and observed negative $g_{_{\text{PR}}}$ signal within a small $\boldsymbol{q}$ vector around $(0,0)$. These signals only exist at the energy window slightly smaller than the gap value, where the impurity scattering exists (Supplementary \prettyref{fig: figPRQPI}{a-c}). These observations place constraints on the SC order parameter of the Fe(Te,Se) thin flakes (Supplementary \prettyref{fig: figPRQPI}{d-i}).

\section{Measurements with lower electron temperature}\label{SI:multipeak} 
The dI/dV measurements with lower electron temperature show fine features around the SC gap edges. We typically observe 
three features corresponding to energies of approximately ±1.4 meV, ±1.9 meV and ±2.4 meV with their visibility declining at higher energies (\prettyref{fig: fig1}{e}, Extended Data \prettyref{fig: figHRSCgap} and Extended Data \prettyref{fig: figlinecutHR}, electron temperature: 1 K, broadening: 0.3 meV). It is not clear whether they originate from SC quasiparticles, gap anisotropy, or $k$-dependence of the Fermi surface. However, these features appear to show in-phase spatial oscillation, the measured high-resolution dI/dV linecut across a few Se$_{+}$ sites (Extended Data \prettyref{fig: figlinecutHR}{a,b}) shows the same behavior as data in the main text obtained for higher electron temperatures. The precise energy positions of the three spectral features were determined from a multipeak extraction algorithm facilitated by a second-derivative method. The extraction details for each spectrum are shown in Supplementary \prettyref{fig: figMPfit}: the peak searching procedure was applied to the negative second derivative curves (lower panel), while the searching window is defined in between the gray bars (upper panel). The extraction results are indicated as blue dots on each panel. These extracted positions are depicted in Extended Data \prettyref{fig: figlinecutHR}{a}, and displayed individually in Extended Data \prettyref{fig: figlinecutHR}{c-e}. All three features evolve in phase, with a roughly constant gap modulation ratio, about 17.5$\%$ (Extended Data \prettyref{fig: figlinecutHR}{g}).

\section{Extraction of modulation amplitude and phase}\label{SI:2dlockin}
We employed two methods to extract the amplitude and phase from a modulation image, $M(\boldsymbol{r}) = A(\boldsymbol{r})\text{cos}\left[\boldsymbol{\rm Q} \cdot \boldsymbol{r} + \phi(\boldsymbol{r})\right]$. The two dimensional lock-in method \cite{fujita2014direct} converts the real quantity with modulation to be a vector-locked complex quantity, 
\begin{equation}
M_{\mathbf{Q}}(\mathbf{r})=\int \mathrm{d} \mathbf{R} M(\mathbf{R})\ \mathrm{e}^{i \mathbf{Q} \cdot \mathbf{R}} \ \mathrm{e}^{-\frac{(\mathbf{r}-\mathbf{R})^2}{2 \sigma^2}}
\end{equation}
where $\boldsymbol{\rm Q}$ is the vector of modulation, and $\sigma$ is the Gaussian cutoff length, which controls the low-pass frequency of the extraction. To remove the relative oscillation phase ($\boldsymbol{\rm Q} \cdot \boldsymbol{r}$), the Gaussian cutoff $\sigma$ should be larger than $2\pi/\boldsymbol{\rm Q}$. A too small $\sigma$ introduces irrelevant distortion, while a too large $\sigma$ smears out the spatial features. Considering the commensurate length scale (3.8 Å) of our PDM state, we use a reasonably large cutoff ($\sigma=1.3$ nm) to avoid artifacts. 

In practice, the vector-locked complex quantity is calculated in reciprocal space,
\begin{equation}
    M_{\mathbf{Q}}(\mathbf{r})=\mathcal{F}^{-1} M_{\mathbf{Q}}(\mathbf{q})=\mathcal{F}^{-1}\left\{\mathcal{F}\left[M(\mathbf{r}) \mathrm{e}^{\mathrm{i} \mathbf{Q} \cdot \mathbf{r}}\right] \cdot \frac{1}{\sqrt{2 \pi} \sigma_{\mathrm{q}}} \mathrm{e}^{-\frac{q^2}{2 \sigma_{\mathrm{q}}{ }^2}}\right\}
\end{equation}
where $\sigma_{\mathrm{q}}\equiv 1/\sigma$, $\mathcal{F}$ and $\mathcal{F}^{-1}$ are Fourier transform operation and inverse Fourier transform operation, respectively. Then the modulation amplitude and phase can be extracted as,
\begin{equation}\label{eq:12}
    A(\mathbf{r}) = \sqrt{\left[\text{Re}M_{\mathbf{Q}}(\mathbf{r})\right]^{2}+\left[\text{Im}M_{\mathbf{Q}}(\mathbf{r})\right]^{2}}; \   \
    \phi(\mathbf{r}) = \text{tan}^{\text{-1}}\frac{\text{Im}M_{\mathbf{Q}}(\mathbf{r})}{\text{Re}M_{\mathbf{Q}}(\mathbf{r})}
\end{equation}

We checked the validity of 2D lock-in method by a less sophisticated but more direct, moving window method\cite{kim2023imaging}. In this method, the field of view is cropped to a small 1.3 $\times$ 1.3 nm$^{\text{2}}$ window. The relative modulation amplitude (or phase) can be extracted from FT magnitude (or FT phase) within the window at the modulation vector $\mathbf{Q}$. Moving the window pixel by pixel throughout the field of view, a spatial map of modulation amplitude (or modulation phase) can be obtained. Although the small window size leads to low signal-to-noise-ratio of the extraction maps, the moving window method qualitatively matches results obtained by the 2D lock-in method (Supplementary \prettyref{fig: figpll}).

\section{Discussion of drift and shearing corrections}\label{SI:distortion}
Atoms slightly deviating from their perfect positions induce the so-called slowly varying lattice distortion, which breaks the perfect periodicity of the lattice. This distortion can be corrected by Lawler-Fujita algorithm \cite{lawler2010intra}. A perfect square lattice takes the form,
\begin{equation}\label{eq:14}
T_0(\boldsymbol{r}_0)
=\sum_{\boldsymbol{\rm{Q}} = \boldsymbol{\rm P}_{\rm X},\boldsymbol{\rm P}_{\rm Y}}\left|T_{\boldsymbol{\rm Q}}\right| \ \text{cos} \left[\boldsymbol{{\rm Q}} \cdot \boldsymbol{r}_0+\Bar{\phi}_{\boldsymbol{{\rm Q}}}^{T}\right]    
\end{equation}
the measured lattice topography is always imperfect with position-dependent lattice phase,
\begin{equation}\label{eq:15}
T(\boldsymbol{r})
=\sum_{\boldsymbol{\rm{Q}} = \boldsymbol{\rm P}_{\rm X},\boldsymbol{\rm P}_{\rm Y}}\left|T_{\boldsymbol{{\rm Q}}}\right| \ \text{cos} \left[\boldsymbol{{\rm Q}} \cdot \boldsymbol{r}+\phi_{\boldsymbol{{\rm Q}}}^{T}(\boldsymbol{r})\right]    
\end{equation}
where the two wavevectors are $\mathbf{P}_{\rm X} = [({\mathbf{P}_{\rm X}})_{x},\ ({\mathbf{P}_{\rm X}})_{y}]$ and $\mathbf{P}_{\rm Y} = [({\mathbf{P}_{\rm Y}})_{x},\ ({\mathbf{P}_{\rm Y}})_{y}]$. We further define the distortion field as, 
\begin{equation}
    \boldsymbol{u}(\boldsymbol{r})\equiv {\boldsymbol{r}_0-\boldsymbol{r}} =
    \left(\begin{array}{l}
    u_{x}(\boldsymbol{r}) \\
    u_{y}(\boldsymbol{r}) \\
    \end{array}\right)
\end{equation} 
which is the local displacement of the real lattice from the perfect lattice. Therefore, the relationship between perfect and distorted lattice is,
\begin{equation}
\begin{pmatrix}
    ({\mathbf{P}_{\rm X}})_{x} & ({\mathbf{P}_{\rm X}})_{y} \\
    ({\mathbf{P}_{\rm Y}})_{x} & ({\mathbf{P}_{\rm Y}})_{y}
\end{pmatrix}
\left(\begin{array}{l}
u_{x}(\boldsymbol{r}) \\
u_{y}(\boldsymbol{r}) \\
\end{array}\right)
= \left(\begin{array}{l}
\phi_{\mathbf{P}_{\rm X}}^{T}(\boldsymbol{r})-\Bar{\phi}_{\mathbf{P}_{\rm X}}^{T} \\
\phi_{\mathbf{P}_{\rm Y}}^{T}(\boldsymbol{r})-\Bar{\phi}_{\mathbf{P}_{\rm Y}}^{T} \\
\end{array}\right)
\end{equation}
Since $\Bar{\phi}_{\boldsymbol{{\rm Q}}}^{T}$ only introduces a rigid shift of the entire image, for simplicity, we set them to be zero in the following. By applying matrix inversion, $\boldsymbol{u}(\boldsymbol{r})$ can be calculated as,
\begin{equation}\label{eq:17}
    u_{x}(\boldsymbol{r}) = \frac{({\mathbf{P}_{\rm Y}})_{y}\phi_{\mathbf{P}_{\rm X}}^{T}(\boldsymbol{r})-({\mathbf{P}_{\rm X}})_{y}\phi_{\mathbf{P}_{\rm Y}}^{T}(\boldsymbol{r})}{({\mathbf{P}_{\rm X}})_{x}({\mathbf{P}_{\rm Y}})_{y}-({\mathbf{P}_{\rm X}})_{y}({\mathbf{P}_{\rm Y}})_{x}}
\end{equation}
\begin{equation}
     u_{y}(\boldsymbol{r}) = \frac{({\mathbf{P}_{\rm X}})_{x}\phi_{\mathbf{P}_{\rm Y}}^{T}(\boldsymbol{r})-({\mathbf{P}_{\rm Y}})_{x}\phi_{\mathbf{P}_{\rm X}}^{T}(\boldsymbol{r})}{({\mathbf{P}_{\rm X}})_{x}({\mathbf{P}_{\rm Y}})_{y}-({\mathbf{P}_{\rm X}})_{y}({\mathbf{P}_{\rm Y}})_{x}}
\end{equation}
The slowly varying phase, $\phi_{\boldsymbol{{\rm Q}}}^{T}(\boldsymbol{r})$, can be extracted from the 2D lock-in method. Therefore a Lawler-Fujita corrected image can be obtained by making the following transform, $T_0(\boldsymbol{r}_0)
=T\left[\boldsymbol{r}+\boldsymbol{u}(\boldsymbol{r})\right]$. 

Shear distortion artifact, which breaks the $C_{4}$ rotational symmetry of the lattice, appears, for example, when unidirectional drift happens during SI-STM measurement. Following the literature\cite{liu2021discovery}, we describe the image shear with an X-shearing matrix $\mathbf{S}$ and a shear angle, $\theta_{\text{shear}}$, which is the angle between the shearing axis and the horizontal axis of the image (Supplementary \prettyref{fig: figShear}{g}). Then the relationship between sheared ($x$,~$y$) and unsheared ($x_0$,~$y_0$) coordinate can be expressed as,
\begin{equation}\label{eq:19}
\left(\begin{array}{l}
x \\
y
\end{array}\right)=\mathbf{R}({\theta_{\text{shear}}})\mathbf{S}\mathbf{R}({-\theta_{\text{shear}}})\left(\begin{array}{c}
x_0 \\
y_0
\end{array}\right)=\mathbf{R}({\theta_{\text{shear}}})\left(\begin{array}{cc}
1 & Sh \\
0 & 1
\end{array}\right)\mathbf{R}({-\theta_{\text{shear}}})\left(\begin{array}{l}
x_0 \\
y_0
\end{array}\right)
\end{equation}
where, $Sh$ is the shearing strength, $\mathbf{R}(\theta)$ is the standard 2D rotation matrix. In reciprocal space, it shears the two adjacent Bragg vectors (we note the perfect Bragg vectors as $\mathbf{Q}^{0}_{1}$, $\mathbf{Q}^{0}_{2}$). The coordinates of sheared Bragg vectors $\mathbf{Q}_{1}$, $\mathbf{Q}_{2}$ are,
\begin{equation}\label{eq:20}
    \left(\begin{array}{l}
(\mathbf{Q}_{i})_{x} \\
(\mathbf{Q}_{i})_{y}
\end{array}\right)=\left(\begin{array}{l}
|q| \cos (\alpha_{\text{lat}}^{i})+ Sh |q| \sin ( \theta_{\text{shear}}) \cos (\theta_{\text{shear}}-\alpha_{\text{lat}}^{i}) \\
|q| \sin (\alpha_{\text{lat}}^{i})-Sh |q| \cos( \theta_{\text{shear}}) \cos (\theta_{\text{shear}}-\alpha_{\text{lat}}^{i})
\end{array}\right),\ i=1,2
\end{equation}
where $|q|=|\mathbf{Q}^{0}_{1}|=|\mathbf{Q}^{0}_{2}|$, $\alpha_{\text{lat}}^{i} = \alpha_{\text{lat}}, \alpha_{\text{lat}}+\pi/2$ for $i=1,2$ respectively, $\alpha_{\text{lat}}$ is the lattice angle, defined as the angle between $\mathbf{Q}^{0}_{1}$ and horizontal axis of the image. The $\mathbf{Q}_{1}$, $\mathbf{Q}_{2}$ can be measured from the sheared image, with these inputs, another four parameters of Eq.~\ref{eq:20} can be solved numerically by global fitting procedure. Finally, the shear corrected image can be obtained by following transform,
\begin{equation}
   \boldsymbol{r}_0=\mathbf{R}({\theta_{\text{shear}}})\mathbf{S}\mathbf{R}({-\theta_{\text{shear}}})\boldsymbol{r} 
\end{equation}
For example, we corrected data with a severe drift by our shear correction code (Supplementary \prettyref{fig: figShear}{b}). Our code captured the parameters of the image shear and with these parameters we reproduced the sheared image as shown in Supplementary \prettyref{fig: figShear}{a,c}. 

We note that the lattice distortion is negligible in our SI-STM dataset shown in the main figures and to avoid unnecessary data processing, we used the raw data for further analysis (the corresponding corrected topographies utilizing Lawler-Fujita and shearing correction code are shown in Supplementary \prettyref{fig: figcorrect}a-c, demonstrating the negligible lattice distortion). 

\section{Curvature methods}\label{SI:curvature}
We employed the curvature method\cite{zhang2011precise} to improve the feature visualization in the dI/dV linecuts measured on vortex core (Extended Data \prettyref{fig: figlinecutfield}), high impurity concentration area (Extended Data \prettyref{fig: figlinecutdirty}) as well as the spectra measured on the thick flake (Supplementary \prettyref{fig: figthickflake}). The curvature method is effectively similar to the second-derivative method but improves the localization of the extrema and reduces the peak broadening, resulting in better visualization of spectral features on intensity image plots.

\section{Simulations of dislocation and gap modulation}\label{SI:naivesimu} 
Phenomenological simulations $\text{\#}$1: the topological defects shown in Extended Data \prettyref{fig: topodefect}. The spatial winding of the modulation phase can be described as  
\begin{equation}
\phi_{\text{D}}(x,y) = m\ \text{tan}^{\text{-1}}\left(\frac{x\ \text{sin}\,\alpha_{\text{dis}}-y\ \text{cos}\,\alpha_{\text{dis}}}{-x\ \text{cos}\,\alpha_{\text{dis}}-y\ \text{sin}\,\alpha_{\text{dis}}}\right)+\phi_{0}    
\end{equation}
where $\alpha_{\text{dis}}$ is the orientation angle of the dislocation, $m$ is 1 or 0.5 for integer or half-integer topological defects, $\phi_0$ is global phase shift. Then the dislocation was simulated as, 
\begin{equation}
    D(x,y) = D_{0}\text{cos}\left[\mathbf{Q}\cdot \boldsymbol{r}+\phi_{\text{D}}(x,y)\right]
\end{equation} 
where $\mathbf{Q}$ is the wavevector of the ordered state. Moreover, simple simulation of a vortex-antivortex pair is shown in Supplementary \prettyref{fig: figvortexpair}. Following the literature\cite{lee2016structure}, phase field of the pair is simulated by 
\begin{equation}
    \phi(\textit{x},\textit{y}) = \text{tan}^{\text{-1}}\left[\frac{4l_{v}\textit{y}\,C(\textit{y})}{4\textit{x}^{2}+4\textit{y}^{2}-l_{v}^{2}}\right]+\pi
\end{equation}
where $l_{v}$ is the distance between vortex and antivortex, and the shape factor is 
\begin{equation}
    C(\textit{y}) = \left|1+A-\frac{|\textit{y}|}{l_{v}}\right|^{\frac{1}{B}}
\end{equation}

Phenomenological simulations $\text{\#}$2: PDM linecut and induced LDOS modulation shown in Supplementary \prettyref{fig: figPDMinduceCDW} (see also the next section). The coherence peaks were simulated by a simple Gaussian function, 
\begin{equation}
   G_{0}(E,x) = \frac{1}{\sqrt{2\pi}\sigma}[\mathrm{e}^{\frac{-(E-E_{0})^{2}}{2\sigma^{2}}}+\mathrm{e}^{\frac{-(E+E_{0})^{2}}{2\sigma^{2}}}] 
\end{equation}
where,
\begin{equation}
    E_{0}(x) = \Delta_{0}+\Delta_{\text{p}}\text{cos}({\rm Q}_{x}x +\phi)
\end{equation} 
is the spatial-modulated SC gap. The background above the SC gap was simulated by a modified Fermi-Dirac function,
\begin{equation}
    G_{1}(E,x) = 1/[\mathrm{e}^{(E+E_0)/k_{\text{B}}T}+1]
\end{equation}
and,
\begin{equation}
    G_{2}(E,x) = 1-1/[\mathrm{e}^{(E-E_0)/k_{\text{B}}T}+1]
\end{equation}
Therefore, the simulated PDM linecut can be produced by adding the three terms, 
\begin{equation}
    G(E,x) = G_{0}(E,x) + G_{1}(E,x) + G_{2}(E,x)
\end{equation}
Then the induced LDOS modulation was simulated by applying Fourier transform on the PDM linecut (Supplementary \prettyref{fig: figPDMinduceCDW}b,d).

\section{PDM state induced LDOS modulation}\label{SI:induce}
In order to show that the observed LDOS imbalance at Fe$_{x}$/Fe$_{y}$ sublattices (\prettyref{fig: fig4}) is not a consequence of modulating SC coherence peak, 
we performed the corresponding data analysis in the bias range well outside the peak regions. Here we further justify this approach. Gap modulation can, in principle, induce LDOS modulation at the energies around SC coherence peak. Assuming a perfect gap modulation following $\Delta(\boldsymbol{r}) = \Delta_{0}+\Delta_{1}\cos (\boldsymbol{\rm Q} \cdot \boldsymbol{r})$, the induced LDOS modulation can be demonstrated by investigating the spatial distribution of LDOS at some typical energies (Supplementary \prettyref{fig: figPDMinduceCDW}a). For example, at the energies of gap maximum ($E_1 =\Delta_{0}+\Delta_{1}$), the LDOS is maximized at positions where $\Delta(\boldsymbol{r})=E_1$ (one time per $2\pi/\boldsymbol{\rm Q}$), thus giving rise to LDOS modulation with the same vector $\boldsymbol{\rm Q}$ as the gap modulation (the same situation also happens for $E_2 = \Delta_{0}-\Delta_{1}$). Moreover, at the energy of gap average ($E_0=\Delta_{0}$), the LDOS is maximized at positions where $\Delta(\boldsymbol{r})=E_0$ (two times per $2\pi/\boldsymbol{\rm Q}$), thus give arise to LDOS modulation with a wavevector equal to 2$\boldsymbol{\rm Q}$.

Phenomenological simulations (see SI Section \ref{SI:naivesimu}) of these induced LDOS modulation in the PDM state are shown in Supplementary \prettyref{fig: figPDMinduceCDW}a-e, which are fully consistent with our measurements (Supplementary \prettyref{fig: figPDMinduceCDW}{f-i}). The induced LDOS modulation is prominent at the energies around the SC coherent peaks, relatively weak deep within the superconducting gap, and completely disappears at the energies well above the gap. The energy threshold for the existence of this modulation is affected by the peak broadening, leading to the threshold being slightly larger than the gap maximum. 

Note that the Fe$_{x}$-Fe$_{y}$ sublattice LDOS imbalance (or, equivalently, the normal state nematicity, \prettyref{fig: fig4}) is measured sufficiently far above the SC gap, and is not affected by the induced LDOS modulation discussed here. Therefore, in the data analysis of the LDOS imbalance (\prettyref{fig: fig4} and Extended Data \prettyref{fig: figLseg}), we use the dI/dV map at energies above the threshold for vanishing of the induced modulations (which is $\pm$4 meV in our case), to ensure the extracted features do not originate from this induced LDOS modulation.

\section{Extraction of modulation ratio}\label{SI:extraction}

As discussed in the main text, the total gap is expressed as
\begin{equation}
\Delta(\boldsymbol{r}) = \Delta_{0}(\boldsymbol{r})+ \sum_{\boldsymbol{\rm{Q}} =  \boldsymbol{\rm P}_{\rm X},\boldsymbol{\rm P}_{\rm Y}}\left|\Delta_{\boldsymbol{{\rm Q}}}(\boldsymbol{r})\right| \text{cos} \left[\boldsymbol{{\rm Q}} \cdot \boldsymbol{r}+\phi_{\boldsymbol{{\rm Q}}}^{\Delta}(\boldsymbol{r})\right]    
\end{equation}
Thus the unidirectional gap modulation ratio is defined as the ratio between the amplitudes of modulating and non-modulating components,
\begin{equation}            \left|\Delta_{\boldsymbol{{\rm Q}}}(\boldsymbol{r})\right|/\Delta_{0}(\boldsymbol{r})
\end{equation}
In Extended Data \prettyref{fig: figcompete}a, we summarized the unidirectional gap modulation ratio among multiple materials in the literature. When the gap modulation was measured by the spatial variation of the energy positions of SC coherence peaks, the ratio $\left|\Delta_{\boldsymbol{{\rm Q}}}(\boldsymbol{r})\right|/\Delta_{0}(\boldsymbol{r})$ could be directly extracted. In references\cite{hamidian2016detection,liu2021discovery,chen2022identification}, the modulation was identified by measuring the spatial variation of Josephson current, which is proportional to Cooper pair density ($n_{\rm s}$). In these cases, the gap modulation ratio $\left|\Delta_{\boldsymbol{{\rm Q}}}(\boldsymbol{r})\right|/\Delta_{0}(\boldsymbol{r})$ can not be directly read out from the data, but simple conversion method can be used for the extraction. Considering the relationship $n_{\rm s} \propto (I_{\rm c}R_{\rm N})^{2}$ and  $I_{\rm c}R_{\rm N} = [\frac{\pi\Delta}{2e}\text{tanh}(\frac{\Delta}{2k_{\rm B}T})]$, we got 
\begin{equation}
    \sqrt{n_{\rm s}}(\boldsymbol{r})\propto\Delta(\boldsymbol{r})
\end{equation}
at low temperatures. The data points extracted through this method are shown as open symbols in Extended Data \prettyref{fig: figcompete}a.

Finally, we discuss the derivation of the gap difference between the neighboring Fe$_{x}$ and Fe$_{y}$ atoms, defined as,
\begin{equation}
   p_{\Delta}(\boldsymbol{r}) \equiv \frac{|\Delta_{\text{Fe}_{x}}-\Delta_{\text{Fe}_{y}}|}{\Delta_{\text{Fe}_{x}}+\Delta_{\text{Fe}_{y}}}(\boldsymbol{r})
\end{equation}
where $\Delta_{\text{Fe}_{x}}$ and $\Delta_{\text{Fe}_{y}}$ are the SC gaps on the two neighboring iron atoms Fe$_{x}$/Fe$_{y}$. Since the non-modulating SC component can be expressed as, $\Delta_{0}=(\Delta_{\text{Fe}_{x}}+\Delta_{\text{Fe}_{y}})/2$ and the energy different between the SC gaps of the two iron sites can be expressed as $|\Delta_{\text{Fe}_{x}}-\Delta_{\text{Fe}_{y}}|=2(|\Delta_{\boldsymbol{{\rm P}_{\text{X}}}}|+|\Delta_{\boldsymbol{{\rm P}_{\text{Y}}}}|)$, the $p_{\Delta}$ can be calculated by the sum of the two unidirectional gap modulation ratios,
\begin{equation}
   p_{\Delta}(\boldsymbol{r}) \equiv \frac{|\Delta_{\text{Fe}_{x}}-\Delta_{\text{Fe}_{y}}|}{\Delta_{\text{Fe}_{x}}+\Delta_{\text{Fe}_{y}}}(\boldsymbol{r}) = \frac{|\Delta_{\boldsymbol{{\rm P}}_{\rm X}}(\boldsymbol{r})|+|\Delta_{\boldsymbol{{\rm P}}_{\rm Y}}(\boldsymbol{r})|}{\Delta_{0}(\boldsymbol{r})}
\end{equation}
The extracted map of $p_{\Delta}$ of the SI-STM data of the main text is shown in Extended Data \prettyref{fig: figcompete}d.

\section{Other supporting display items}

\noindent (S01) List of Symbols (Supplementary Table \ref{table: tabs1}).

\begin{figure}[p]
\begin{center}
    \includegraphics[width=15cm]{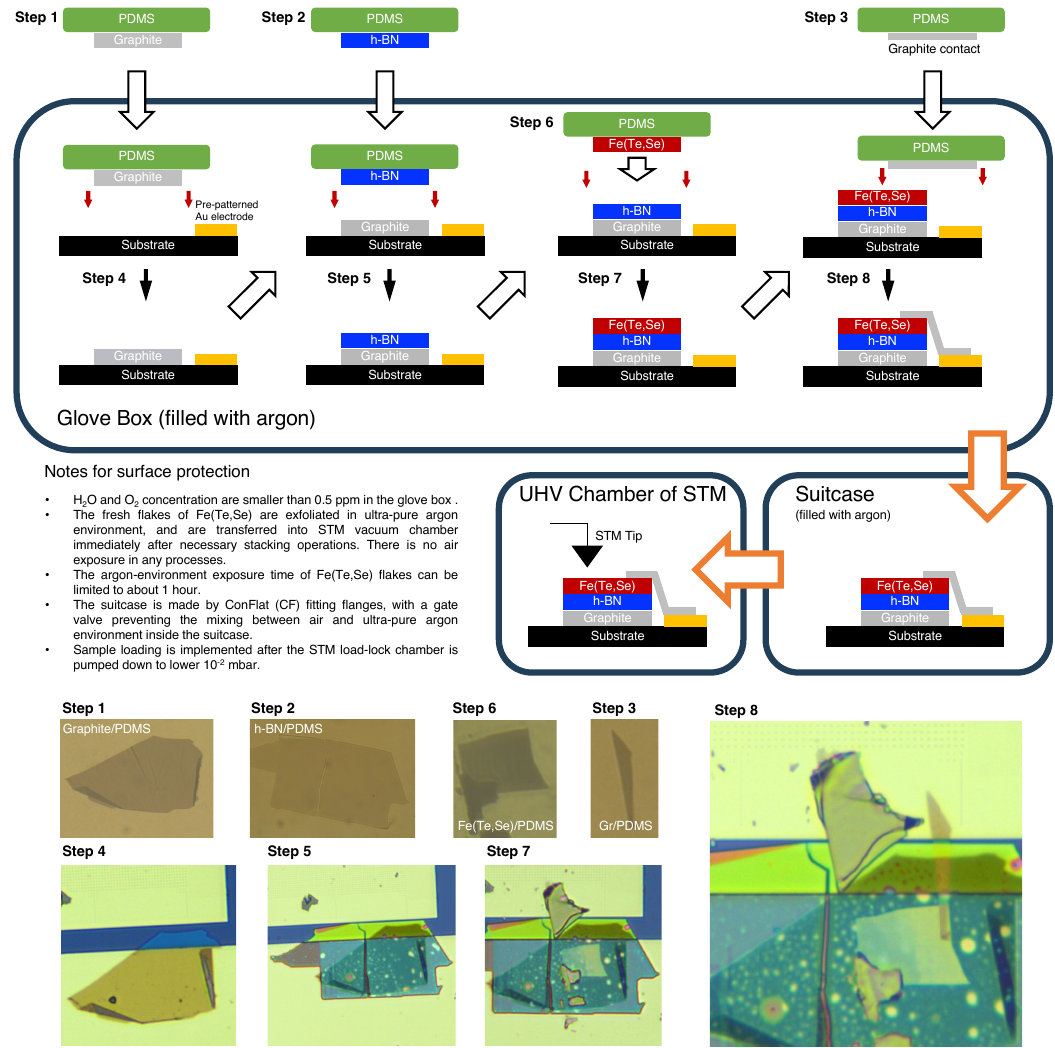}
\end{center}
\caption{{\bf Sample fabrication.}
Thin flakes of bottom graphite (Step {\bf 1}) and hexagonal boron nitride (h-BN) (Step {\bf 2}) were prepared in advance on polydimethylsiloxane (PDMS). After transferring them onto pre-patterned gold markers (Steps {\bf 4} and Steps {\bf 5}), a freshly exfoliated flake of Fe(Te,Se) on PDMS (Step {\bf 6}) was transferred onto h-BN (Step {\bf 7}) inside an argon-filled glovebox with oxygen and moisture concentration below 0.5 ppm, followed by placing a top graphite contact (Step {\bf 8}), which is also prepared in advance (Step {\bf 3}). The lower panels show optical images corresponding to each step of sample fabrication. The van der Waals stack was transferred into ultra-high vacuum (UHV) chamber immediately through a home-made air-tight suitcase. The total dwell time of the Fe(Te,Se) thin flakes under ultra-pure argon environment ($t_{\text{Ar}}$) was about 1 hour. The short argon exposure time ($t_{\text{Ar}}$) is crucial for achieving sufficiently high surface quality. A history of device quality optimization can be found in Supplementary \prettyref{fig: fighistory}. 
}
\label{fig: figfab}
\end{figure}

\noindent (S02) Sample fabrication (Supplementary \prettyref{fig: figfab}).

\begin{figure}[p]
\begin{center}
    \includegraphics[width=\linewidth]{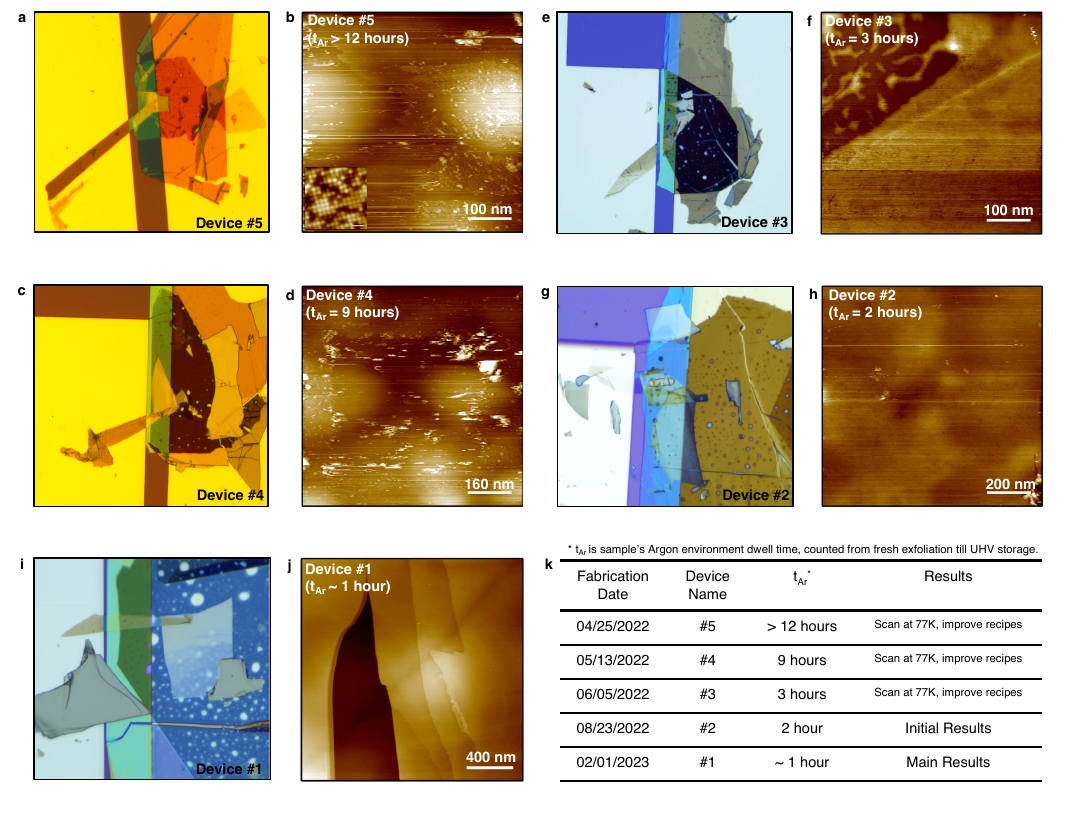}
\end{center}
\caption{
{\bf  History of device quality optimization.} All five devices were fabricated inside an argon-filled glovebox, with oxygen and moisture concentrations below 0.5 ppm. The details and steps of sample fabrication procedure are shown in Supplementary \prettyref{fig: figfab}. In all cases, the Fe(Te,Se) flakes are never exposed to air. {\bf a, b}, Optical image and micrometer-sized STM topography of device $\text{\#}$5. {\bf c, d}, Optical image and micrometer-sized STM topography of device $\text{\#}$4. {\bf e, f}, Optical image and micrometer-sized STM topography of device $\text{\#}$3. {\bf g, h}, Optical image and micrometer-sized STM topography of device $\text{\#}$2. The spectroscopic measurements on this device are shown in Extended Data \prettyref{fig: figlinecutdevice2}. {\bf i, j}, Optical image and micrometer-sized STM topography of device $\text{\#}$1. Most of the data of this work is from this device. {\bf k}, Summary of the fabrication information of the five devices. The argon dwell time ($t_{\text{Ar}}$) is defined as the time span between the exfoliation of Fe(Te,Se) flakes and the moment that the sample was transferred into the STM UHV chamber. $t_{\text{Ar}}$ is found crucial for surface quality. The devices $\text{\#}$5 and $\text{\#}$4 were fabricated before optimizing the fabrication procedures, and a significant amount of impurities is observed on STM topography ({\bf b, d}). In device $\text{\#}$4, a square region in the middle of the field of view was cleaned up by STM tip through constant-current-mode scanning. In device $\text{\#}$5, this cleaning procedure did not 
work. Note that even in the worst scenario (for device $\text{\#}$5), we could find small clean regions after intensive 
search on the surface [see atom-resolved topography in inset of ({\bf b}), (a 5.5$\times$5.5 nm$^{\text{2}}$ scan, scale bar is 1 nm)].}
\label{fig: fighistory}
\end{figure}

\noindent (S03) History of device quality optimization (Supplementary \prettyref{fig: fighistory}).

\begin{figure}[p]
\begin{center}
    \includegraphics[width=15cm]{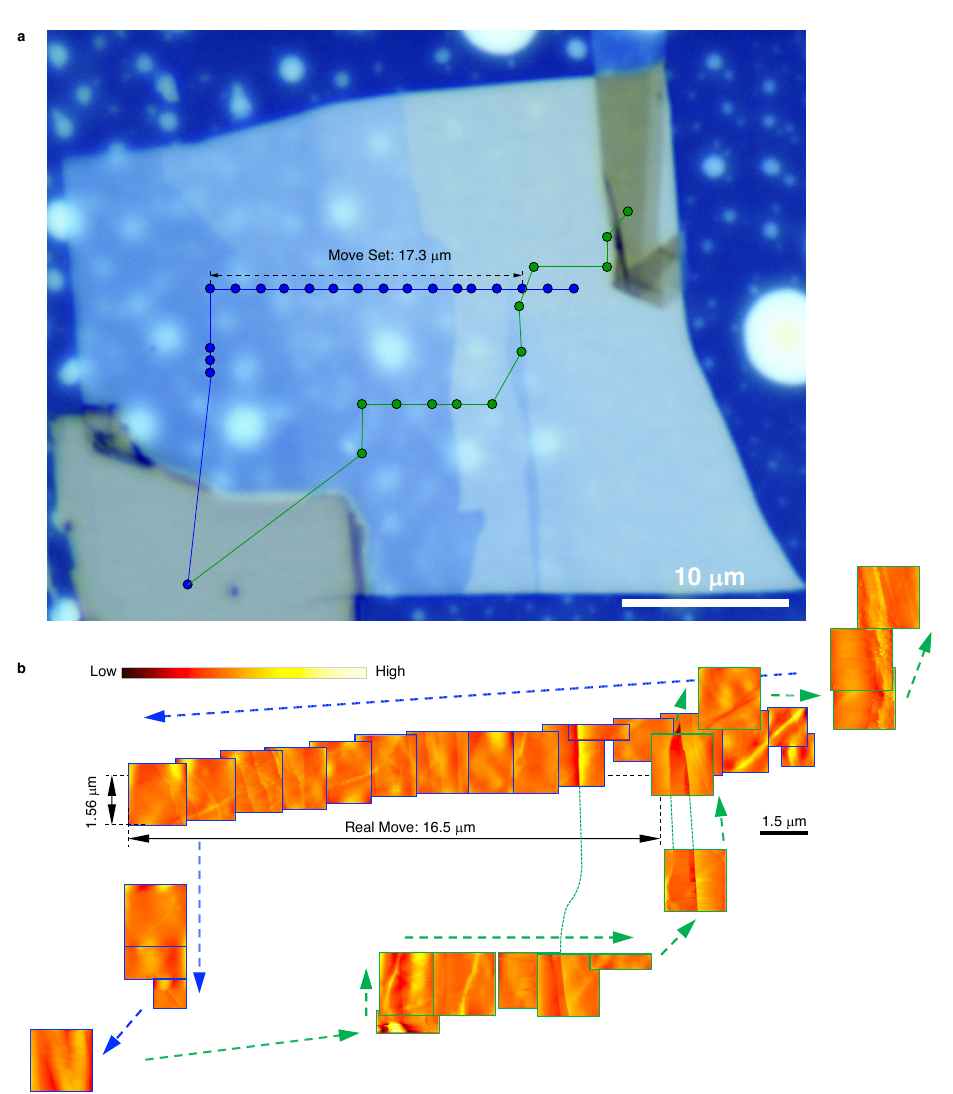}
\end{center}
\caption{{\bf Precise tip navigation on Fe(Te,Se) flakes.}
{\bf a}, High resolution optical image of the Fe(Te,Se) flake of device $\text{\#}$1. The blue and green paths show the tip trajectory driven by walking piezos, and the solid dots indicate the positions where STM topographies were taken. {\bf b}, Micrometer-sized STM images of the corresponding areas. The precise navigation allows us to calibrate walking piezos precision to about 5$\%$ accuracy. 
}
\label{fig: fignevig}
\end{figure}

\noindent (S04) Precise tip navigation on Fe(Te,Se) flakes (Supplementary \prettyref{fig: fignevig}).

\begin{figure}[p]
\begin{center}
    \includegraphics[width=15cm]{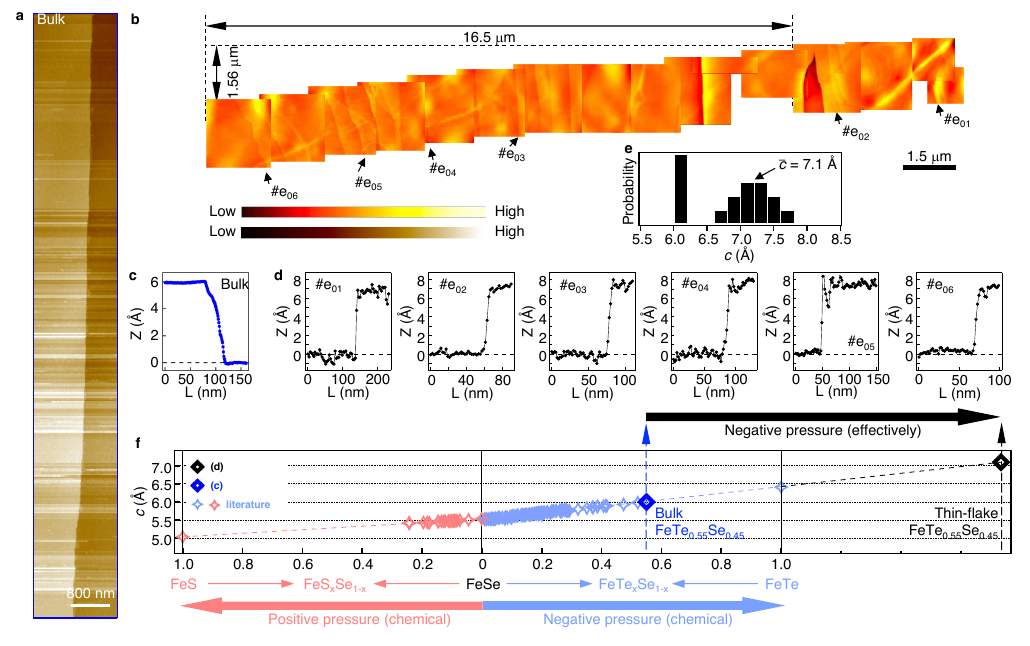}
\end{center}
\caption{{\bf Step height on Fe(Te,Se) thin flakes.}
{\bf a}, STM topography of an in-situ cleaved Fe(Te,Se) bulk single crystal that contains a monolayer step edge. {\bf b}, Micrometer-sized STM topographies of the blue path of Supplementary \prettyref{fig: fignevig}. 
{\bf c}, Averaged horizontal line profile from ({\bf a}), showing step height of around 6 Å in agreement with previous works. {\bf d}, Examples of the height line profiles on the thin flakes. The six step edges are marked in ({\bf b}). We observed expanded \textit{c}- lattice constant (step height), which effectively results in a negative pressure compared to the bulk. 
{\bf e}, Histogram of the \textit{c}- lattice constants among all the step edges shown in ({\bf b}). The central peak of the histogram is around 7.1 Å, which is 18$\%$ larger than that of bulk materials. {\bf f}, Summary of \textit{c}- lattice constant among Fe(Te,Se) and Fe(S,Se) bulk materials\cite{ishida2022pure}. The Te (S) substitution acts as negative (positive) chemical pressure. The \textit{c}- lattice constant of our exfoliated thin-flake Fe(Te,Se) sample was indicated as the black symbol. }
\label{fig: figcheight}
\end{figure}

\noindent (S05) Step height on Fe(Te,Se) thin flakes (Supplementary \prettyref{fig: figcheight}).

\begin{figure}[p]
\begin{center}
    \includegraphics[width=15 cm]{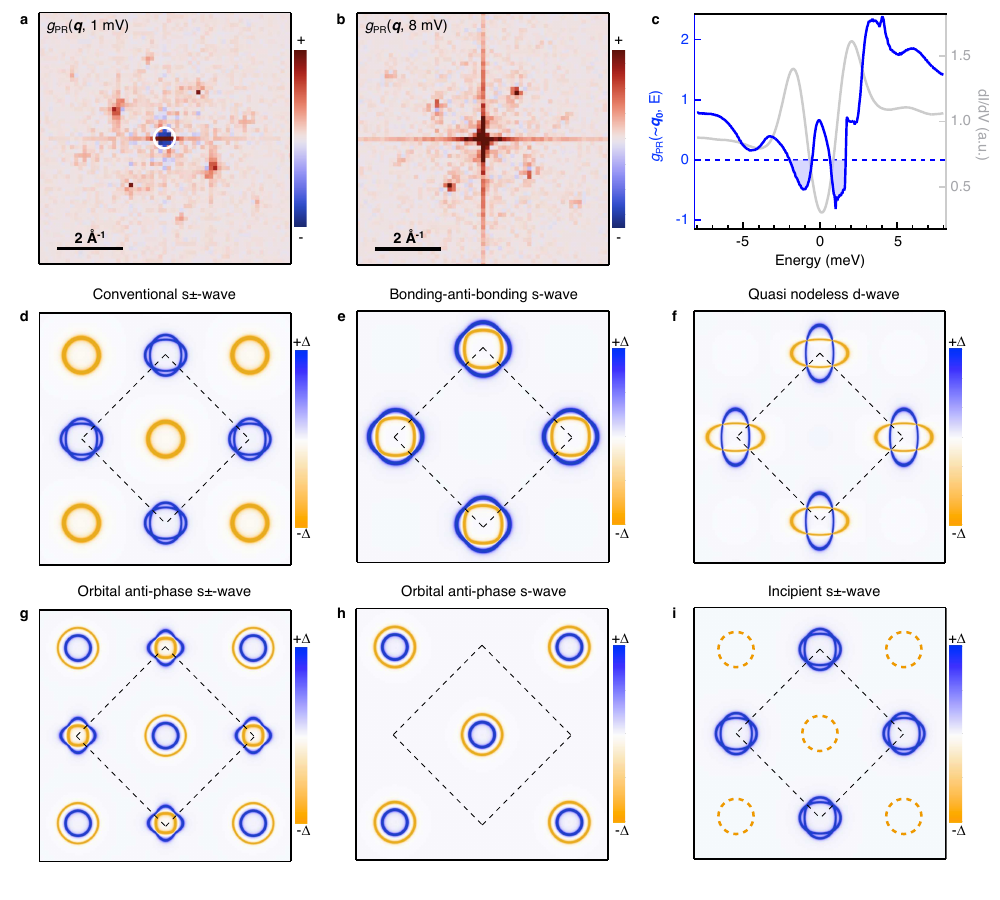}
\end{center}
\caption{
{\bf Possible symmetries of SC order parameter.} {\bf a}, Phase-referenced quasiparticle interference pattern at 1 mV. The negative signal appears around (0,0), supporting the opposite sign of SC order parameter for pockets separated by small scattering wavevector $\boldsymbol{q}$\cite{hirschfeld2015robust,sprau2017discovery,chi2017extracting,gu2019directly,chen2019direct}. {\bf b}, Phase-referenced quasiparticle interference pattern at 8 mV for comparison. There is no pair breaking scattering beyond the SC gap, thus the absence of negative $g_{\text{PR}}(\boldsymbol{q})$ is expected. {\bf c}, Energy-dependent $g_{\text{PR}}(E)$ integrated over the $\boldsymbol{q}$ inside the white circle shown in ({\bf a}). A dI/dV spectrum (gray curve) is appended for reference. The negative $g_{\text{PR}}$ around (0,0) only appears within $\pm$[0.5,1.5] meV. {\bf d-i}, The candidates for SC pairing symmetry. ({\bf d}) and ({\bf g}) correspond to s±-wave pairing in bulk FeSCs, where the sign change is between ${\Gamma}$ and M Fermi surface pockets\cite{hirschfeld2011gap}. Considering orbital dependence, an equal-sign condition among all the orbitals leads to conventional s± wave ({\bf d}), an opposite-sign between ${d_{\text{xy}}}$ and ${d_{\text{xz/yz}}}$ orbitals leads to orbital anti-phase s±-wave ({\bf g})\cite{yin2014spin}. ({\bf i}) is the incipient s±-wave pairing, the sign change is between ${\Gamma}$ and M, but in contrast to the case of normal s± wave, here the Fermi surface only has either the ${\Gamma}$ or the M pockets\cite{chen2015electron,gao2016hidden}. Our quasiparticle interference measurements support a small $\boldsymbol{q}$ sign-changing pairing with only electron or hole pockets, that restricts a possible scenario of pairing symmetry to either bonding-antibonding s-wave ({\bf e}), quasi-nodeless d-wave ({\bf f}), or orbital anti-phase s-wave ({\bf h}) pairings. For the case with only electron Fermi surfaces, theoretically, the ({\bf e}) favors strong interpocket hybridization and weak band anisotropy, while the ({\bf f}) favors the opposite\cite{khodas2012interpocket}. The ({\bf h}) is the case with only ${\Gamma}$ pockets.}
\label{fig: figPRQPI}
\end{figure}

\noindent (S06) Possible symmetries of SC order parameter (Supplementary \prettyref{fig: figPRQPI}).

\begin{figure}[p]
\begin{center}
    \includegraphics[width=15cm]{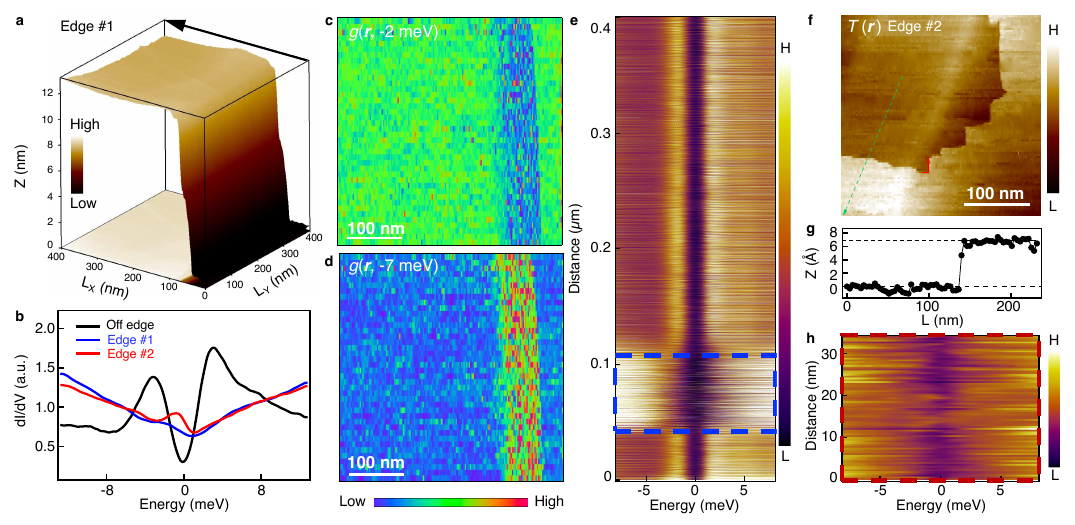}
\end{center}
\caption{{\bf Breakdown of superconductivity at step edges.}
{\bf a, c-e}, STM/S measurements at a 12 nm-height step edge $\text{\#}$1. {\bf a}, STM topography across the step edge. {\bf c, d}, Differential conductance map of the step edge $\text{\#}$1 at -2 mV (around SC coherence peak) and -7 mV (continuum of electronic states). SC is suppressed everywhere along the edge. {\bf e}, dI/dV linecut measured perpendicular to the step edge $\text{\#}$1 [black arrow in ({\bf a})]. {\bf f-h}, STM/S measurements at a monolayer step edge $\text{\#}$2. {\bf f}, The corresponding STM topography of the step edge $\text{\#}$2. {\bf g}, Height line profile along the green dashed line in ({\bf f}). The interlayer spacing is about 7 Å. {\bf h}, dI/dV linecut measured along the step edge $\text{\#}$2 [red arrow in ({\bf f})]. {\bf b}, Summary of spectral behavior on and off the step edge. SC is suppressed on both step edges. An in-gap state appears on step edge $\text{\#}$2 (red curve). Since the step edges are nonmagnetic, the observation of SC breakdown indicates an unconventional sign-changing SC order parameter on the Fe(Te,Se) thin flakes.}
\label{fig: figedgekillsc}
\end{figure}

\noindent (S07) Breakdown of superconductivity at step edges (Supplementary \prettyref{fig: figedgekillsc}).

\begin{figure}[p]
\begin{center}
    \includegraphics[width=12cm]{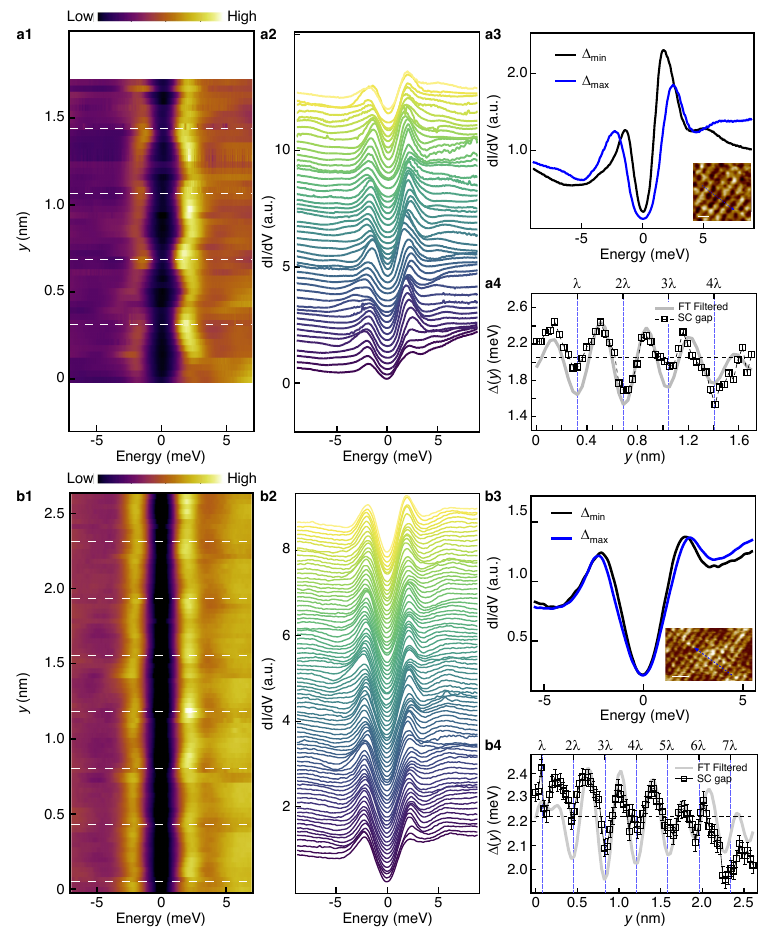}
\end{center}
\caption{{\bf More examples of periodic gap modulation}. Panels {\bf a1-a4} and {\bf b1-b4} show similar measurements as the one shown in \prettyref{fig: fig2} and Extended Data \prettyref{fig: figlinecut2}. {\bf a}, The scale bar in the inset of ({\bf a3}): 0.5 nm. The modulation ratio $|\Delta_{\boldsymbol{\rm P}_{\rm Y}}|/\bar{\Delta}$ is 13.5$\%$. {\bf b}, The scale bar in the inset of ({\bf b3}): 1 nm. The modulation ratio $|\Delta_{\boldsymbol{\rm P}_{\rm Y}}|/\bar{\Delta}$ is 4.5$\%$.}
\label{fig: figlinecut3}
\end{figure}

\noindent (S08) More examples of periodic gap modulation (Supplementary \prettyref{fig: figlinecut3}).

\begin{figure}[p]
\begin{center}
    \includegraphics[width=15cm]{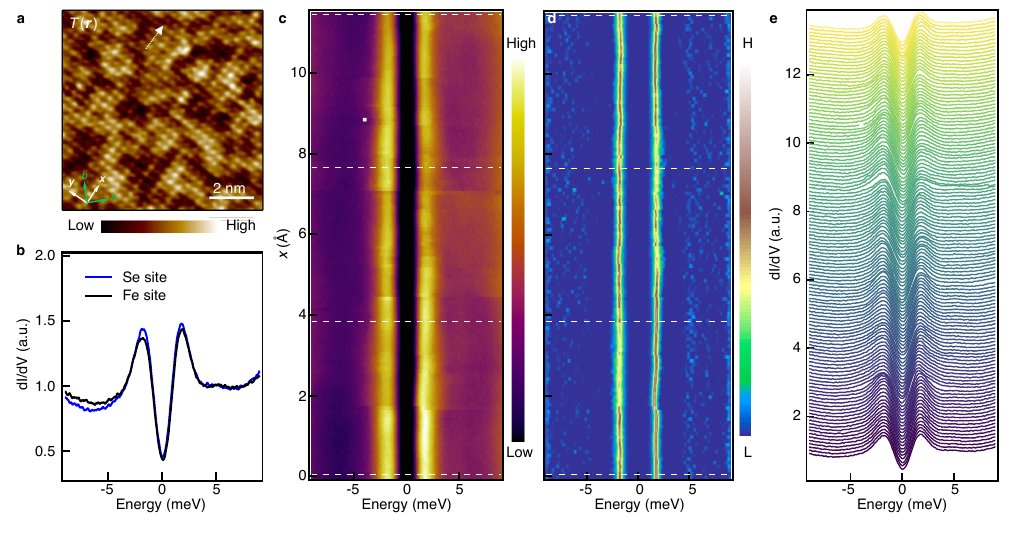}
\end{center}
\caption{{\bf Absence of the gap modulation on 210-nm thick flake.}
{\bf a}, Atom-resolved STM topography. {\bf b}, Spatially-averaged dI/dV spectra at Se$_{+}$ site (black) and Fe$_{x}$ (blue) sites. {\bf c}, False-color plot of dI/dV linecut. {\bf d}, Negative curvature plot of ({\bf c}), see SI Section \ref{SI:curvature}. {\bf e}, Waterfall spectrum plot of ({\bf c}).}
\label{fig: figthickflake}
\end{figure}

\noindent (S09) Absence of gap modulation on 210-nm thick flake (Supplementary \prettyref{fig: figthickflake}).

\begin{figure}[p]
\begin{center}
    \includegraphics[width=\linewidth]{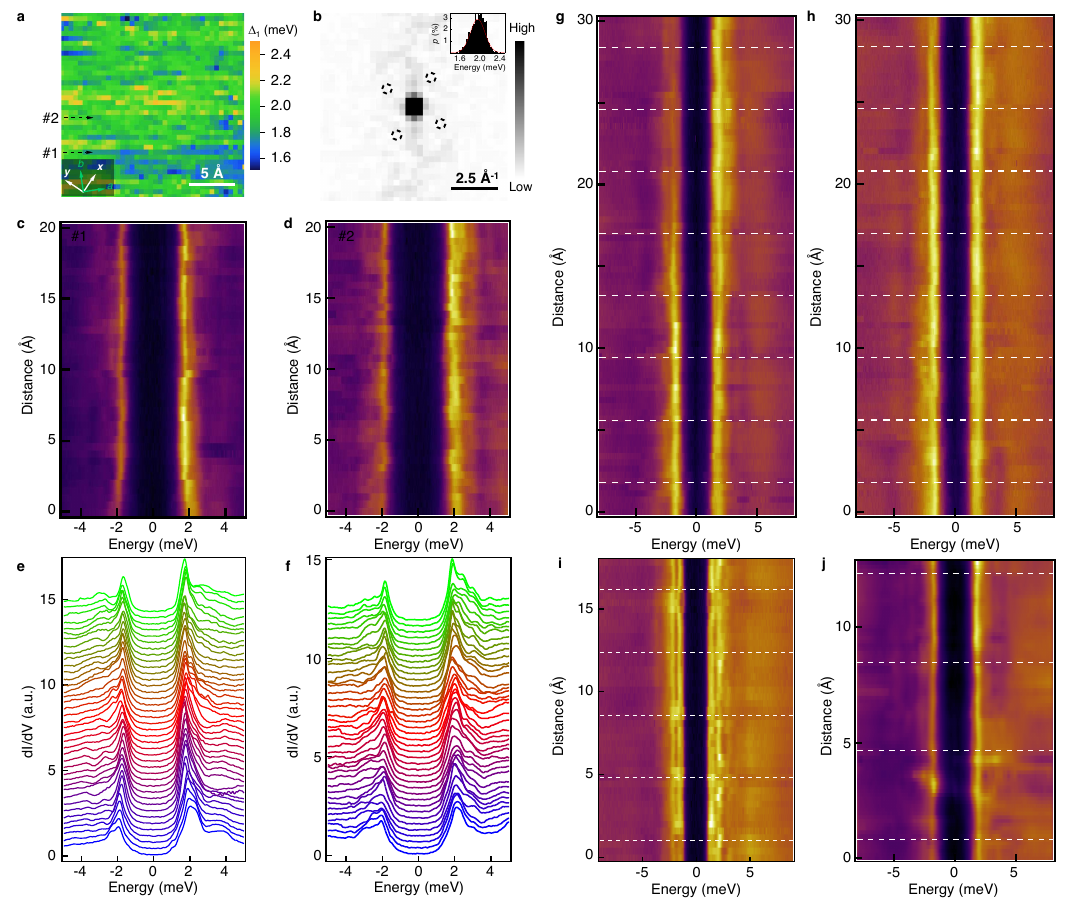}
\end{center}
\caption{
{\bf  Absence of gap modulation on the surface of bulk crystals.} {\bf a}, SC gap map in a 2$\times$2 nm$^{\text{2}}$ area. {\bf b}, FT magnitude of ({\bf a}). Inset: histogram of ({\bf a}). {\bf c, d}, Horizontal dI/dV linecuts; the position of these linecuts are indicated in ({\bf a}), as $\text{\#}$1 and $\text{\#}$2 respectively. {\bf e, f}, Waterfall spectrum plot of ({\bf c}) and ({\bf d}). {\bf g-j}, More examples of dI/dV linecuts measured along $\textit{x}$-axis or $\textit{y}$-axis on the surface of bulk crystals. The white dashed lines indicate the position of Se$_{+}$.}
\label{fig: figbulk}
\end{figure}

\noindent (S10) Absence of gap modulation on bulk crystals (Supplementary \prettyref{fig: figbulk}).

\begin{figure}[p]
\begin{center}
    \includegraphics[width=\linewidth]{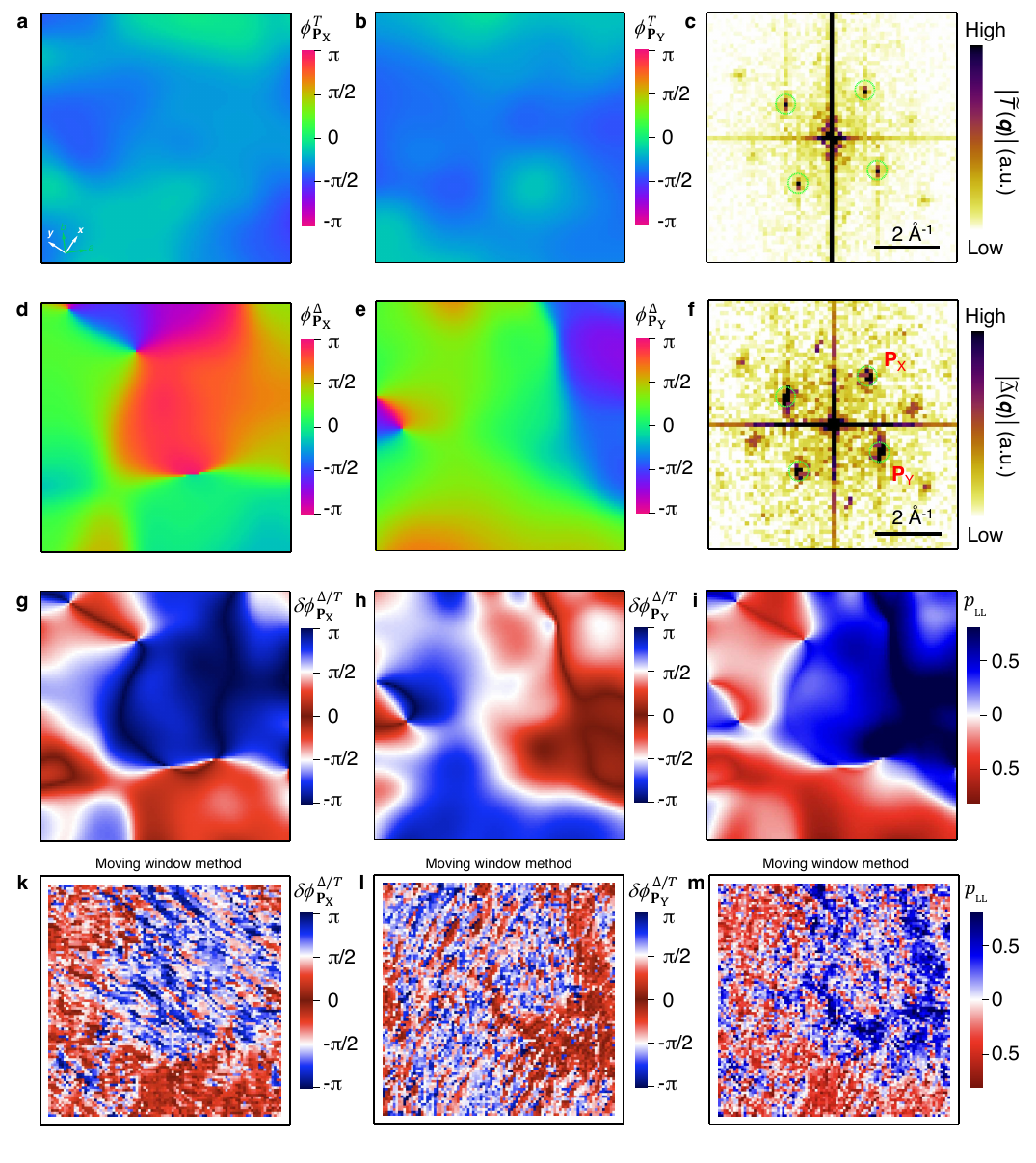}
\end{center}
\caption{
{\bf  Calculation of the PDM lattice-lock-in polarization.} {\bf a, b, d, e} are the phases of the topography and the gap map in $\pm\boldsymbol{\rm P}_{\rm X}$ and $\pm\boldsymbol{\rm P}_{\rm Y}$ directions (see the definition of each quantity in Supplementary Table \ref{table: tabs1}). {\bf c, f}, The FT magnitude of the raw data of topography and gap map. The FT magnitudes on background-subtracted images are shown in \prettyref{fig: fig3}{d}, where the intensity around (0,0) was removed. {\bf g-i} show the phase difference along the two directions and the resulting PDM lattice-lock-in polarization $p_{\rm_{LL}}$ [same panel as ({\bf i}) is shown in \prettyref{fig: fig3}{g}]. {\bf k-m}, Same quantities as ({\bf g-i}), but calculated by using a different, moving window method (see SI Section \ref{SI:2dlockin}). The two methods are qualitatively equivalent to each other ({\bf g-i}).  The Gaussian cut-off length ($\sigma$) of FT was set to 1.3 nm for ({\bf a, b, d, e, g-i}). The window size was set to 1.3 nm for ({\bf k-m}).}
\label{fig: figpll}
\end{figure}

\noindent (S11) Calculation of the PDM lattice-lock-in polarization (Supplementary \prettyref{fig: figpll}).

\begin{figure}[p]
\begin{center}
    \includegraphics[width=\linewidth]{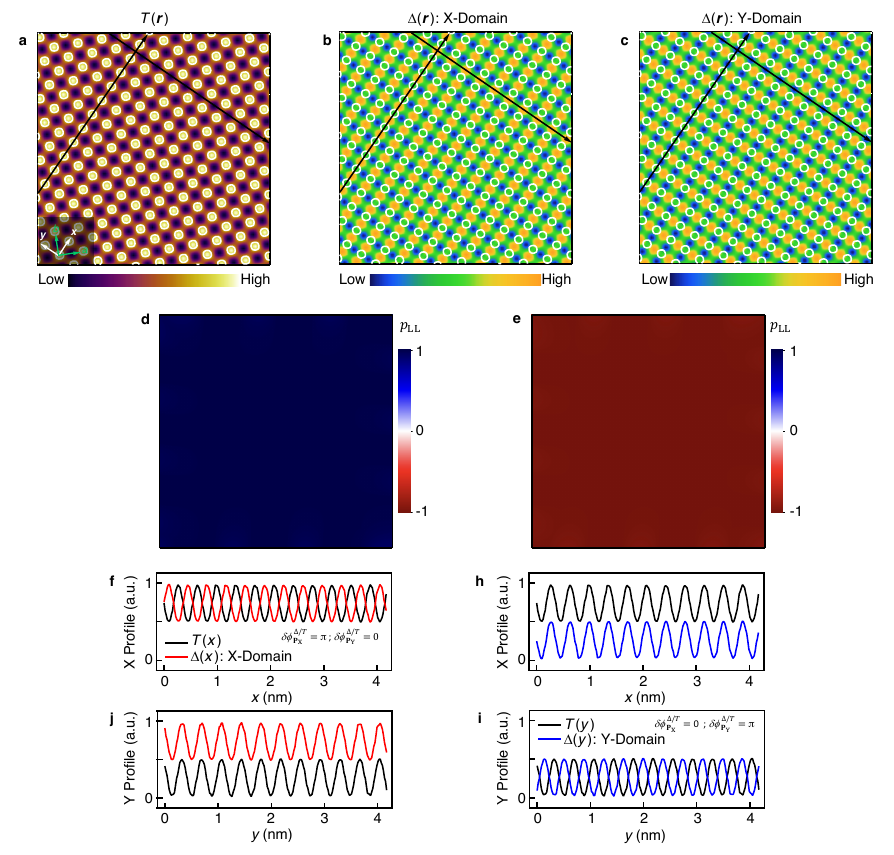}
\end{center}
\caption{
{\bf  Simulation of the PDM lattice-lock-in polarization.} {\bf a}, A simulated Se$_{+}$ lattice. {\bf b}, A simulated gap map in a X-Domain, where the gap maxima coincide with Fe$_{x}$ sites. {\bf c}, A simulated gap map in a Y-Domain, where gap maxima coincide with Fe$_{y}$ sites. The white circles in the three panels mark the positions of Se$_{+}$ atoms. {\bf d}, Calculated lattice-lock-in polarization of X-Domain, which is equal to +1. {\bf e}, Same as ({\bf d}), but for Y-Domain, the calculated polarization is  -1. {\bf f, j}, Line profile of $T(\boldsymbol{r})$ and $\Delta(\boldsymbol{r})$ in a X-Domain, along $\textit{x}$ and $\textit{y}$-axis respectively. The positions of the line profile are indicated as the black lines in ({\bf a-c}). When gap maxima lock to Fe$_{x}$ perfectly, the $\textit{x}$ profile shows a $\pi$ phase difference between $T(\boldsymbol{r})$ and $\Delta(\boldsymbol{r})$, $\left|\delta \phi_{\boldsymbol{{\rm P_{X}}}}^{\Delta/T}(\boldsymbol{r})\right|=\pi$, while the $\textit{y}$ profile shows a zero phase different, $\left|\delta \phi_{\boldsymbol{{\rm P_{Y}}}}^{\Delta/T}(\boldsymbol{r})\right|=0$. Following the definition of lattice-lock-in polarization, $p_{_{\text{LL}}}(\boldsymbol{r})\equiv \left[\left|\delta \phi_{\boldsymbol{{\rm P_{X}}}}^{\Delta/T}(\boldsymbol{r})\right|-\left|\delta \phi_{\boldsymbol{{\rm P_{Y}}}}^{\Delta/T}(\boldsymbol{r})\right|\right]/\pi$, $p_{_{\text{LL}}}(\boldsymbol{r})$ = +1 in the X-Domain. {\bf h, j}, Same as ({\bf f, i}), but calculated for the Y-Domain, -1 polarization can be derived.}
\label{fig: figpll_demo}
\end{figure}

\noindent (S12) Simulation of the PDM lattice-lock-in polarization (Supplementary \prettyref{fig: figpll_demo}).

\begin{figure}[p]
\begin{center}
    \includegraphics[width=15cm]{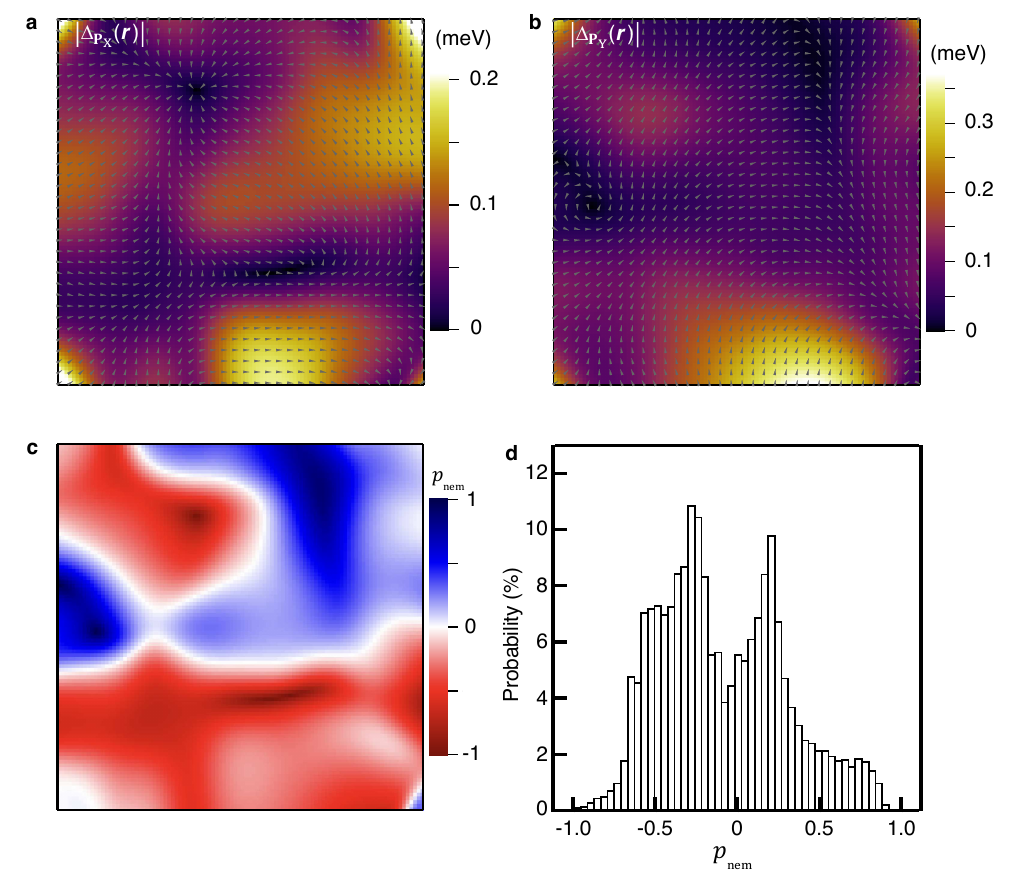}
\end{center}
\caption{{\bf PDM nematicity.} {\bf a, b}, PDM modulation amplitude along the two directions $\pm$$\boldsymbol{\rm P}_{\rm X}$ and  $\pm$$\boldsymbol{\rm P}_{\rm Y}$ of the SI-STM data in \prettyref{fig: fig3}. Small gray arrows superimposed on data depict the local PDM modulation phase (Extended Data \prettyref{fig: topodefect}a,b).
{\bf c}, Nematic polarization ($p\rm_{_{nem}}$) of the PDM state (see definition in Table \ref{table: tabs1}). $p\rm_{_{nem}}$ = +1, if $|\Delta_{\boldsymbol{{\rm P}}_{\rm X}}(\boldsymbol{r})|$$\gg$$|\Delta_{\boldsymbol{{\rm P}}_{\rm Y}}(\boldsymbol{r})|$, and vice versa.   {\bf d}, Histogram of ({\bf c}), the well-separated peaks identify the true dichotomy of differently polarized nematicity on the PDM state. The central nematic strength at each domain is about 30$\%$. ({\bf a, b}) were extracted by the 2D lock-in method, with a Gaussian cut-off 1.3 nm. }
\label{fig: figHist_pnem}
\end{figure}

\noindent (S13) PDM nematicity (Supplementary \prettyref{fig: figHist_pnem}).

\begin{figure}[p]
\begin{center}
    \includegraphics[width=12.5cm]{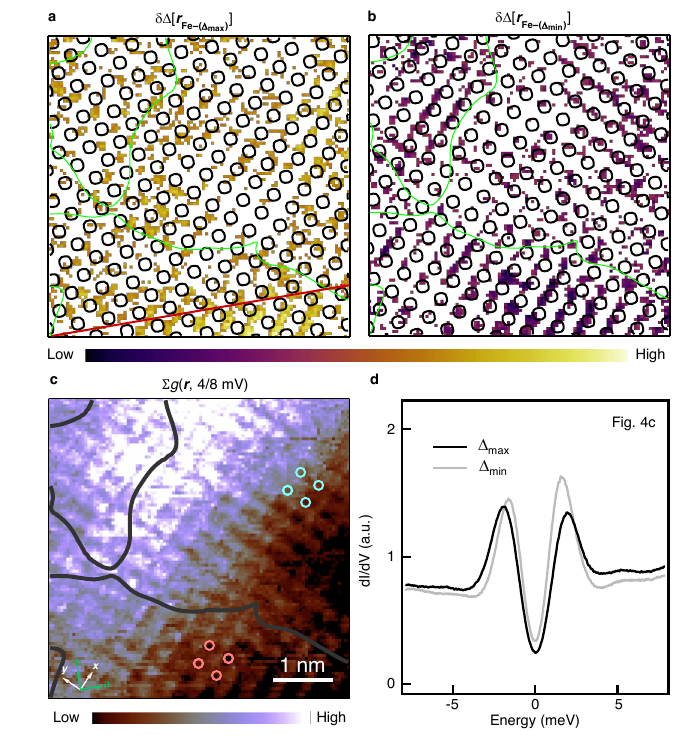}
\end{center}
\caption{
{\bf  SC gap and LDOS behavior on iron sublattices.} {\bf a, b}, Color plots indicating the positions of superconducting gap local maxima and minima. The segregation was performed on the superconducting gap map as described in Methods. 
The areas with $\delta\Delta >$ 0.2$(\delta\Delta)_{\text{max}}$ or $\delta\Delta <$ -0.2$(\delta\Delta)_{\text{min}}$ are depicted by colors in ({\bf a}) or ({\bf b}) respectively, while other areas are depicted in white color. This demonstrates that gap maxima lie on Fe$_{x}$ in one domain and on Fe$_{y}$ in the other domain. 
The red line in ({\bf a}) indicates the positions of the linecut shown in \prettyref{fig: fig4}{c}. This linecut is along $\textit{a}$- axis, avoiding the positions of Se$_{+}$/Se$_{-}$ atoms. The black circles in ({\bf a}) and ({\bf b}) mark the positions of Se$_{+}$ sites. {\bf c}, A similar data as \prettyref{fig: fig4}{d}, but integrated at positive energies. The pink and cyan circles in ({\bf c}) mark the positions of Se$_{+}$ sites. {\bf d}, A similar data as \prettyref{fig: fig4}{b}, but averaged only among the dI/dV spectra in the linecut shown in \prettyref{fig: fig4}{c}.}
\label{fig: figLseg2}
\end{figure}

\noindent (S14) SC gap and LDOS behavior on iron sublattices (Supplementary \prettyref{fig: figLseg2}).

\begin{figure}[p]
\begin{center}
    \includegraphics[width=15 cm]{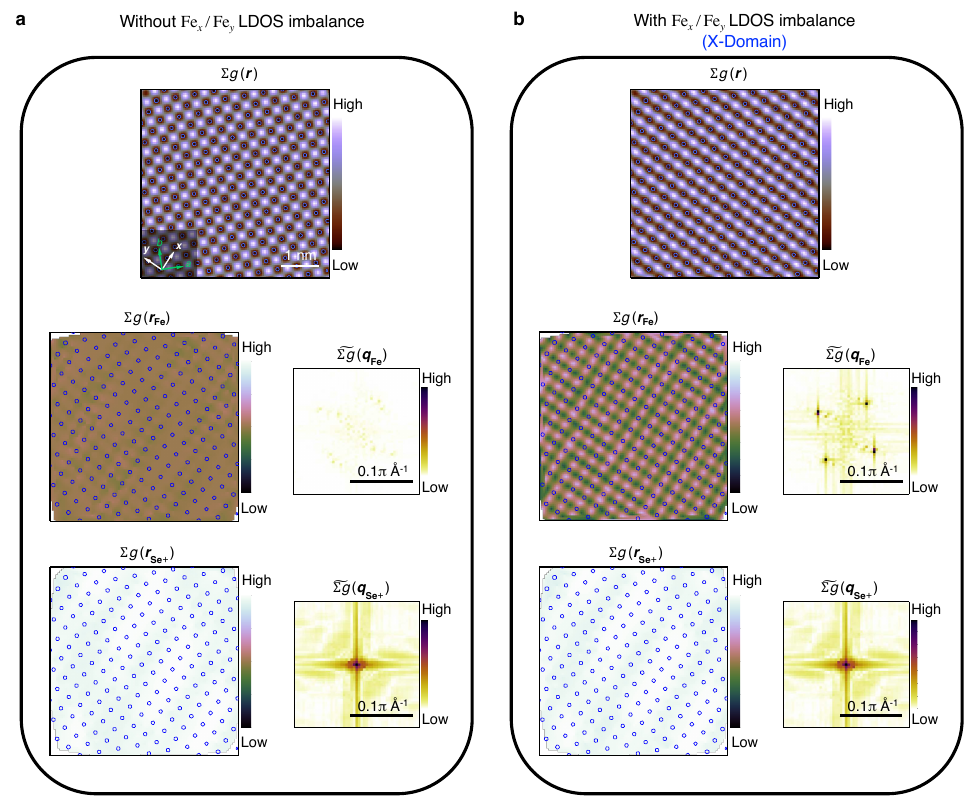}
\end{center}
\caption{
{\bf  Examples of lattice segregation procedure.} The lattice segregation algorithm was described in Methods, and the segregation results of the experimental data were shown in Extended Data \prettyref{fig: figLseg}. In order to validate it, here we simulate two images of LDOS (first row), with ({\bf b}) or without ({\bf a}) LDOS imbalance on Fe$_{x}$/Fe$_{y}$ sublattice, and apply the segregation procedure used in the main text to these simulated images. The LDOS of segregated Fe$_{x}$/Fe$_{y}$ (or Se$_{+}$) sublattice, as well as their corresponding FT magnitude, are shown in the second (third) row. The 
LDOS simulations (first row) are performed by preparing images where LDOS maxima positions are matched with the Se$_{-}$ sites (same as our experimental observations, see \prettyref{fig: fig1}g). In ({\bf b}), we assume a X-Domain throughout the area. The stripe feature was generated by taking the Fe$_{x}$ sites having higher LDOS than the Fe$_{y}$ sites. While the segregated Se$_{+}$ sublattice does not show modulations (third row), the segregated Fe$_{x}$/Fe$_{y}$ sublattice (second row) exhibits clear lattice periodicity when the LDOS imbalance exists ({\bf b}), but no modulation when the imbalance is absent ({\bf a}). }
\label{fig: segsimu}
\end{figure}

\noindent (S15) Examples of lattice segregation procedure (Supplementary \prettyref{fig: segsimu}).

\begin{figure}[p]
\begin{center}
    \includegraphics[width=14 cm]{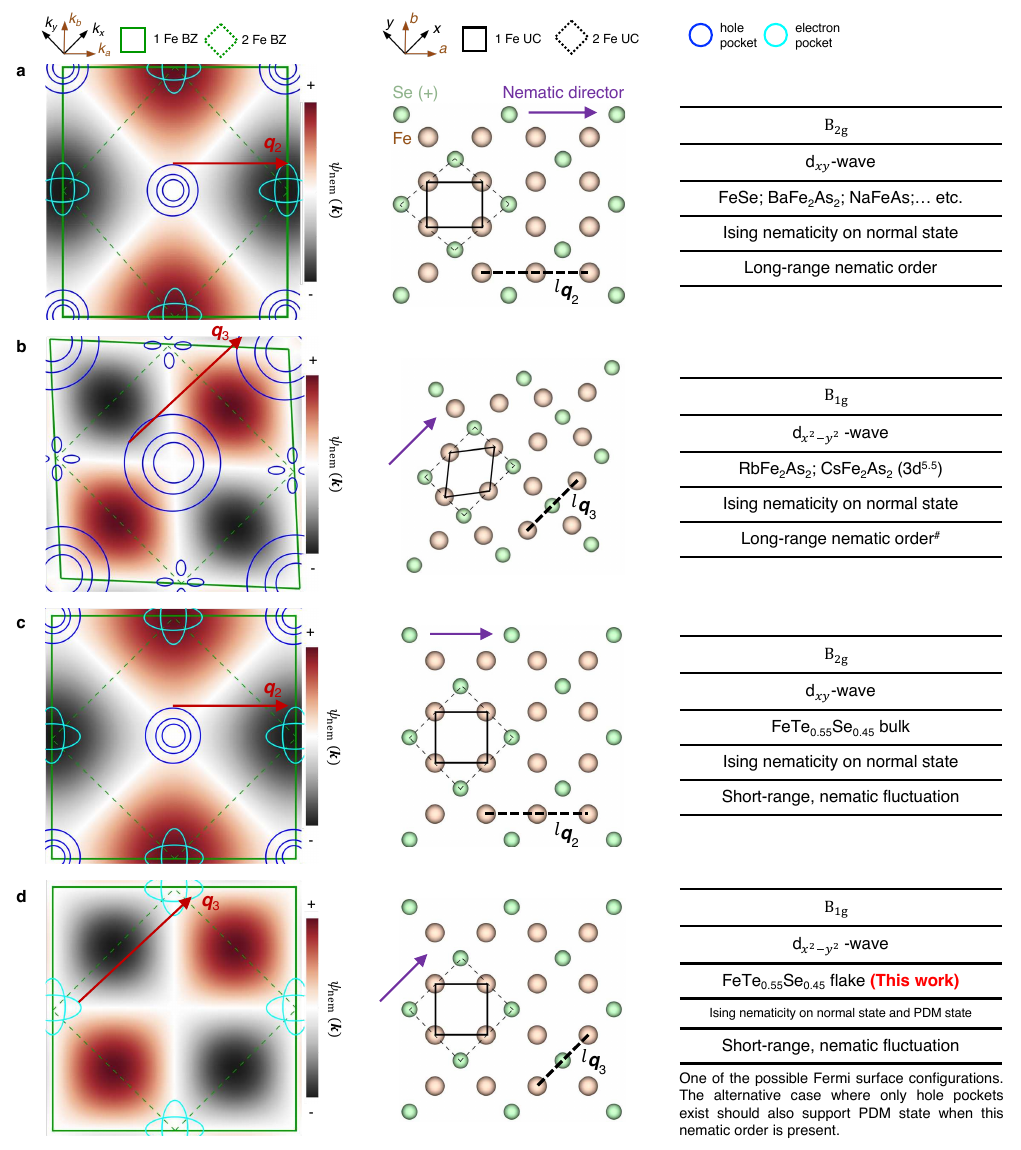}
\end{center}
\caption{
{\bf  Summary of the possible nematic distortions in FeSCs.} Left column: Schematic of Fermi surface and symmetry of nematic order parameter. The case with a distorted Brillouin zone indicates the existence of a long-range nematic order. The red arrows indicate the dominant vectors of Fermi surface nesting, which also coincide with the directions of the nematic director (indicated by purple arrows) in all the cases summarized here. Middle column: Corresponding lattice structure. Lattice distortion is shown for the cases with long-range order. Right column: summary of the relevant information about nematicity. {\bf a}, Common case of nematic order in the bulk FeSCs. {\bf b}, A rare case of nematic order in the bulk FeSCs. {\bf c}, The absence of long-range nematic order in the FeTe$_{\text{0.55}}$Se$_{\text{0.45}}$ bulk material, which is near the phase boundary of the nematic phase in Fe(Te,Se) phase diagram. {\bf d}, Nematicity in our thin flakes. In contrast to the bulk, the nematic director is rotated by 45$^{\circ}$. The $\boldsymbol{q}_2$ ($\pi$,0) and $\boldsymbol{q}_3$ ($\pi$,$\pi$) are the same vectors shown in \prettyref{fig: fig1}c and their corresponding length scale is indicated on the panels in the middle column. We stress that the nematic distortion along the nearest neighbor Fe-Fe bond (as in panel {\bf a} and {\bf c}) will not lead to the PDM state in our theory.
}
\label{fig: fignematicFeSC}
\end{figure}

\noindent (S16) Summary of the possible nematic distortions in FeSCs (Supplementary \prettyref{fig: fignematicFeSC}).

\begin{figure}[p]
\begin{center}
    \includegraphics[width=15cm]{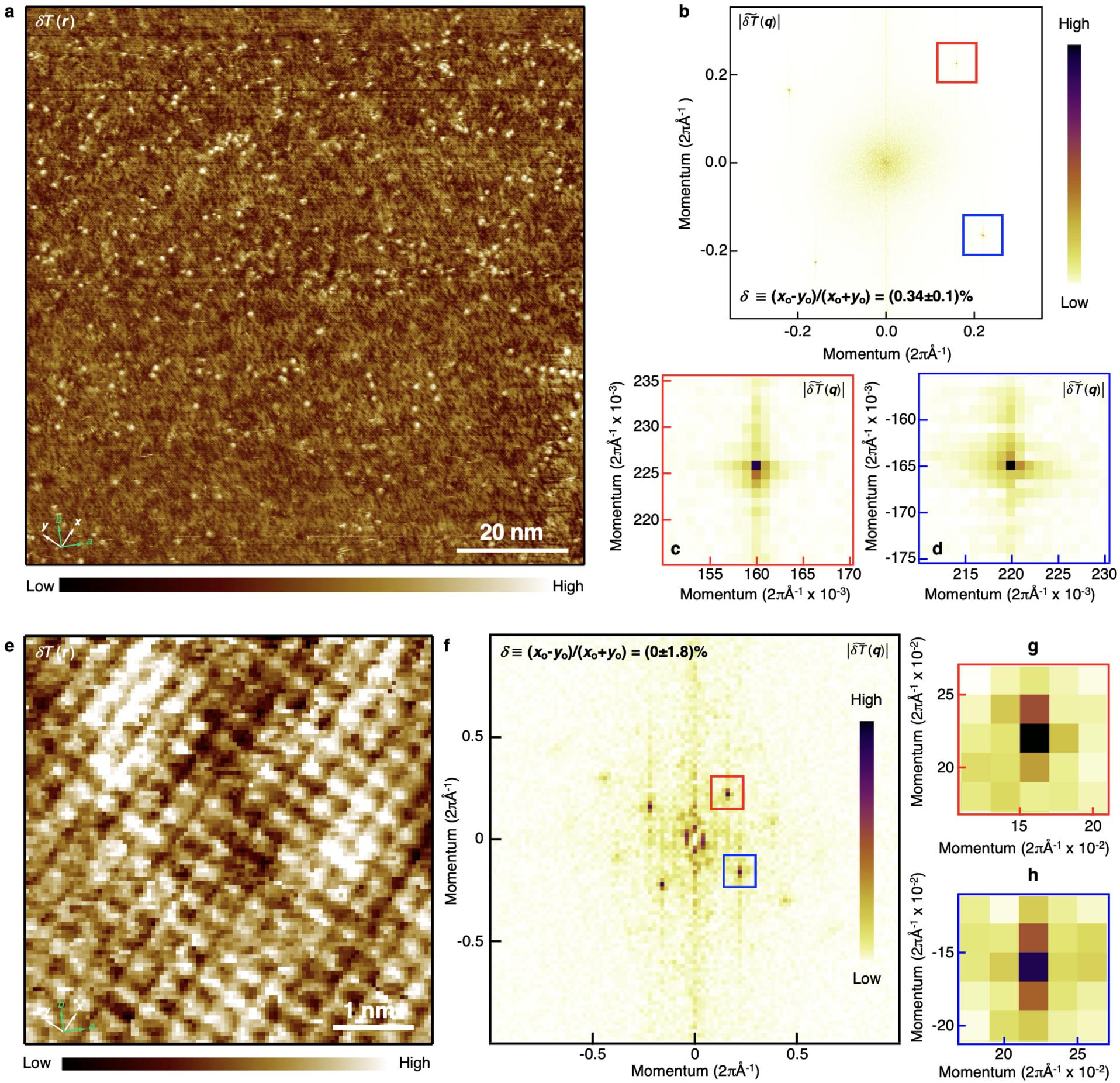}
\end{center}
\caption{{\bf Lattice orthorhombic distortion of Fe(Te,Se) thin flake.} The orthorhombic distortion is defined as $\delta \equiv (x_{\rm o}-y_{\rm o})/(x_{\rm o}+y_{\rm o})$ where $x_{\rm o}$ and $y_{\rm o}$ are lattice constants of orthorhombic unit cell (2-Fe unit cell in our case). In order to check if the nematic distortion is lattice-driven, high momentum resolution or equivalently large area scans are required. {\bf a}, 100$\times$100 nm$^{\text{2}}$ atom-resolved topography of device $\text{\#}$1. 
{\bf b}, FT analysis showing a nearly perfect square lattice, corresponding to orthorhombic distortion $\delta = (\text{0.34}\pm\text{0.1})\%$, 
with the deviation calculated as the uncertainty from the size of one reciprocal pixel. This is comparable to the value of nematoelastic coupling induced distortion, which can be up to 0.6$\%$ in the literature\cite{prozorov2009intrinsic,rossler2022nematic}, indicating an electronic-driven nematic distortion. {\bf c, d}, Zoom-in on the Bragg peaks along $\textit{x}$- and $\textit{y}$- direction respectively. {\bf e-h}, Small scale topography (5$\times$5 nm$^{\text{2}}$) showing distortion of $\delta = (\text{0}\pm\text{1.8})\%$, where SI-STM data in the main text is obtained.}
\label{fig: figorthtopo}
\end{figure}

\noindent (S17) Lattice orthorhombic distortion of Fe(Te,Se) thin flake (Supplementary \prettyref{fig: figorthtopo}).

\begin{figure}[p]
\begin{center}
    \includegraphics[width=\linewidth]{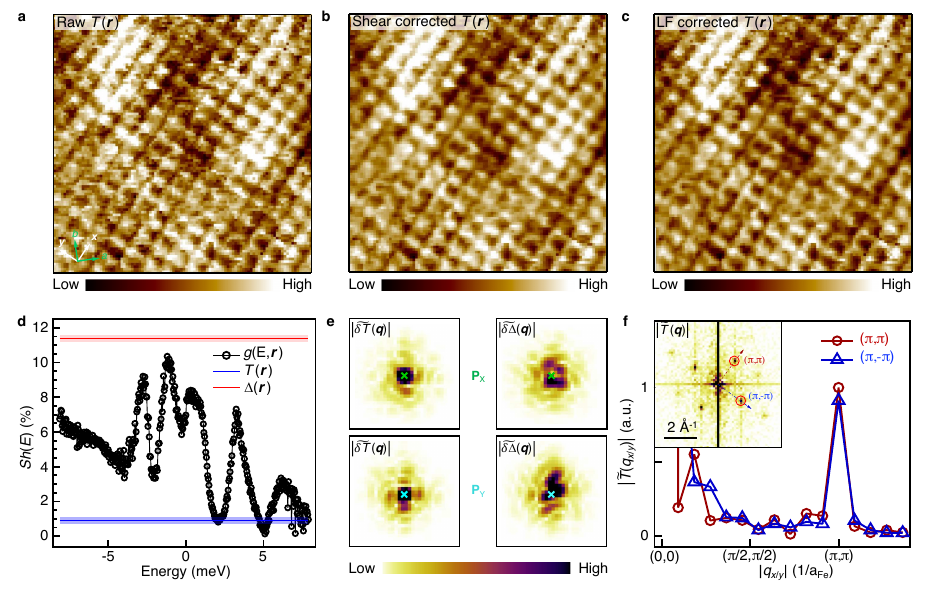}
\end{center}
\caption{
{\bf Drift and shear correction of the SI-STM data.} {\bf a}, Raw data of STM topography (same data was shown in \prettyref{fig: fig3}{b}). {\bf b}, A shear-corrected image of ({\bf a}). The shear correction method is shown in SI Section \ref{SI:distortion} and Supplementary \prettyref{fig: figShear}. {\bf c}, A drift-corrected image of ({\bf a}) by Lawler-Fujita picometer-drift correction algorithm\cite{lawler2010intra}. ({\bf a-c}) show a negligible distortion in this topography. To avoid unnecessary data processing, we used the raw data for the analysis. {\bf d}, Black dots: Energy-dependent shear strength of the differential conductance map ${\textit{Sh}(E)}$. Blue and red lines indicate shear strength of topography [0.88$\%$, $T({\boldsymbol{r}})$ in \prettyref{fig: fig3}{b}] and gap map [11.38$\%$, $\Delta(\boldsymbol{r})$ in \prettyref{fig: fig3}{a}], respectively. While the lattice distortion is negligible, our measurements indicate considerably larger distortions present on the electronic state and, concomitantly, on the PDM state, consistent with the observations of small domain size and the appearance of the topological defects. {\bf e}, Zoom in (1.2$\times$1.2 Å$^{\text{-2}}$) around the PDM vector $\boldsymbol{\rm P}_{\rm X}$ (upper row) and $\boldsymbol{\rm P}_{\rm Y}$ (lower row). While the FT magnitude of topography exhibits well ordered Bragg peaks, the case of gap map shows disordered quasi-Bragg peaks, demonstrating the PDM phase with no long-range order. {\bf f}, Line profile of FT magnitude of $T({\boldsymbol{r}})$. The red and blue lines are extracted along $(\pi,\pi)$ and $(\pi,-\pi)$ directions respectively (inset). The difference of FT magnitude of the Bragg peaks is less than 8$\%$, consistent with a nearly isotropic STM tip\cite{du2020imaging}.}
\label{fig: figcorrect}
\end{figure}

\noindent (S18) Drift and shear correction of the SI-STM data (Supplementary \prettyref{fig: figcorrect}).

\begin{figure}[p]
\begin{center}
    \includegraphics[width=\linewidth]{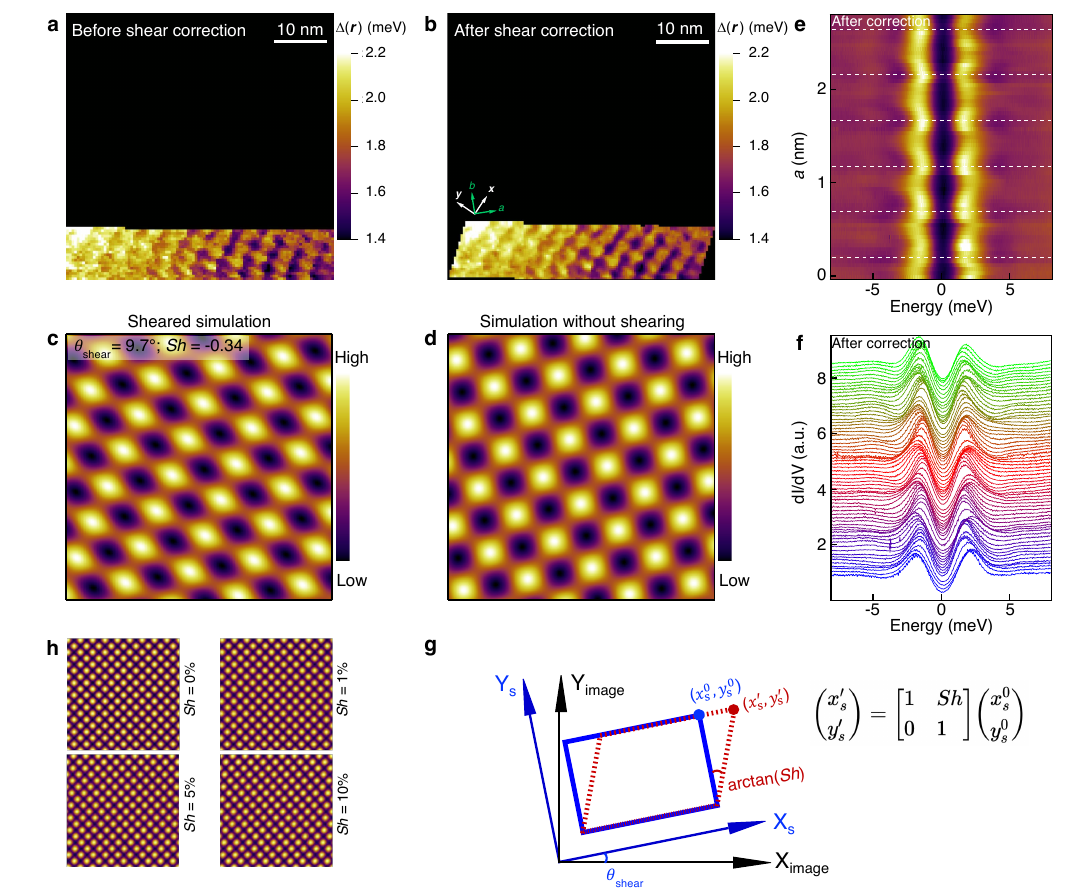}
\end{center}
\caption{
{\bf  Shear correction of a dataset with severe thermal drift.} {\bf a}, Superconducting gap map $\Delta(\boldsymbol{r})$ of a severely drifted SI-STM data before correction. {\bf b}, Shear-corrected image of ({\bf a}). The shear parameters are fitted through the corresponding $T(\boldsymbol{r})$ (not shown) measured simultaneously with the SI-STM data. {\bf c, d}, The simulated image before and after shear correction, using the parameters fitted through our shear correction codes. {\bf e, f}, False-color plot and waterfall spectrum plot of a dI/dV linecut extracted along ${\textit{a}}$-axis from the corrected SI-STM dataset. Prominent and well-ordered gap oscillation was observed between Fe$_{y}$ and Fe$_{x}$ sites. {\bf g}, An illustration of image shearing. The sheared coordinates $(x_{\text{s}}^{'},y_{\text{s}}^{'})$ can be derived by applying the shearing matrix to the original coordinates $(x_{\text{s}}^{0},y_{\text{s}}^{0})$. $\textit{Sh}$ is the shearing strength, and $\theta_{\text{shear}}$ is the shearing angle measured between the shear and the image axis. {\bf h}, Simulated lattice under different shear strengths (other parameters: lattice angle, $\alpha_{\text{lat}} = \text{45}^{\circ}, \theta_{\text{shear}}= \text{0}^{\circ}$).}
\label{fig: figShear}
\end{figure}

\noindent (S19) Shear correction of a dataset with severe thermal drift (Supplementary \prettyref{fig: figShear}).

\begin{figure}[p]
\begin{center}
    \includegraphics[width=15cm]{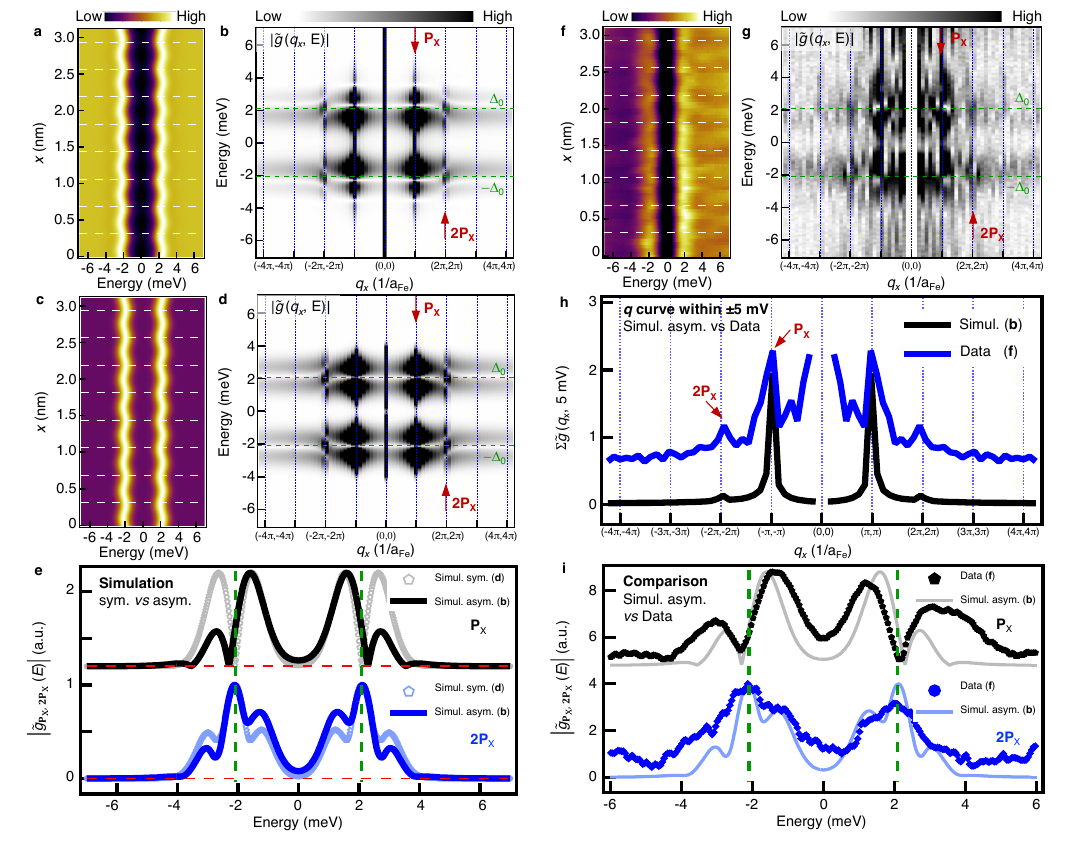}
\end{center}
\caption{{\bf PDM state induced LDOS modulation.}
{\bf a, c}, Simulation of coherence peak modulation under a PDM vector $\pm$$\boldsymbol{\rm P}_{\rm X}$, with and without asymmetric background, respectively. {\bf b, d}, Energy-dependent FT magnitude of ({\bf a}) and ({\bf c}). The most prominent LDOS modulation follows the $\pm$$\boldsymbol{\rm P}_{\rm X}$ vector of PDM state. 
There is also an additional LDOS modulation appearing at $\pm$$2\boldsymbol{\rm P}_{\rm X}$. 
{\bf e}, Energy-dependent linecut of FT magnitude at  $\boldsymbol{\rm P}_{\rm X}$ and  2$\boldsymbol{\rm P}_{\rm X}$ from ({\bf b}) and ({\bf d}). Without asymmetric background, at $\Delta_{0}$, the $\pm\boldsymbol{\rm P}_{\rm X}$ LDOS modulation is the weakest while the $\pm$2$\boldsymbol{\rm P}_{\rm X}$ LDOS modulation is the strongest (gray and blue dot lines). The background above SC gap shifts the position of local minima of $\pm\boldsymbol{\rm P}_{\rm X}$ LDOS modulation slightly away from $\Delta_{0}$, while the strongest $\pm$2$\boldsymbol{\rm P}_{\rm X}$ LDOS modulation is still right at $\Delta_{0}$ (black and blue solid lines). The red horizontal dashed lines are zero intensity of each curve. The induced LDOS modulation disappears at high energy. {\bf f}, A dI/dV linecut measured on a 35-nm thick flake (see details in Extended Data \prettyref{fig: figlinecut2}{b}). {\bf g}, Energy-dependent FT magnitude of ({\bf f}). {\bf h}, Energy-integrated FT magnitude within $\pm$5 mV from simulation ({\bf d}) and experiment ({\bf g}). Both $\pm\rm\boldsymbol{\rm P}_{\rm X}$ and $\pm$2$\rm\boldsymbol{\rm P}_{\rm X}$ LDOS modulations are observed in experiments. {\bf i}, Comparison of energy-dependent linecut of FT magnitude at  $\boldsymbol{\rm P}_{\rm X}$ (top) and  2$\boldsymbol{\rm P}_{\rm X}$ (bottom) between simulation ({\bf d}) and experiment ({\bf g}). See more details of the simulation in SI Section \ref{SI:naivesimu}. The parameters used in simulation are $\Delta_{0}$ = 2.09 meV; $|\Delta_{\boldsymbol{\rm P}_{\rm X}}|$ = 0.13 meV; ${\boldsymbol{\rm P}_{\rm X}}$ = 2$\pi$/3.8 Å$^{-1}$; $\phi_{\boldsymbol{{\rm P_{X}}}}^{\Delta}$ = $-\pi$/2; FWHM of the coherence peak (simulated by simple Gaussian function): 1.3 meV; the asymmetric background was simulated by a Fermi-Dirac function with simultaneously modulated Fermi energy, temperature used in the simulation is 4 K.}
\label{fig: figPDMinduceCDW}
\end{figure}

\noindent (S20) PDM state induced LDOS modulation (Supplementary \prettyref{fig: figPDMinduceCDW}).

\begin{figure}[p]
\begin{center}
    \includegraphics[width=11cm]{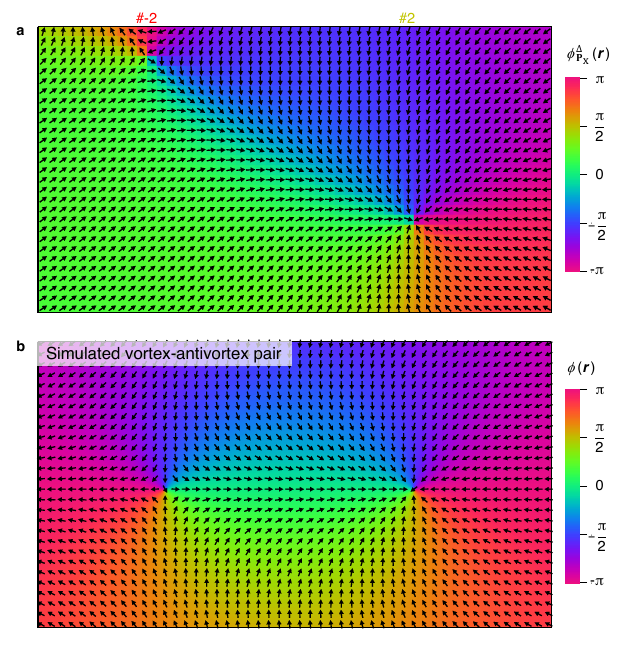}
\end{center}
\caption{{\bf Phase of a dislocation-antidislocation pair of the PDM state.}
{\bf a}, Vector field plot of the phase of a dislocation-antidislocation pair $\text{\#}$2 and $\text{\#}$-2 shown in Extended Data \prettyref{fig: topodefect}{a}. The direction of the arrows indicate the value of $\phi_{\boldsymbol{{\rm P_{X}}}}^{\Delta}$. {\bf b}, Simulation of a vortex-antivortex pair (right and left) with vorticity $m = +1$ (see details in SI Section~\ref{SI:naivesimu}).}
\label{fig: figvortexpair}
\end{figure}

\noindent (S21) Phase of a dislocation-antidislocation pair of the PDM state (Supplementary \prettyref{fig: figvortexpair}).

\begin{figure}[p]
\begin{center}
    \includegraphics[width=8cm]{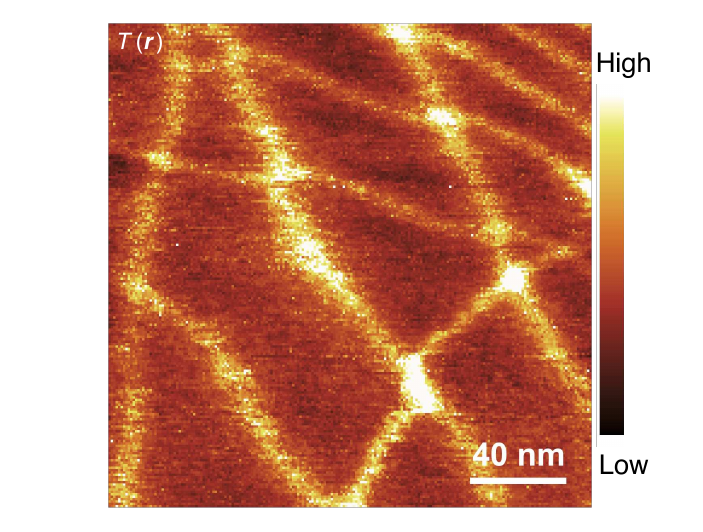}
\end{center}
\caption{
{\bf  Topography of a highly strained area.} An example of a highly strained area. In the main experiment, such areas were avoided. Setpoints: $V_{\text{bias}}$ = -100 mV, $I_{\text{t}}$ = 20 pA.}
\label{fig: figstraintopo}
\end{figure}

\noindent (S22) Topography of a highly strained area (Supplementary \prettyref{fig: figstraintopo}).

\begin{figure}[p]
\begin{center}
    \includegraphics[width=\linewidth]{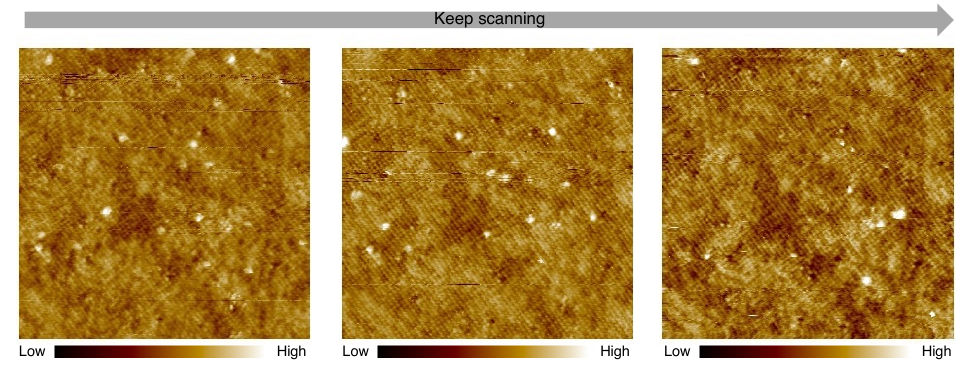}
\end{center}
\caption{
{\bf  Cleaning the surface by constant-current-mode scanning.}  Recording the motion of impurities in a 25$\times$25 nm$^{\text{2}}$ area by simple STM scanning. As shown by subsequent images, some of the 
observed impurities are very mobile. Most likely, these are introduced during sample fabrication and are different from interstitial iron impurities that are frequently observed in unannealed bulk materials.}
\label{fig: figremovedirt}
\end{figure}

\noindent (S23) Cleaning the surface by constant-current-mode scanning (Supplementary \prettyref{fig: figremovedirt}).

\begin{figure}[p]
\begin{center}
    \includegraphics[width=\linewidth]{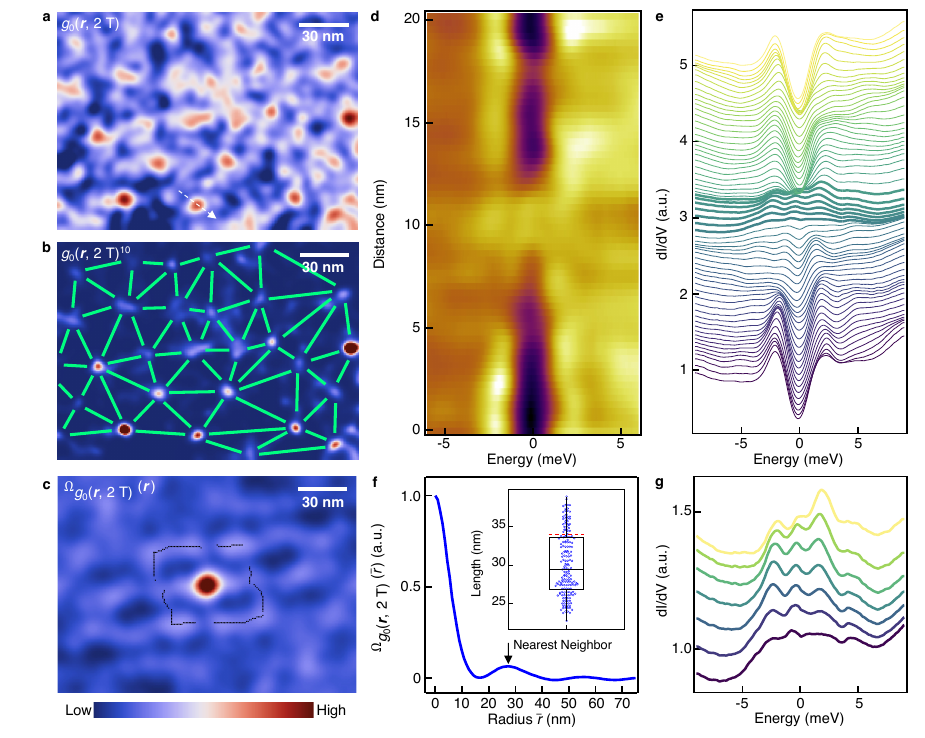}
\end{center}
\caption{
{\bf  Distorted vortex lattice on device $\text{\#}$1.} {\bf a}, Zero-bias conductance map at out-of-plane magnetic field ($B$ = 2 T). {\bf b}, To increase visibility of the lattice, the 10th power of ({\bf a}) is shown, which suppresses the background intensity and superimposed with standard Delaunay triangulation. {\bf c}, Autocorrelation of ({\bf a}). The black dots are an extraction of the positions of the nearest ring from the image center, approximately corresponding to the average vortex distance. {\bf d}, A typical dI/dV linecut measured along the white dashed arrow shown in ({\bf a}), crossing a vortex core. {\bf e}, Waterfall spectrum plot of ({\bf d}). {\bf f}, Angle-averaged radial distribution curve of ({\bf c}). Inset: Box plot of the dots in ({\bf c}). The data analysis leads to an averaged vortex distance $l_{\text{v}}$ = 28 nm, which is close to the value calculated from perfect Abrikosov vortex lattice, $l_{\text{v}}\equiv(2 \Phi_0 / \sqrt{3} B)^{\frac{1}{2}} = $ 34 nm, where $\Phi_0 = h/2e$ is magnetic flux quanta, and $B$ is the magnetic field. {\bf g}, Zoom-in into the dI/dV spectra measured at the center of the vortex core [thicker curves in ({\bf e})]. }
\label{fig: figvortexlattice}
\end{figure}

\noindent (S24) Distorted vortex lattice on device $\text{\#}$1 (Supplementary \prettyref{fig: figvortexlattice}).

\begin{figure}[p]
\begin{center}
    \includegraphics[width=\linewidth]{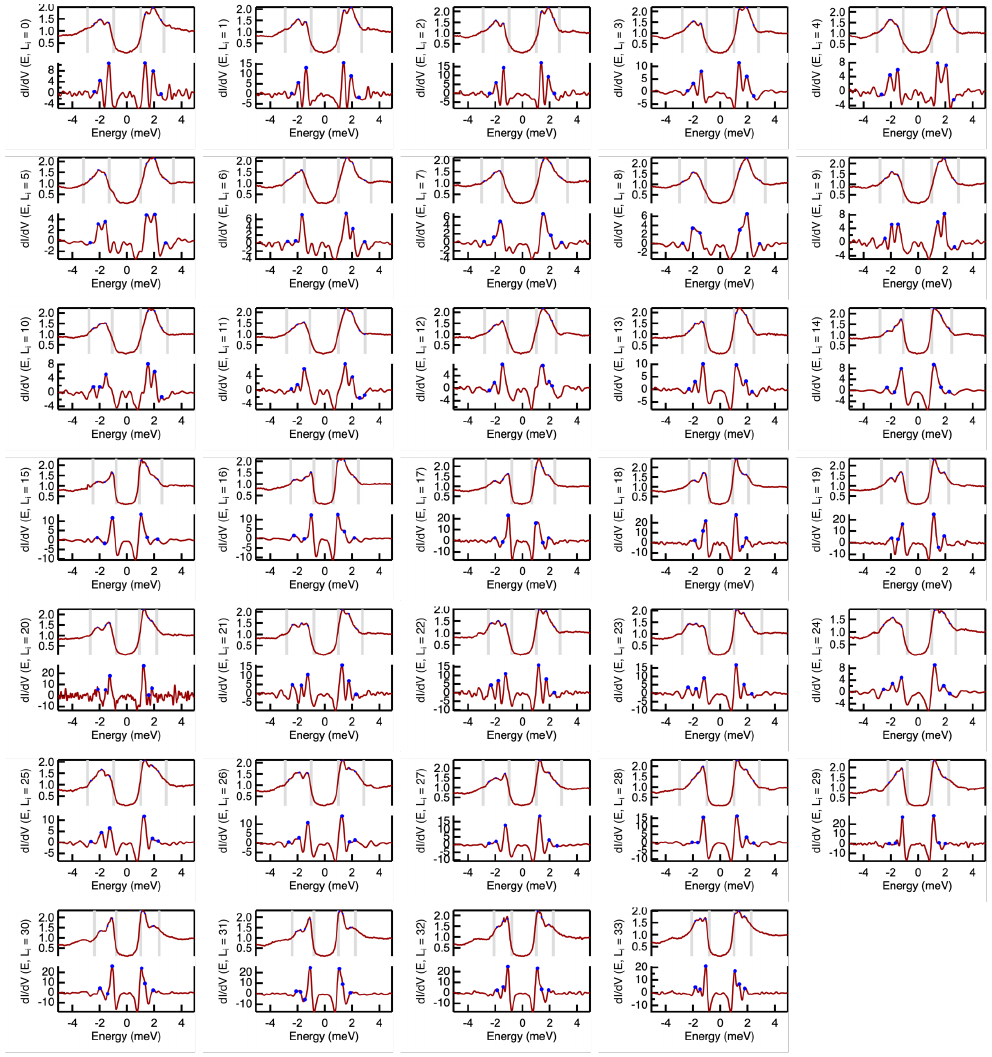}
\end{center}
\caption{
{\bf  Details of the multipeak extraction.} Extraction of energies corresponding to the spectral features near the coherence peaks measured at low electron temperature (Extended Data \prettyref{fig: figlinecutHR}). The upper curves are the raw dI/dV data, the lower curves show the negatives of the second derivative of the dI/dV spectrum. Gray solid bars indicate the energy window for the peak fitting procedure. The blue dots indicate the fitting results.}
\label{fig: figMPfit}
\end{figure}

\noindent (S25) Details of the multipeak extraction (Supplementary \prettyref{fig: figMPfit}).

\clearpage

\clearpage

\clearpage

\clearpage

\clearpage

\clearpage

\clearpage

\clearpage

\clearpage

\clearpage

\clearpage

\clearpage

\clearpage

\clearpage

\clearpage

\clearpage

\clearpage

\clearpage

\clearpage

\clearpage

\clearpage

\clearpage

\clearpage

\renewcommand{\tablename}{Supplementary Table}

\begin{table*}[]
 \begin{center}
  \caption{List of Symbols}
  
\begin{tabular} { c    l   }

      \toprule
      Symbols/Variables & Meanings \\
      \midrule
     
      Se$_{+}$, Se$_{-}$ & Top, Bottom selenium atoms.    \\
      
      Fe$_{x}$, Fe$_{y}$ &  Iron atoms at the \textit{x}-, \textit{y}- bonds of Se$_{+}$ atoms.   \\

      \(\textit{x}, \textit{y}\)  &  Lattice axis of 2-Fe unit cell (Se$_{+}$-Se$_{+}$ direction).    \\
      
      \(\textit{a}, \textit{b}\)   &  Lattice axis of 1-Fe unit cell (Fe$_{x}$-Fe$_{y}$ direction).    \\
      
      \(I_{\text{t}}\), \(V_{\text{b}}\)  &  Setpoints: tunneling current \text{\&} bias voltage.   \\

       \(\boldsymbol{{\rm P}}_{\text{X}}\), \(\boldsymbol{{\rm P}}_{\text{Y}}\) &  Modulation vectors (Bragg) of the PDM state.    \\

      \(d_{\text{t}}\),  \(t_{\text{Ar}}\) & Flake thickness \& Argon dwell time of the sample.  \\
     
     \midrule
      
      \(T(\boldsymbol{r})\),  \(\delta T(\boldsymbol{r}) \equiv T(\boldsymbol{r})-T_{0}(\boldsymbol{r})\)  &  Topography w/o or w/ background subtraction. \\

    \(|\widetilde{T}(\boldsymbol{q})|\), \(|\widetilde{\delta T}(\boldsymbol{q})|\)  &  FT Magnitude of  \( T(\boldsymbol{r})\), \(\delta T(\boldsymbol{r})\). \\

           \(\phi_{\boldsymbol{{\rm P}}_{\rm X}}^{T}(\boldsymbol{r})\),  \(\phi_{\boldsymbol{{\rm P}}_{\rm Y}}^{T}(\boldsymbol{r})\)  &  Modulation phase of topography.  \\

      \(\delta T_{_{(\boldsymbol{{\rm P}}_{\text{1}}, \boldsymbol{{\rm P}}_{\text{2}})}}(\boldsymbol{r})\equiv \mathcal{F}lt_{_{(\boldsymbol{{\rm P}}_{\text{1}}, \boldsymbol{{\rm P}}_{\text{2}})}}[\delta T(\boldsymbol{r})]\) & Filtered image of \(\delta  T(\boldsymbol{r})\) with \(\pm\boldsymbol{{\rm P}}_{\text{1,2}}\) locked.\\
    
     \midrule  
      \(g(\boldsymbol{r},E) \equiv \frac{dI}{dV}(\boldsymbol{r},E)\)   &  Constant energy map of differential conductance.    \\
      
      \(\left|\widetilde{g}\left(\boldsymbol{q}, E\right)\right|\) \footnote{\(\left|\widetilde{g}\left(\boldsymbol{q}, E\right)\right|=\sqrt{[\text{Re}\widetilde{g}(\boldsymbol{q},E)]^{2}+[\text{Im}\widetilde{g}(\boldsymbol{q},E)]^{2}}\), in which \(\widetilde{g}(\boldsymbol{q},E) \equiv \mathcal{F}[g(\boldsymbol{q},E)]\). }  &  FT Magnitude of \(g(\boldsymbol{r},E)\).   \\
      
      \(g_{_{\text{PR}}}\left(\boldsymbol{q}, E\right) \equiv\left|\widetilde{g}\left(\boldsymbol{q}, E\right)\right| \cos \left(\theta_{\boldsymbol{q}, E} -\theta_{\boldsymbol{q}, -E} \right)\) \footnote{\(\theta_{\boldsymbol{q}, E} = \text{arctan}\left[\text{Im}\widetilde{g}(\boldsymbol{q},E)/\text{Re}\widetilde{g}(\boldsymbol{q},E)\right]\), is the phase of Fourier transform of  \(g(\boldsymbol{r},E)\).}  &  Phase-referenced quasi-particle interference pattern.     \\

       \(\Sigma \widetilde{g}(\boldsymbol{q}, E) \equiv \sum_{-E}^{+E} |\widetilde{g}(\boldsymbol{q}, E)|\)  &  Energy-integrated \(\left|\widetilde{g}\left(\boldsymbol{q}, E\right)\right|\) within \([-E,E]\).     \\

    \(g_{0}(x,0)\), \(|\widetilde{g_{0}}(\boldsymbol{q}_{x},0)|\)   &  Line profile of zero-bias \(\frac{dI}{dV}\) and its FT magnitude. \\

    \(H(x)\), \(|\widetilde{H}(\boldsymbol{q}_{x})|\)   &  \(\frac{dI}{dV}\) line profile of gap peak, and its FT magnitude. \\

    \(\Sigma g(\boldsymbol{r}, E_{1}/E_{2})\equiv\sum_{E_{1}}^{E_{2}} g(\boldsymbol{r}, E)\) &  Energy-integrated \(g\left(\boldsymbol{r}, E\right)\) within \([E_{1},E_{2}]\).   \\

    \(|\widetilde{\Sigma g}(\boldsymbol{q},E_{1}/E_{2})|\)   &  FT Magnitude of \(\Sigma g(\boldsymbol{r}, E_{1}/E_{2})\).  \\

    \(\Sigma g(\boldsymbol{r_{\text{Fe}}}, E_{1}/E_{2})\), \(\Sigma g(\boldsymbol{r_{\text{Se}_{+}}}, E_{1}/E_{2})\) &  Lattice segregation of \(\Sigma g(\boldsymbol{r}, E_{1}/E_{2})\).  \\

     \(|\widetilde{\Sigma g}(\boldsymbol{q_{\text{Fe}}}, E_{1}/E_{2})|\), \(|\widetilde{\Sigma g}(\boldsymbol{q_{\text{Se}_{+}}}, E_{1}/E_{2})|\) &  FT Magnitude of \(\Sigma g(\boldsymbol{r_{\text{Fe/Se}_{+}}}, E_{1}/E_{2})\).  \\

    \(\Sigma g_{_{(\boldsymbol{{\rm P}}_{\text{1}}, \boldsymbol{{\rm P}}_{\text{2}})}}(\boldsymbol{r}, E_{1}/E_{2})\equiv \mathcal{F}lt_{_{(\boldsymbol{{\rm P}}_{\text{1}}, \boldsymbol{{\rm P}}_{\text{2}})}}[\Sigma g(\boldsymbol{r}, E_{1}/E_{2})]\) &  Filtered image of \(\Sigma g(\boldsymbol{r}, E_{1}/E_{2})\) with \(\pm\boldsymbol{{\rm P}}_{\text{1,2}}\) locked. \\

     \midrule  
      \(\Delta_{0}(\boldsymbol{r})\)  &  Non-modulating SC gap.    \\
    
      \(|\Delta_{\boldsymbol{{\rm P}}_{\rm X}}(\boldsymbol{r})|\), \(|\Delta_{\boldsymbol{{\rm P}}_{\rm Y}}(\boldsymbol{r})|\)  &  Modulation amplitude of the PDM state.    \\
      \(\phi_{\boldsymbol{{\rm P}}_{\rm X}}^{\Delta}(\boldsymbol{r})\),  \(\phi_{\boldsymbol{{\rm P}}_{\rm Y}}^{\Delta}(\boldsymbol{r})\)  &  Modulation phase of the PDM state.    \\

  \(\delta \Delta(\boldsymbol{r}) \equiv \sum_{_{\boldsymbol{{\rm Q}}= \boldsymbol{{\rm P}}_{\rm X},\boldsymbol{{\rm P}}_{\rm Y}}}\left|\Delta_{\boldsymbol{{\rm Q}}}(\boldsymbol{r})\right| \cos \left[\boldsymbol{{\rm Q}} \cdot \boldsymbol{r}+\phi_{\boldsymbol{{\rm Q}}}^{\Delta}(\boldsymbol{r})\right]\)  &  Modulating component of the PDM state. \\

    \(\Delta(\boldsymbol{r}) \equiv \Delta_{0}(\boldsymbol{r})+\delta \Delta(\boldsymbol{r})\)  & Map of the total superconducting gap.    \\

    \(|\widetilde{\Delta}(\boldsymbol{q})|\), \(|\widetilde{\delta \Delta}(\boldsymbol{q})|\)  &  FT Magnitude of  \( \Delta(\boldsymbol{r})\), \(\delta \Delta(\boldsymbol{r})\). \\

      \(\delta \Delta_{_{(\boldsymbol{{\rm P}}_{\text{1}}, \boldsymbol{{\rm P}}_{\text{2}})}}(\boldsymbol{r})\equiv \mathcal{F}lt_{_{(\boldsymbol{{\rm P}}_{\text{1}}, \boldsymbol{{\rm P}}_{\text{2}})}}[\delta \Delta(\boldsymbol{r})]\) &  Filtered image of \(\delta\Delta(\boldsymbol{r})\) with \(\pm\boldsymbol{{\rm P}}_{\text{1,2}}\) locked. \\

      \(\delta \Delta_{_{\boldsymbol{{\rm P}}_{\text{1}}}}(\boldsymbol{r})\equiv \mathcal{F}lt_{_{\boldsymbol{{\rm P}}_{\text{1}}}}[\delta \Delta(\boldsymbol{r})]\) &  Filtered image of \(\delta\Delta(\boldsymbol{r})\) with only \(\pm\boldsymbol{{\rm P}}_{\text{1}}\) locked. \\

    \midrule  

    \(p_{_{\text{LL}}}(\boldsymbol{r})\) & Lattice-lock-in polarization of the PDM state. \\
    
    \(p_{_{\text{nem}}}\equiv\frac{|\Delta_{\boldsymbol{{\rm P}}_{\rm X}}(r)|-|\Delta_{\boldsymbol{{\rm P}}_{\rm Y}}(r)|}{|\Delta_{\boldsymbol{{\rm P}}_{\rm X}}(r)|+|\Delta_{\boldsymbol{{\rm P}}_{\rm Y}}(r)|}\) & Nematic polarization of the PDM state.\\

    \(\Omega_{|\Delta_{\boldsymbol{{\rm P}}_{\rm X}}|}(\boldsymbol{r})\), \(\Omega_{|\Delta_{\boldsymbol{{\rm P}}_{\rm Y}}|}(\boldsymbol{r})\), \(\Omega_{|p_{_{\text{LL}}}|}(\boldsymbol{r})\) & Auto-correlation of the quantities in the subscript.\\ 

    \(\Omega_{A}(\Bar{r})\), \(g(\Bar{r})\) & Angle-averaged radial distribution curves.\\

    $\left|\Delta_{\boldsymbol{{\rm Q}}}(\boldsymbol{r})\right|/\Delta_{0}(\boldsymbol{r}) $ & Gap modulation ratio.\\

    $p_{g}(E)\equiv \frac{g\left[\bar{\boldsymbol{r}}_{\text{Fe}({\Delta_{\text{max}}})},E\right]-g\left[\bar{\boldsymbol{r}}_{\text{Fe}({\Delta_{\text{min}}})},E\right]}{g\left[\bar{\boldsymbol{r}}_{\text{Fe}({\Delta_{\text{max}}})},E\right]+g\left[\bar{\boldsymbol{r}}_{\text{Fe}({\Delta_{\text{min}}})},E\right]}$ & LDOS imbalance ratio between Fe$_{x/y}$ and Fe$_{y/x}$. \\

    $p_{_{\Delta}}(\boldsymbol{r})\equiv \frac{\left|\Delta_{\text{Fe}_{x}}-\Delta_{\text{Fe}_{y}}\right|}{\Delta_{\text{Fe}_{x}}+\Delta_{\text{Fe}_{y}}}(\boldsymbol{r}) = \frac{|\Delta_{\boldsymbol{{\rm P}}_{\rm X}}(\boldsymbol{r})|+|\Delta_{\boldsymbol{{\rm P}}_{\rm Y}}(\boldsymbol{r})|}{\Delta_{0}(\boldsymbol{r})}$ & Gap difference between Fe$_{x}$ and Fe$_{y}$. \\

      \bottomrule
    \end{tabular}
    \label{table: tabs1}
  \end{center}
\end{table*}

\end{document}